\documentclass[]{article}
\usepackage{fullpage} %more standardized margins

\usepackage{multirow}
\usepackage[bottom]{footmisc}% places footnotes at page bottom
\usepackage[round]{natbib}
\usepackage[small]{caption}
\usepackage{amsmath} % the standard math package
\usepackage{amsfonts} % the standard math package
\def\uupsilon{\pmb{\upsilon}}
\def\llambda{\pmb{\lambda}}
\def\bbeta{\pmb{\beta}}
\def\aalpha{\pmb{\alpha}}
\def\zzeta{\pmb{\zeta}}

\def\xixi{\mbox{\boldmath $\xi$}}

\def\LAM{\mbox{\boldmath $\Lambda$}}
\def\LAMm{\mathbb{L}}

\def\OMG{\mbox{\boldmath $\Omega$}}

\def\UPS{\mbox{\boldmath $\Upsilon$}}

\def\AA{\mbox{$\mathbf A$}}	\def\aa{\mbox{$\mathbf a$}} \def\Ab{\mbox{$\mathbf F$}} \def\Aa{\mbox{$\mathbf h$}} 
\def\BB{\mbox{$\mathbf B$}}	\def\bb{\mbox{$\mathbf b$}} \def\Bb{\mbox{$\mathbf J$}} \def\Ba{\mbox{$\mathbf L$}} \def\Bm{\UPS}
\def\CC{\mbox{$\mathbf C$}}	\def\cc{\mbox{$\mathbf c$}}
 
\def\DD{\mbox{$\mathbf D$}}	\def\dd{\mbox{$\mathbf d$}}
\def\EE{\mbox{$\mathbf E$}}	
\def\E{\,\textup{\textrm{E}}}	
\def\EXy{\,\textup{\textrm{E}}_{\text{{\bf XY}}}}
\def\FF{\mbox{$\mathbf F$}} \def\ff{\mbox{$\mathbf f$}}
\def\GG{\mbox{$\mathbf G$}}	\def\gg{\mbox{$\mathbf g$}}
\def\HH{\mbox{$\mathbf H$}}	\def\hh{\mbox{$\mathbf h$}}
\def\II{\mbox{$\mathbf I$}}
\def\IIm{\mbox{$\mathbf I$}}
\def\JJ{\mbox{$\mathbf J$}}
\def\KK{\mbox{$\mathbf K$}}
\def\LL{\mbox{$\mathbf L$}}	\def\ll{\mbox{$\mathbf l$}}
\def\MM{\mbox{$\mathbf M$}}  \def\mm{\mbox{$\mathbf m$}}

\def\MVN{\,\textup{\textrm{MVN}}}
\def\OO{\mbox{$\mathbf O$}}
\def\PP{\mbox{$\mathbf P$}}  \def\pp{\mbox{$\mathbf p$}}
\def\QQ{\mbox{$\mathbf Q$}}	 \def\qq{\mbox{$\mathbf q$}} \def\Qb{\mbox{$\mathbf G$}}  \def\Qm{\mathbb{Q}}
\def\RR{\mbox{$\mathbf R$}}	 \def\rr{\mbox{$\mathbf r$}} \def\Rb{\mbox{$\mathbf H$}}	\def\Rm{\mathbb{R}}
\def\SS{\mbox{$\mathbf S$}}
\def\UU{\mbox{$\mathbf U$}}	\def\uu{\mbox{$\mathbf u$}} \def\Ub{\mbox{$\mathbf C$}} \def\Ua{\mbox{$\mathbf g$}} 
\def\VV{\mbox{$\mathbf V$}}	\def\vv{\mbox{$\mathbf v$}}
\def\WW{\mbox{$\mathbf W$}}	\def\ww{\mbox{$\mathbf w$}}
\def\XX{\mbox{$\pmb{X}$}}	\def\xx{\mbox{$\pmb{x}$}}
\def\YY{\mbox{$\pmb{Y}$}}	\def\yy{\mbox{$\pmb{y}$}}
\def\ZZ{\mbox{$\mathbf Z$}}		\def\Zb{\mbox{$\mathbf M$}} \def\Za{\mbox{$\mathbf N$}} 

\def\vec{\,\textup{\textrm{vec}}}
\def\var{\,\textup{\textrm{var}}}
\def\cov{\,\textup{\textrm{cov}}}

\def\trace{\,\textup{\textrm{trace}}}
\def\hatxt{\widetilde{\mbox{$\mathbf x$}}_t}
\def\hatxone{\widetilde{\mbox{$\mathbf x$}}_1}

\def\hatxtm{\widetilde{\mbox{$\mathbf x$}}_{t-1}}

\def\hatyt{\widetilde{\mbox{$\mathbf y$}}_t}

\def\hatyone{\widetilde{\mbox{$\mathbf y$}}_1}

\def\hatOt{\widetilde{\OO}_t}
\def\hatWt{\widetilde{\WW}_t}
\def\hatYXt{\widetilde{\mbox{$\mathbf{y}\mathbf{x}$}}_t}
\def\hatYXttm{\widetilde{\mbox{$\mathbf{y}\mathbf{x}$}}_{t,t-1}}
\def\hatPt{\widetilde{\PP}_t}
\def\hatPtm{\widetilde{\PP}_{t-1}}

\def\hatPttm{\widetilde{\PP}_{t,t-1}}

\def\hatPtmt{\widetilde{\PP}_{t-1,t}}
\def\hatVt{\widetilde{\VV}_t}
\def\hatVttm{\widetilde{\VV}_{t,t-1}}

\def\IR{\nabla}
\usepackage[round]{natbib} % to get references that are like in ecology papers
\makeatletter
\renewcommand*\env@matrix[1][*\c@MaxMatrixCols c]{%
  \hskip -\arraycolsep
  \let\@ifnextchar\new@ifnextchar
  \array{#1}}
\makeatother
\begin{document}
\author{Elizabeth Eli Holmes\footnote{Northwest Fisheries Science Center, NOAA Fisheries, Seattle, WA 98112, 
       eli.holmes@noaa.gov, http://faculty.washington.edu/eeholmes}}
\title{Derivation of an EM algorithm for constrained and unconstrained multivariate autoregressive state-space (MARSS) models}
\maketitle
\begin{abstract}
This report presents an Expectation-Maximization (EM) algorithm for estimation of the maximum-likelihood parameter values of constrained multivariate autoregressive Gaussian state-space (MARSS) models.  The MARSS model can be written: x(t)=Bx(t-1)+u+w(t), y(t)=Zx(t)+a+v(t), where w(t) and v(t) are multivariate normal error-terms with variance-covariance matrices Q and R respectively.  MARSS models are a class of dynamic linear model and vector autoregressive model state-space model. Shumway and Stoffer presented an unconstrained EM algorithm for this class of models in 1982, and a number of researchers have presented EM algorithms for specific types of constrained MARSS models since then.  In this report, I present a general EM algorithm for constrained MARSS models, where the constraints are on the elements within the paramater matrices (B,u,Q,Z,a,R). The constraints take the form vec(M)=f+Dm, where M is the parameter matrix, f is a column vector of fixed values, D is a matrix of multipliers, and m is the column vector of estimated values.  This allows a wide variety of constrained parameter matrix forms.  The presentation is for a time-varying MARSS model, where time-variation enters through the fixed (meaning not estimated) f(t) and D(t) matrices for each parameter.  The algorithm allows missing values in y and partially deterministic systems where 0s appear on the diagonals of Q or R.
\end{abstract}
\vfill
{\noindent \tiny citation: Holmes, E. E. 2012. Derivation of an EM algorithm for constrained and unconstrained multivariate autoregressive state-space (MARSS) models. }
 \newpage
\section{Overview}

EM algorithms extend maximum-likelihood estimation to models with hidden states and are widely used in engineering and computer science applications. This report presents an EM algorithm for a general class of Gaussian constrained multivariate autoregressive state-space (MARSS) models, with a hidden multivariate autoregressive process (state) model and a multivariate observation model.  This is an important class of time-series model used in many different scientific fields. The reader is referred to \citet{McLachlanKrishnan2008} for general background on EM algorithms and to \citet{Harvey1989} for a discussion of EM algorithms for time-series data.  \citet{Borman2009} has a nice tutorial on the EM algorithm.  

Before showing the derivation for the constrained case, I first show a derivation of the EM algorithm  for unconstrained\footnote{``unconstrained'' means that each element in the parameter matrix is estimated and no elements are fixed or shared.} MARSS model. This EM algorithm was published by \citet{ShumwayStoffer1982}, but my derivation is more similar to Ghahramani et al's \citep{GhahramaniHinton1996, RoweisGhahramani1999} slightly different presentation.  One difference in my presentation and all these previous presentations, however, is that I treat the data as a random variable throughout; this means that there are no ``special" update equations for the missing values case.  Another difference is that I present the update equations for both stochastic initial states and fixed initial states.  I then extend the derivation to constrained MARSS models where there are fixed and shared elements in the parameter matrices and to the case of degenerate MARSS models where some processes in the model are deterministic rather than stochastic. See also \citet{Wuetal1996} and \citet{Zuuretal2003a} for other examples of the EM algorithm for different classes of constrained MARSS models.

When working with MARSS models, one should be cognizant that misspecification of the prior on the initial hidden states can have catastrophic and difficult to detect effects on the parameter estimates.  There is often no sign that something is amiss with the MLE estimates output by an EM algorithm.  There has been much work on how to avoid these initial conditions effects; see especially literature on vector autoregressive state-space models in the economics literature.  The trouble often occurs when the prior on the initial states is inconsistent with the distribution of the initial states that is implied by the maximum-likelihood model.  This often happens when the model implies a specific covariance structure on the initial states, but since the maximum-likelihood parameters are unknown, this covariance structure is unknown.  Using a diffuse prior does not help since your diffuse prior still has some covariance structure (often independence is being imposed).  In some ways the EM algorithm is less sensitive to a mis-specified prior because it uses the smoothed states conditioned on all the data.  However, if the prior is inconsistent with the model, the EM algorithm will not (cannot) find the MLEs.  It is very possible however that it will find parameter estimates that are closer to what you intend (estimates uninfluenced by the prior), but they will not be MLEs.  The derivation presented here allows one to circumvent these problems by treating the initial states as fixed (and estimated) parameters.  The problematic initial state variance-covariance matrix is removed from the model, albeit at the cost of additional estimated parameters.

Finally, when working with MARSS models, one needs to ensure that the model is identifiable, i.e.  a unique solution exists.  For a given MARSS model,  some of the parameter elements will need to be fixed (not estimated) in order to produce a model with one solution.  How to do that depends on the MARSS model being fitted and is up to the user.

\subsection{The MARSS model}

The linear MARSS model with a stochastic initial state\footnote{`Stochastic' means the initial state has a distribution rather than a fixed value. Because the process must start somewhere, one needs to specify the initial state. In equation \ref{eq:MARSS}, I show the initial state specified as a distribution.  However, the derivation will also discuss the case where the initial state is specified as an unknown fixed parameter.} is
\begin{subequations}\label{eq:MARSS}
\begin{gather}
\xx_t = \BB\xx_{t-1} + \uu + \ww_t, \text{ where } \WW_t \sim \MVN(0,\QQ) \label{eq:MARSSx}\\
\yy_t = \ZZ\xx_t + \aa + \vv_t, \text{ where } \VV_t \sim \MVN(0,\RR) \label{eq:MARSSy}\\
\XX_0 \sim \MVN(\xixi,\LAM) \label{eq:MARSSx1}
\end{gather}
\end{subequations}
The $\yy$ equation is called the observation process, and $\yy_t$ is a $n \times 1$ vector.  The $\xx$ equation is called the state or process equation, and $\xx_t$ is a $m \times 1$ vector. The equation for $\xx$ describes a multivariate autoregressive process (also called a random walk or Markov process). $\ww$ are the process errors and are specific realizations of the random variable $\WW$; $\vv$ is defined similarly.  The initial state can either defined at $t=0$, as is done in equation \ref{eq:MARSS}, or at $t=1$.  When presenting the MARSS model, I use $t=0$ but the derivations will show the EM algorithm for both cases. $\QQ$ and $\RR$ are variance-covariance matrices that specify the stochasticity in the observation and state equations.  

In the MARSS model, $\xx$ and $\yy$ equations describe two stochastic processes.  By tradition, one conditions on observations of $\yy$, and $\xx$ is treated as completely hidden, hence the name `hidden Markov process' of which a MARSS model is a special type.  However, you could condition on (partial) observations of $\xx$ and treat $\yy$ as a (partially) hidden process---with as usual proper constraints to ensure identifiability.  Nonetheless in this report, I follow tradition and treat $\xx$ as hidden and $\yy$ as (partially) observed.  If $\xx$ is partially observed then the update equations stay the same but the expectations shown in section \ref{sec:compexpectations} would be computed conditioned on the partially observed $\xx$.

The first part of this report will review the derivation of an EM algorithm for the time-constant MARSS model (equation \ref{eq:MARSS}).  However the main objective of this report is to show the derivation of an EM algorithm to solve a much more general MARSS model  (section \ref{sec:tvMARSS}), which is a MARSS model with linear constraints on time-varying parameters: 
\begin{equation}\label{eq:MARSS.ex}
\begin{gathered}
\xx_t = \BB_t\xx_{t-1} + \uu_t + \GG_t\ww_t, \text{ where } \WW_t \sim \mathrm{MVN}(0,\QQ_t)\\
\yy_t = \ZZ_t\xx_t + \aa_t + \HH_t\vv_t, \text{ where } \VV_t \sim \mathrm{MVN}(0,\RR_t)\\
\xx_{t_0} = \xixi + \FF\ll, \text{ where }  \ll \sim \mathrm{MVN}(0,\LAM)
\end{gathered}
\end{equation}
The linear constraints appear as the vectorization of each parameter ($\BB$, $\uu$, $\QQ$, $\ZZ$, $\aa$, $\RR$, $\xixi$, $\LAM$) is described by the relation $\ff_t+\DD_t\mm$. This relation  specifies linear constraints of the form $\beta_i + \beta_{a,i} a + \beta_{b,i} b + \dots$ on the elements in each MARSS parameter matrix. Equation \eqref{eq:MARSS.ex} is a much broader class of MARSS models that includes MARSS models with exogenous variable (covariates), AR-p models, moving average models, constrained MARSS models and models that are combinations of these. The derivation also includes partially deterministic systems where $\GG_t$, $\HH_t$ and $\FF$ may have all zero rows.

\subsection{The joint log-likelihood function}
Denote the set of all $y$'s and $x$'s from $t=1$ to $T$ by $\yy$ and $\xx$. The joint log-likelihood\footnote{This is not the log likelihood output by the Kalman filter.  The log likelihood output by the Kalman filter is the $\log\LL(\yy;\Theta)$ (notice $\xx$ does not appear), which is known as the marginal log likelihood.} of $\yy$ and $\xx$ can then be written then as follows\footnote{The log-likelihood function is shown here for the MARSS with non-time varying parameters (equation \ref{eq:MARSS}).}, where $\XX_t$ denotes the random variable and $\xx_t$ is a realization from that random variable (and similarly for $\YY_t$):\footnote{To alleviate clutter, I have left off subscripts on the $f$'s.  To emphasize that the $f$'s represent different density functions, one would often use a subscript showing what parameters are in the functions, i.e. $f(\xx_t|\XX_{t-1}=\xx_{t-1})$ becomes $f_{B,u,Q}(\xx_t|\XX_{t-1}=\xx_{t-1})$.}
\begin{equation}
f(\yy,\xx) = f(\yy|\XX=\xx)f(\xx),
\end{equation}
where
\begin{equation}
\begin{split}
f(\xx)&=f(\xx_0)\prod_{t=1}^T f(\xx_t|\XX_1^{t-1}=\xx_1^{t-1})\\
f(\yy|\XX=\xx) &= \prod_{t=1}^T f(\yy_t|\XX=\xx)
\end{split}
\end{equation}
Thus,
\begin{equation}\label{eq:jointL}
\begin{split}f(\yy,\xx) &= \prod_{t=1}^T f(\yy_t|\XX=\xx) \times f(\xx_0)\prod_{t=1}^T f(\xx_t|\XX_1^{t-1}=\xx_1^{t-1}) \\
&=\prod_{t=1}^T f(\yy_t|\XX_t=\xx_t) \times f(\xx_0)\prod_{t=1}^T f(\xx_t|\XX_{t-1}=\xx_{t-1}).
\end{split}
\end{equation}
Here $\xx_{t1}^{t2}$ denotes the set of $\xx_t$ from $t=t1$ to $t=t2$ (and thus $\xx$ is shorthand for $\xx_1^T$).  The third line follows because conditioned on $\xx$, the $\yy_t$'s are independent of each other (because the $\vv_t$ are independent of each other).  In the last line, $\xx_1^{t-1}$ becomes $\xx_{t-1}$ from the Markov property of the equation for $\xx_t$ (equation \ref{eq:MARSSx}), and $\xx$ becomes $\xx_t$ because $\yy_t$ depends only on $\xx_t$ (equation \ref{eq:MARSSy}).

Since $(\XX_t|\XX_{t-1}=\xx_{t-1})$ is multivariate normal and $(\YY_t|\XX_t=\xx_t)$ is multivariate normal (equation \ref{eq:MARSS}), we can write down the joint log-likelihood function using the likelihood function for a multivariate normal distribution \citep[sec. 4.3]{JohnsonWichern2007}.  
\begin{equation}\label{eq:logL}
\begin{split}
&\log\LL(\yy,\xx ; \Theta) = -\sum_1^T \frac{1}{2}(\yy_t - \ZZ \xx_t - \aa)^\top \RR^{-1} (\yy_t - \ZZ \xx_t - \aa) -\sum_1^T\frac{1}{2} \log |\RR|\\
&\quad  -\sum_1^T \frac{1}{2} (\xx_t - \BB \xx_{t-1} - \uu)^\top \QQ^{-1} (\xx_t - \BB \xx_{t-1} - \uu) - \sum_1^T\frac{1}{2}\log |\QQ|\\
&\quad  -\frac{1}{2}(\xx_0 - \xixi)^\top \LAM^{-1}(\xx_0 - \xixi) - \frac{1}{2}\log |\LAM| -   \frac{n}{2}\log 2\pi 
\end{split}
\end{equation}
$n$ is the number of data points. This is the same as equation 6.64 in \citet{ShumwayStoffer2006}. The above equation is for the case where $\xx_0$ is stochastic (has a known distribution).  However, if we instead treat $\xx_0$ as fixed but unknown (section 3.4.4 in Harvey, 1989), it is then a parameter and there is no $\LAM$.  The likelihood then is slightly different.  $\xx_0$ is defined as a parameter $\xixi$ and
\begin{equation}\label{eq:logL.V0.is.0}
\begin{split}
&\log\LL(\yy,\xx ; \Theta) = -\sum_1^T \frac{1}{2}(\yy_t - \ZZ \xx_t - \aa)^\top \RR^{-1} (\yy_t - \ZZ \xx_t - \aa) -\sum_1^T\frac{1}{2} \log |\RR|\\
&\quad  -\sum_1^T \frac{1}{2} (\xx_t - \BB \xx_{t-1} - \uu)^\top \QQ^{-1} (\xx_t - \BB \xx_{t-1} - \uu) - \sum_1^T\frac{1}{2}\log |\QQ|
\end{split}
\end{equation}
Note that in this case, $\xx_0$ is no longer a realization of a random variable $\XX_0$; it is a fixed (but unknown) parameter.  Equation \ref{eq:logL.V0.is.0} is written as if all the $\xx_0$ are fixed, however when the general derivation is presented, it allowed that some $\xx_0$ are fixed ($\LAM$=0) and others are stochastic.

If $\RR$ is constant through time, then $\sum_1^T\frac{1}{2} \log |\RR|$ in the likelihood equation reduces to $\frac{T}{2}\log |\RR|$, however sometimes one needs to includes time-dependent weighting on $\RR$\footnote{If for example, one wanted to include a temporally dependent weighting on $\RR$ replace $|\RR|$ with $|\alpha_t\RR|=\alpha_t^n|\RR|$, where $\alpha_t$ is the weighting at time $t$ and is fixed not estimated.}.  The same applies to $\sum_1^T\frac{1}{2}\log |\QQ|$.

All bolded elements are column vectors (lower case) and matrices (upper case).  $\AA^\top$ is the transpose of matrix $\AA$, $\AA^{-1}$ is the inverse of $\AA$, and $|\AA|$ is the determinant of $\AA$.  Parameters are non-italic while elements that are slanted are realizations of a random variable ($\xx$ and $\yy$ are slated)\footnote{In matrix algebra, a capitol bolded letter indicates a matrix.  Unfortunately in statistics, the capitol letter convention is used for random variables.  Fortunately, this derivation does not need to reference random variables except indirectly when using expectations.  Thus, I use capitols to refer to matrices not random variables.  The one exception is the reference to $\XX$ and $\YY$. In this case a bolded {\it slanted} capitol is used.}
 
\subsection{Missing values}\label{sec:missing}
In Shumway and Stoffer and other presentations of the EM algorithm for MARSS models \citep{ShumwayStoffer2006,Zuuretal2003a}, the missing values case is treated separately from the non-missing values case.  In these derivations, a series of modifications are given for the EM update equations when there are missing values.  In my derivation, I present the missing values treatment differently, and there is only one set of update equations and these equations apply in both the missing values and non-missing values cases. My derivation does this by keeping $\E[\YY_t|\text{data}]$ and $\E[\YY_t\XX_t^\top|\text{data}]$ in the update equations (much like $\E[\XX_t|\text{data}]$ is kept in the equations) while Shumway and Stoffer replace these expectations involving $\YY_t$ by their values, which depend on whether or not the data are a complete observation of $\YY_t$ with no missing values.  Section \ref{sec:compexpectations} shows how to compute the expectations involving $\YY_t$ when the data are an incomplete observation of $\YY_t$.

\section{The EM algorithm}
The EM algorithm cycles iteratively between an expectation step (the integration in the equation) followed by a maximization step (the arg max in the equation):
\begin{equation}\label{eq:EMalg}
\Theta_{j+1} = \arg \underset{\Theta}{\max} \int_{\xx}{\int_{\yy}{\log\LL(\xx,\yy;\Theta) f(\xx,\yy|\YY(1)=\yy(1),\Theta_j)d\xx d\yy}}
\end{equation}
$\YY(1)$ indicates those $\YY$ that have an observation and $\yy(1)$ are the actual observations. Note that $\Theta$ and $\Theta_j$ are different.  If $\Theta$ consists of multiple parameters, we can also break this down into smaller steps.  Let $\Theta=\{\alpha,\beta\}$, then
\begin{equation}\label{eq:EMalg.j}
\alpha_{j+1} = \arg \underset{\alpha}{\max} \int_{\xx}{\int_{\yy}{\log\LL(\xx,\yy,\beta_j;\alpha) f(\xx,\yy|\YY(1)=\yy(1),\alpha_j,\beta_j)d\xx d\yy}}
\end{equation}
Now the maximization is only over $\alpha$, the part that appears after the ``;'' in the log-likelihood.

\textbf{Expectation step} The integral that appears in equation \eqref{eq:EMalg} is an expectation. The first step in the EM algorithm is to compute this expectation.  This will involve computing expectations like $\E[\XX_t\XX_t^\top|\YY_t(1)=\yy_t(1),\Theta_j]$ and $\E[\YY_t\XX_t^\top|\YY_t(1)=\yy_t(1),\Theta_j]$. The $j$ subscript on $\Theta$ denotes that these are the parameters at iteration $j$ of the algorithm.

\textbf{Maximization step}: A new parameter set $\Theta_{j+1}$ is computed by finding the parameters that maximize the \textit{expected} log-likelihood function (the part in the integral) with respect to $\Theta$.  The equations that give the parameters for the next iteration ($j+1$) are called the update equations and this report is devoted to the derivation of these update equations.

After one iteration of the expectation and maximization steps, the cycle is then repeated. New expectations  are computed using $\Theta_{j+1}$, and then a new set of parameters $\Theta_{j+2}$ is generated.  This cycle is continued until the likelihood no longer increases more than a specified tolerance level.   This algorithm is guaranteed to increase in likelihood at each iteration (if it does not, it means there is an error in one's update equations).  The algorithm must be started from an initial set of parameter values $\Theta_1$.  The algorithm is not particularly sensitive to the initial conditions but the surface could definitely be multi-modal and have local maxima.  See section \ref{sec:implementation} on using Monte Carlo initialization to ensure that the global maximum is found.

\subsection{The expected log-likelihood function}\label{sec:expLL}
The function that is maximized in the ``M'' step is the expected value of the log-likelihood function. This expectation is conditioned on two things: 1) the observed $\YY$'s which are denoted $\YY(1)$ and which are equal to the fixed values $\yy(1)$ and 2) the parameter set $\Theta_j$.  Note that since there may be missing values in the data, $\YY(1)$ can be a subset of $\YY$, that is, only some $\YY$ have a corresponding $\yy$ value at time $t$.  Mathematically what we are doing is $\EXy[g(\XX,\YY)|\YY(1)=\yy(1),\Theta_j]$.  This is a multivariate conditional expectation because $\XX,\YY$ is multivariate (a $m \times n \times T$ vector). The function $g(\Theta)$ that we are taking the expectation of is $\log\LL(\YY,\XX ; \Theta)$. Note that $g(\Theta)$ is a random variable involving the random variables, $\XX$ and $\YY$, while $\log\LL(\yy,\xx ; \Theta)$ is not a random variable but rather a specific value since $\yy$ and $\xx$ are a set of specific values.

We denote this expected log-likelihood by $\Psi$. The goal is to find the $\Theta$ that maximize $\Psi$ and this becomes the new $\Theta$ for the  $j+1$ iteration of the EM algorithm.  The equations to compute the new $\Theta$ are termed the update equations.  Using the log likelihood equation \eqref{eq:logL} and expanding out all the terms, we can write out $\Psi$  in verbose form as:
\begin{equation}\label{eq:expLL}
\begin{split}
&\EXy[\log\LL(\YY,\XX ; \Theta);\YY(1)=\yy(1),\Theta_j] = \Psi = \\
&\quad -\frac{1}{2}\sum_1^T\bigg( \E[\YY_t^\top \RR^{-1} \YY_t] - \E[\YY_t^\top \RR^{-1}\ZZ\XX_t] - \E[(\ZZ\XX_t)^\top\RR^{-1}\YY_t] - \E[\aa^\top\RR^{-1}\YY_t] - \E[\YY_t^\top\RR^{-1}\aa]\\
&\quad   + \E[(\ZZ\XX_t)^\top\RR^{-1}\ZZ\XX_t] + \E[\aa^\top\RR^{-1}\ZZ\XX_t] + \E[(\ZZ\XX_t)^\top\RR^{-1}\aa] + \E[\aa^\top\RR^{-1}\aa]\bigg) 
 - \frac{T}{2}\log|\RR|\\
&\quad - \frac{1}{2}\sum_1^T\bigg(\E[\XX_t^\top\QQ^{-1}\XX_t] - \E[\XX_t^\top\QQ^{-1}\BB\XX_{t-1}]  - \E[(\BB\XX_{t-1})^\top\QQ^{-1}\XX_t]\\ 
&\quad - \E[\uu^\top\QQ^{-1}\XX_t] - \E[\XX_t^\top\QQ^{-1}\uu] + \E[(\BB\XX_{t-1})^\top\QQ^{-1}\BB\XX_{t-1}]\\
&\quad  + \E[\uu^\top\QQ^{-1}\BB\XX_{t-1}] + \E[(\BB\XX_{t-1})^\top\QQ^{-1}\uu] + \uu^\top\QQ^{-1}\uu\bigg) - \frac{T}{2}\log|\QQ| \\
&\quad - \frac{1}{2}\bigg(\E[\XX_0^\top\VV_0^{-1}\XX_0] - \E[\xixi^\top\LAM^{-1}\XX_0] - \E[\XX_0^\top\LAM^{-1}\xixi] + \xixi^\top\LAM^{-1}\xixi\bigg) - \frac{1}{2}\log|\LAM|
-\frac{n}{2}\log\pi
\end{split}
\end{equation}
All the $\E[\quad]$ appearing here denote $\EXy[g()|\YY(1)=\yy(1),\Theta_j]$.  In the rest of the derivation, I drop the conditional and the $XY$ subscript on $\E$ to remove clutter, but it is important to remember that whenever $\E$ appears, it refers to a specific conditional multivariate expectation.  If $\xx_0$ is treated as fixed, then $\XX_0=\xixi$ and the last two lines involving $\LAM$ are dropped.

Keep in mind that $\Theta$ and $\Theta_j$ are different.  $\Theta$ is a parameter appearing in function $g(\XX,\YY,\Theta)$ (i.e. the parameters in equation \ref{eq:logL}).  $\XX$ and $\YY$ are random variables which means that $g(\XX,\YY,\Theta)$ is a random variable.  We take the expectation of $g(\XX,\YY,\Theta)$, meaning we take integral over the joint distribution of $\XX$ and $\YY$.  We need to specify what that distribution is and the conditioning on $\Theta_j$ (meaning the $\Theta_j$ appearing to the right of the $|$ in $\E(g()|\Theta_j)$) is specifying this distribution. This conditioning affects the value of the expectation of $g(\XX,\YY,\Theta)$, but it does not affect the value of $\Theta$, which are the $\RR$, $\QQ$, $\uu$, etc. values on the right side of equation \eqref{eq:expLL}.  We will first take the expectation of $g(\XX,\YY,\Theta)$ conditioned on $\Theta_j$ (using integration) and then take the differential of that expectation with respect to $\Theta$.

\subsection{The expectations used in the derivation}\label{sec:expectations}
The following expectations appear frequently in the update equations and are given special names\footnote{This notation is different than what you see in Shumway and Stoffer (2006), section 6.2.  What I call $\hatVt$, they refer to as $P_t^n$, and my $\hatPt$ would be $P_t^n + \hatxt \hatxt^\prime$ in their notation.}:
\begin{subequations}\label{eq:expectations}
\begin{align}
&\hatxt = \EXy[\XX_t | \YY(1)=\yy(1), \Theta_j]\\
&\hatyt = \EXy[\YY_t | \YY(1)=\yy(1), \Theta_j]\\
&\hatPt=\EXy[\XX_t\XX_t^\top | \YY(1)=\yy(1), \Theta_j]\\
&\hatPttm=\EXy[\XX_{t}\XX_{t-1}^\top | \YY(1)=\yy(1), \Theta_j]\\
&\hatVt = \var_{XY}[\XX_t|\YY(1)=\yy(1), \Theta_j] = \hatPt-\hatxt\hatxt^\top\label{eq:hatVt}\\
&\hatOt=\EXy[\YY_t\YY_t^\top | \YY(1)=\yy(1), \Theta_j]\\
&\hatWt = \var_{XY}[\YY_t|\YY(1)=\yy(1), \Theta_j] = \hatOt-\hatyt\hatyt^\top\label{eq:hatWt}\\
&\hatYXt = \EXy[\YY_t\XX_t^\top| \YY(1)=\yy(1), \Theta_j]\\
&\hatYXttm = \EXy[\YY_t\XX_{t-1}^\top| \YY(1)=\yy(1), \Theta_j]
\end{align}
\end{subequations}
The subscript on the expectation, $\E$, denotes that this is a multivariate expectation taken over $\XX$ and $\YY$.  The right sides of equations \eqref{eq:hatVt} and \eqref{eq:hatWt} arise from the computational formula for variance and covariance: 
\begin{align}\label{eq:comp.formula.variance}
\var[X] &= \E[XX^\top] - \E[X]\E[X]^\top\\
\cov[X,Y] &= \E[XY^\top] - \E[X]\E[Y]^\top.	
\end{align}
Section \ref{sec:compexpectations} shows how to compute the expectations in equation \ref{eq:expectations}.

\begin{table}
	\caption{Notes on multivariate expectations.  For the following examples, let $\XX$ be a vector of length three, $X_1,X_2,X_3$. $f()$ is the probability distribution function (pdf). $C$ is a constant (not a random variable).}
	\label{tab:MultivariateExpectations}
\begin{center}\begin{tabular}{lr}
\hline\\
$\E_X[g(\XX)]=\int{\int{\int{g(\xx)f(x_1,x_2,x_3) dx_1 dx_2 dx_3}}}$\\
$\E_X[X_1]=\int{\int{\int{x_1 f(x_1,x_2,x_3) dx_1 dx_2 dx_3}}}=\int{x_1 f(x_1) dx_1}=\E[X_1]$ \\
$\E_X[X_1+X_2]=\E_X[X_1]+\E_X[X_2]$\\
$\E_X[X_1+C]=\E_X[X_1]+C$\\
$\E_X[C X_1]=C\E_X[X_1]$\\
$\E_X[\XX|\XX=\xx]=\xx$ \\
\\
\hline
\end{tabular}
\end{center}
\end{table}

\section{The unconstrained update equations}\label{sec:generalupdate}
In this section, I show the derivation of the update equations when all elements of a parameter matrix are estimated and are all allowed to be different, i.e. the unconstrained case. These are similar to the update equations one will see in \citet{ShumwayStoffer2006}.  Section \ref{sec:constrained} shows the update equations when there are unestimated (fixed) or estimated but shared values in the parameter matrices, i.e. the constrained update equations.   

To derive the update equations, one must find the $\Theta$, where $\Theta$ is comprised of the MARSS parameters $\BB$, $\uu$, $\QQ$, $\ZZ$, $\aa$, $\RR$, $\xixi$, and $\LAM$,  that maximizes $\Psi$ (equation \ref{eq:expLL}) by partial differentiation of $\Psi$  with respect to $\Theta$.  However, I will be using the EM equation where one maximizes each parameter matrix in $\Theta$ one-by-one (equation \ref{eq:EMalg.j}).  In this case, the parameters that are not being maximized are set at their iteration $j$ values, and then one takes the derivative of $\Psi$ with respect to the parameter of interest.  Then solve for the parameter value that sets the partial derivative to zero.  The partial differentiation is with respect to each individual parameter element, for example each $u_{i,j}$ in matrix $\uu$. The idea is to single out those terms in equation \eqref{eq:expLL} that involve $u_{i,j}$ (say), differentiate by $u_{i,j}$, set this to zero and solve for $u_{i,j}$.  This gives the new $u_{i,j}$ that maximizes the partial derivative with respect to $u_{i,j}$ of the expected log-likelihood.  Matrix calculus gives us a way to jointly maximize $\Psi$ with respect to all elements (not just element $i,j$) in a parameter matrix. 

\subsection{Matrix calculus need for the derivation}\label{sec:MatrixDerivatives}
Before commencing, some definitions from matrix calculus will be needed.  The partial derivative of a scalar ($\Psi$ is a scalar) with respect to some column vector $\bb$ (which has elements $b_1$, $b_2$ . . .) is 
\begin{equation*}
\frac{\partial\Psi}{\partial\bb}=
\begin{bmatrix}
\dfrac{\partial\Psi}{\partial b_1}& \dfrac{\partial\Psi}{\partial b_2}& \cdots& \dfrac{\partial\Psi}{\partial b_n}
\end{bmatrix}
\end{equation*}
 Note that the derivative of a column vector $\bb$ is a row vector. The partial derivatives of a scalar with respect to some $n \times n$ matrix $\BB$ is
\begin{equation*}
\frac{\partial\Psi}{\partial\BB}=
\begin{bmatrix}
\dfrac{\partial\Psi}{\partial b_{1,1}}& \dfrac{\partial\Psi}{\partial b_{2,1}}& \cdots& \dfrac{\partial\Psi}{\partial b_{n,1}}\\
\\
\dfrac{\partial\Psi}{\partial b_{1,2}}& \dfrac{\partial\Psi}{\partial b_{2,2}}& \cdots& \dfrac{\partial\Psi}{\partial b_{n,2}}\\
\\
\cdots&  \cdots&  \cdots&  \cdots\\
\\
\dfrac{\partial\Psi}{\partial b_{1,n}}& \dfrac{\partial\Psi}{\partial b_{2,n}}& \cdots& \dfrac{\partial\Psi}{\partial b_{n,n}}\\
\end{bmatrix}
\end{equation*} 
Note that the indexing is interchanged; $\partial\Psi/\partial b_{i,j}=\big[\partial\Psi/\partial\BB\big]_{j,i}$. For $\QQ$ and $\RR$, this is unimportant because they are variance-covariance matrices and are symmetric. For $\BB$ and $\ZZ$, one must be careful because these may not be symmetric. 

A number of derivatives of a scalar with respect to vectors and matrices will be needed in the derivation and are shown in table \ref{tab:MatrixDerivatives}.  In the table, both the vectorized and non-vectorized versions are shown. The vectorized version of a matrix $\DD$ with dimension $n \times m$ is
\begin{gather*}
\vec(\DD_{n,m})\equiv
\begin{bmatrix}
d_{1,1}\\
\cdots\\
d_{n,1}\\
d_{1,2}\\
\cdots\\
d_{n,2}\\
\cdots\\
d_{1,m}\\
\cdots\\
d_{n,m}
\end{bmatrix}\\
\end{gather*}

\begin{table}
	\caption{Derivatives of a scalar with respect to vectors and matrices.  In the following $\aa$ and $\cc$ are $n \times 1$ column vectors, $\bb$ and $\dd$ are $m \times 1$ column vectors, $\DD$ is a $n \times m$ matrix, $\CC$ is a $n \times n$ matrix, and $\AA$ is a diagonal $n \times n$ matrix (0s on the off-diagonals).  $\CC^{-1}$ is the inverse of $\CC$, $\CC^\top$ is the transpose of $\CC$, $\CC^{-\top} = \big(\CC^{-1}\big)^\top = \big(\CC^\top\big)^{-1}$, and $|\CC|$ is the determinant of $\CC$. Note, all the numerators in the differentials reduce to scalars.  Although the matrix names may be the same as in the text, these matrices are dummy matrices to show the matrix derivative relations.}
	\label{tab:MatrixDerivatives}
\begin{center}\begin{tabular}{lr}
\hline
\\
\refstepcounter{equation}\label{eq:derivaTc}
$\partial(\aa^\top\cc)/\partial\aa = \partial(\cc^\top\aa)/\partial\aa = \cc^\top$
& (\theequation) \\
\\
\refstepcounter{equation}\label{eq:derivaTDb}
$\partial(\aa^\top\DD\bb)/\partial\DD = \partial(\bb^\top\DD^\top\aa)/\partial\DD = \bb\aa^\top$
& \multirow{2}{*}{(\theequation)} \\
$\partial(\aa^\top\DD\bb)/\partial\vec(\DD) = \partial(\bb^\top\DD^\top\aa)/\partial\vec(\DD) = \big(\vec(\bb\aa^\top)\big)^\top$
&\\
\\
\refstepcounter{equation}\label{eq:derivlogDet}
$\partial(\log |\CC|)/\partial\CC = -\partial(\log |\CC^{-1}|)/\partial\CC=(\CC^\top)^{-1} = \CC^{-\top}$
& \multirow{2}{*}{(\theequation)} \\
$\partial(\log |\CC|)/\partial\vec(\CC) = \big(\vec(\CC^{-\top})\big)^\top$& \\
\\
\refstepcounter{equation}\label{eq:derivbDTCDd}
$\partial(\bb^\top\DD^\top\CC\DD\dd)/\partial\DD = \dd\bb^\top\DD^\top\CC + \bb\dd^\top\DD^\top\CC^\top$ 
& \multirow{3}{*}{(\theequation)} \\
$\partial(\bb^\top\DD^\top\CC\DD\dd)/\partial\vec(\DD) = 
\big(\vec(\dd\bb^\top\DD^\top\CC + \bb\dd^\top\DD^\top\CC^\top)\big)^\top $ &\\
If $\bb=\dd$ and $\CC$ is symmetric then the sum reduces to $2\bb\bb^\top\DD^\top\CC$ & \\
\\
\refstepcounter{equation}\label{eq:derivaTCa}
$\partial(\aa^\top\CC\aa)/\partial\aa = \partial(\aa\CC^\top\aa^\top)/\partial\aa = 2\aa^\top\CC$
& (\theequation) \\
\\
\refstepcounter{equation}\label{eq:derivInv}
$\partial(\aa^\top\CC^{-1}\cc)/\partial\CC = -\CC^{-1}\aa\cc^\top\CC^{-1} $
& \multirow{2}{*}{(\theequation)} \\
$\partial(\aa^\top\CC^{-1}\cc)/\partial\vec(\CC) = -\big(\vec(\CC^{-1}\aa\cc^\top\CC^{-1})\big)^\top$ & \\
\\
\hline
\end{tabular}\end{center}
\end{table}

\subsection{The update equation for $\uu$ (unconstrained)}
Take the partial derivative of $\Psi$ with respect to $\uu$, which is a $m \times 1$ matrix. All parameters other than $\uu$ are fixed to constant values (because partial derivation is being done).  Since the derivative of a constant is 0, terms not involving $\uu$ will equal 0 and drop out.  Taking the derivative to equation \eqref{eq:expLL} with respect to $\uu$:
\begin{equation}\label{eq:u.unconstrained1}
\begin{split}
&\partial\Psi/\partial\uu = - \frac{1}{2}\sum_{t=1}^T \bigg(-  \partial(\E[\XX_t^\top\QQ^{-1}\uu])/\partial\uu
- \partial(\E[\uu^\top\QQ^{-1}\XX_t])/\partial\uu \\
&\quad + \partial(\E[(\BB\XX_{t-1})^\top\QQ^{-1}\uu])/\partial\uu 
+ \partial(\E[\uu^\top\QQ^{-1}\BB\XX_{t-1}])/\partial\uu + \partial(\uu^\top\QQ^{-1}\uu)/\partial\uu \bigg)
\end{split}
\end{equation}
The parameters can be moved out of the expectations and then the matrix derivative relations (table \ref{tab:MatrixDerivatives}) are used to take the derivative. 
\begin{equation}\label{eq:u.unconstrained2}
\begin{split}
\partial\Psi/\partial\uu = - \frac{1}{2}\sum_{t=1}^T\bigg(- \E[\XX_t]^\top\QQ^{-1} 
- \E[\XX_t]^\top\QQ^{-1} + (\BB\E[\XX_{t-1}])^\top\QQ^{-1} + (\BB\E[\XX_{t-1}])^\top\QQ^{-1} + 2\uu^\top\QQ^{-1} \bigg)
\end{split}
\end{equation}
This also uses $\QQ^{-1} = (\QQ^{-1})^\top$. This can then be reduced to 
\begin{equation}\label{eq:u.unconstrained3}
\begin{split}
&\partial\Psi/\partial\uu = \sum_{t=1}^T\big(\E[\XX_t]^\top\QQ^{-1} 
 - \E[\XX_{t-1}]^\top\BB^\top\QQ^{-1} -  \uu^\top\QQ^{-1} \big)
\end{split}
\end{equation}
Set the left side to zero (a $p \times m$ matrix of zeros) and transpose the whole equation. $\QQ^{-1}$ cancels out\footnote{$\QQ$ is a variance-covariance matrix and is invertible. $\QQ^{-1}\QQ=\II$, the identity matrix.} by multiplying on the left by $\QQ$ (left since the whole equation was just transposed), giving
\begin{equation}\label{eq:u.unconstrained4}
\mathbf{0} = \sum_{t=1}^T\big(\E[\XX_t] - \BB\E[\XX_{t-1}] - \uu \big)
= \sum_{t=1}^T\big(\E[\XX_t] - \BB\E[\XX_{t-1}] \big) - \uu
\end{equation}
Solving for $\uu$ and replacing the expectations with their names from equation \ref{eq:expectations}, gives us the new $\uu$ that maximizes $\Psi$, 
\begin{equation}\label{eq:uupdate.unconstrained}
\uu_{j+1} =  \frac{1}{T} \sum_{t=1}^T\big(\hatxt - \BB\hatxtm \big)
\end{equation}

\subsection{The update equation for $\BB$ (unconstrained)}
Take the derivative of $\Psi$ with respect to $\BB$.  Terms not involving $\BB$, equal 0 and drop out.  I have put the $\E$ outside the partials by noting that $\partial(\E[h(\XX_t,\BB)])/\partial\BB=\E[\partial(h(\XX_t,\BB))/\partial\BB]$ since the expectation is conditioned on $\BB_j$ not $\BB$. 
\begin{equation}\label{eq:B.unconstrained1}
\begin{split}
&\partial\Psi/\partial\BB = -\frac{1}{2} \sum_{t=1}^T\bigg(-\E[\partial(\XX_t^\top\QQ^{-1}\BB\XX_{t-1})/\partial\BB] \\
&\quad - \E[\partial((\BB\XX_{t-1})^\top\QQ^{-1}\XX_t)/\partial\BB] + \E[\partial((\BB\XX_{t-1})^\top\QQ^{-1}(\BB\XX_{t-1}))/\partial\BB] \\
&\quad +  \E[\partial((\BB\XX_{t-1})^\top\QQ^{-1}\uu)/\partial\BB] 
+ \E[\partial(\uu^\top\QQ^{-1}\BB\XX_{t-1})/\partial\BB]\bigg)\\
&= -\frac{1}{2} \sum_{t=1}^T\bigg(-\E[\partial(\XX_t^\top\QQ^{-1}\BB\XX_{t-1}])/\partial\BB] \\
&\quad - \E[\partial(\XX_{t-1}^\top\BB^\top\QQ^{-1}\XX_t)/\partial\BB] 
+ \E[\partial(\XX_{t-1}^\top\BB^\top\QQ^{-1}(\BB\XX_{t-1}))/\partial\BB] \\
&\quad +  \E[\partial(\XX_{t-1}^\top\BB^\top\QQ^{-1}\uu)/\partial\BB] 
+ \E[\partial(\uu^\top\QQ^{-1}\BB\XX_{t-1})/\partial\BB\bigg)]\\
\end{split}
\end{equation}
After pulling the constants out of the expectations, we use relations \eqref{eq:derivaTDb} and \eqref{eq:derivbDTCDd} to take the derivative and note that $\QQ^{-1} = (\QQ^{-1})^\top$:
\begin{equation}\label{eq:B.unconstrained2}
\begin{split}
&\partial\Psi/\partial\BB = -\frac{1}{2} \sum_{t=1}^T\bigg(-\E[\XX_{t-1}\XX_t^\top]\QQ^{-1} - \E[ \XX_{t-1}\XX_t^\top]\QQ^{-1} \\
&\quad + 2 \E[\XX_{t-1}\XX_{t-1}^\top]\BB^\top\QQ^{-1} + \E[\XX_{t-1}]\uu^\top\QQ^{-1}  + \E[\XX_{t-1}]\uu^\top\QQ^{-1} \bigg) \\
\end{split}
\end{equation}
This can be reduced to
\begin{equation}\label{eq:B.unconstrained3}
\partial\Psi/\partial\BB = -\frac{1}{2} \sum_{t=1}^T\bigg(-2\E[\XX_{t-1}\XX_t^\top]\QQ^{-1}  + 2 \E[\XX_{t-1}\XX_{t-1}^\top ]\BB^\top\QQ^{-1} 
+ 2\E[\XX_{t-1}]\uu^\top\QQ^{-1} \bigg)
\end{equation}
Set the left side to zero (an $m \times m$ matrix of zeros), cancel out $\QQ^{-1}$ by multiplying by $\QQ$ on the right, get rid of the -1/2, and transpose the whole equation to give
\begin{equation}\label{eq:B.unconstrained4}
\begin{split}
&\mathbf{0}  =  \sum_{t=1}^T\big(\E[\XX_t\XX_{t-1}^\top] - \BB \E[\XX_{t-1}\XX_{t-1}^\top] - \uu \E[\XX_{t-1}^\top]\big)\\ 
&\quad =  \sum_{t=1}^T \big( \hatPttm  - \BB \hatPtm - \uu^\top\hatxtm^\top \big)
\end{split}
\end{equation}
The last line replaced the expectations  with their names shown in  equation \eqref{eq:expectations}.
Solving for $\BB$ and noting that $\hatPtm$ is like a variance-covariance matrix and is invertible, gives us the new $\BB$ that maximizes $\Psi$, 
\begin{equation}\label{eq:B.update.unconstrained}
\BB_{j+1}= \bigg( \sum_{t=1}^T \big( \hatPttm  - \uu^\top\hatxtm^\top \big)\bigg) \bigg(\sum_{t=1}^T \hatPtm\bigg)^{-1}
\end{equation}

Because all the equations above also apply to block-diagonal matrices, the derivation immediately generalizes to the case where $\BB$ is an unconstrained block diagonal matrix:
\begin{equation*}
\BB =
\begin{bmatrix}
b_{1,1}&b_{1,2}&b_{1,3}&0&0&0&0&0\\
b_{2,1}&b_{2,2}&b_{2,3}&0&0&0&0&0\\
b_{3,1}&b_{3,2}&b_{3,3}&0&0&0&0&0\\
0&0&0&b_{4,4}&b_{4,5}&0&0&0\\
0&0&0&b_{5,4}&b_{5,5}&0&0&0\\
0&0&0&0&0&b_{6,6}&b_{6,7}&b_{6,8}\\
0&0&0&0&0&b_{7,6}&b_{7,7}&b_{7,8}\\
0&0&0&0&0&b_{8,6}&b_{8,7}&b_{8,8}
\end{bmatrix}
=
\begin{bmatrix}
\BB_1&0&0\\
0&\BB_2&0\\
0&0&\BB_3\\
\end{bmatrix}
\end{equation*} 

For the block diagonal $\BB$,
\begin{equation}\label{eq:B.update.blockdiag}
\BB_{i,j+1}= \bigg( \sum_{t=1}^T \big( \hatPttm  - \uu^\top\hatxtm^\top \big) \bigg)_i \bigg(\sum_{t=1}^T \hatPtm \bigg)_i^{-1}
\end{equation}
where the subscript $i$ means to take the parts of the matrices that are analogous to $\BB_i$; take the whole part within the parentheses not the individual matrices inside the parentheses.  If $\BB_i$ is comprised of rows $a$ to $b$ and columns $c$ to $d$ of matrix $\BB$, then  take rows $a$ to $b$ and columns $c$ to $d$ of the matrices subscripted by $i$ in equation \eqref{eq:B.update.blockdiag}.

\subsection{The update equation for $\QQ$ (unconstrained)}
\label{subsec:Qunconstrained}
The usual way to do this derivation is to use what is known as the ``trace trick'' which will pull the $\QQ^{-1}$ out to the left of the $\cc^\top\QQ^{-1}\bb$ terms which appear in the likelihood \eqref{eq:expLL}.  Here I'm showing a less elegant derivation that plods step by step through each of the likelihood terms.  Take the derivative of $\Psi$ with respect to $\QQ$. Terms not involving $\QQ$ equal 0 and drop out.   Again the expectations are placed outside the partials by noting that $\partial(\E[h(\XX_t,\QQ)])/\partial\QQ=\E[\partial(h(\XX_t,\QQ))/\partial\QQ]$. 
\begin{equation}\label{eq:Q.unconstrained1}
\begin{split}
&\partial\Psi/\partial\QQ = -\frac{1}{2} \sum_{t=1}^T\bigg(
\E[\partial(\XX_t^\top\QQ^{-1}\XX_t)/\partial\QQ]
-\E[\partial(\XX_t^\top\QQ^{-1}\BB\XX_{t-1})/\partial\QQ] \\
&\quad -\E[\partial((\BB\XX_{t-1})^\top\QQ^{-1}\XX_t)/\partial\QQ ]
 - \E[\partial(\XX_t^\top\QQ^{-1}\uu)/\partial\QQ] \\
&\quad - \E[\partial(\uu^\top\QQ^{-1}\XX_t)/\partial\QQ] 
+ \E[\partial((\BB\XX_{t-1})^\top\QQ^{-1}\BB\XX_{t-1})/\partial\QQ] \\
&\quad + \E[\partial((\BB\XX_{t-1})^\top\QQ^{-1}\uu)/\partial\QQ] 
+ \E[\partial(\uu^\top\QQ^{-1}\BB\XX_{t-1})/\partial\QQ]\\
&\quad +\partial(\uu^\top\QQ^{-1}\uu)/\partial\QQ
\bigg) - \partial\bigg(\frac{T}{2}\log |\QQ| \bigg)/\partial\QQ \\
\end{split}
\end{equation}
The relations \eqref{eq:derivInv} and \eqref{eq:derivlogDet} are used to do the differentiation. Notice that all the terms in the summation are of the form $\cc^\top\QQ^{-1}\bb$, and thus after differentiation, all the $\cc^\top\bb$ terms can be grouped inside one set of parentheses.  Also there is a minus that comes from equation \eqref{eq:derivInv} and it cancels out the minus in front of the initial $-1/2$.
\begin{equation}\label{eq:Q.unconstrained2}
\begin{split}
&\partial\Psi/\partial\QQ = \frac{1}{2} \sum_{t=1}^T \QQ^{-1} \bigg( 
 \E[\XX_t\XX_t^\top] -\E[\XX_t(\BB\XX_{t-1})^\top ] - \E[\BB\XX_{t-1}\XX_t^\top ]  - \E[ \XX_t\uu^\top ] - \E[ \uu\XX_t^\top ] \\
&\quad + \E[ \BB\XX_{t-1}(\BB\XX_{t-1})^\top ] + \E[\BB\XX_{t-1}\uu^\top] + \E[ \uu(\BB\XX_{t-1})^\top ] + \uu\uu^\top \bigg)\QQ^{-1} - \frac{T}{2}\QQ^{-1} 
\end{split}
\end{equation}
Pulling the parameters out of the expectations and using $(\BB\XX_t)^\top = \XX_t^\top\BB^\top$, we have
\begin{equation}\label{eq:Q.unconstrained3}
\begin{split}
&\partial\Psi/\partial\QQ = \frac{1}{2} \sum_{t=1}^T \QQ^{-1} \bigg( 
 \E[\XX_t\XX_t^\top] -\E[\XX_t\XX_{t-1}^\top ]\BB^\top - \BB\E[\XX_{t-1}\XX_t^\top ] - \E[ \XX_t ]\uu^\top - \uu \E[ \XX_t^\top ]\\
&\quad + \BB\E[ \XX_{t-1}\XX_{t-1}^\top ]\BB^\top + \BB\E[\XX_{t-1}]\uu^\top  + \uu\E[\XX_{t-1}^\top ]\BB^\top + \uu\uu^\top \bigg)\QQ^{-1} - \frac{T}{2}\QQ^{-1} 
\end{split}
\end{equation}
The partial derivative is then rewritten in terms of the Kalman smoother output:
\begin{equation}\label{eq:Q.unconstrained4}
\begin{split}
&\partial\Psi/\partial\QQ = \frac{1}{2} \sum_{t=1}^T \QQ^{-1} \bigg(  
 \hatPt - \hatPttm \BB^\top - \BB\hatPtmt 
- \hatxt\uu^\top - \uu \hatxt^\top \\
&\quad + \BB\hatPtm\BB^\top + \BB\hatxtm\uu^\top + \uu\hatxtm^\top\BB^\top + \uu\uu^\top \bigg)\QQ^{-1} - \frac{T}{2}\QQ^{-1} 
\end{split}
\end{equation}
Setting this to zero (a $m \times m$ matrix of zeros), $\QQ^{-1}$ is canceled out by multiplying by $\QQ$ twice, once on the left and once on the right and the $1/2$ is removed: 
\begin{equation}\label{eq:Q.unconstrained5}
\begin{split}
T\QQ = \sum_{t=1}^T \bigg(  
 \hatPt - \hatPttm \BB^\top - \BB\hatPtmt 
 - \hatxt\uu^\top - \uu \hatxt^\top 
 + \BB\hatPtm\BB^\top + \BB\hatxtm\uu^\top + \uu\hatxtm^\top\BB^\top 
 + \uu\uu^\top \bigg) 
\end{split}
\end{equation}
This gives us the new $\QQ$ that maximizes $\Psi$, 
\begin{equation}\label{eq:Q.update.unconstrained}
\begin{split}
&\QQ_{j+1} = \frac{1}{T}\sum_{t=1}^T \bigg( 
\hatPt - \hatPttm \BB^\top - \BB\hatPtmt 
 - \hatxt\uu^\top - \uu \hatxt^\top \\
&\quad + \BB\hatPtm\BB^\top + \BB\hatxtm\uu^\top + \uu\hatxtm^\top\BB^\top 
 + \uu\uu^\top \bigg)
\end{split}
\end{equation}

This derivation immediately generalizes to the case where $\QQ$ is a block diagonal matrix:
\begin{equation*}
\QQ =
\begin{bmatrix}
q_{1,1}&q_{1,2}&q_{1,3}&0&0&0&0&0\\
q_{1,2}&q_{2,2}&q_{2,3}&0&0&0&0&0\\
q_{1,3}&q_{2,3}&q_{3,3}&0&0&0&0&0\\
0&0&0&q_{4,4}&q_{4,5}&0&0&0\\
0&0&0&q_{4,5}&q_{5,5}&0&0&0\\
0&0&0&0&0&q_{6,6}&q_{6,7}&q_{6,8}\\
0&0&0&0&0&q_{6,7}&q_{7,7}&q_{7,8}\\
0&0&0&0&0&q_{6,8}&q_{7,8}&q_{8,8}
\end{bmatrix}
=
\begin{bmatrix}
\QQ_1&0&0\\
0&\QQ_2&0\\
0&0&\QQ_3\\
\end{bmatrix}
\end{equation*}
In this case,
\begin{equation}\label{eq:Q.update.blockdiag}
\begin{split}
&\QQ_{i,j+1} = \frac{1}{T}\sum_{t=1}^T \bigg(  
 \hatPt - \hatPttm \BB^\top - \BB\hatPtmt 
 - \hatxt\uu^\top - \uu \hatxt^\top \\
&\quad + \BB\hatPtm\BB^\top + \BB\hatxtm\uu^\top + \uu\hatxtm^\top\BB^\top 
 + \uu\uu^\top \bigg)_i
\end{split}
\end{equation}
where the subscript $i$ means take the elements of the matrix (in the big parentheses) that are analogous to $\QQ_i$; take the whole part within the parentheses not the individual matrices inside the parentheses).  If $\QQ_i$ is comprised of rows $a$ to $b$ and columns $c$ to $d$ of matrix $\QQ$, then take rows $a$ to $b$ and columns $c$ to $d$ of matrices subscripted by $i$ in equation \eqref{eq:Q.update.blockdiag}.

By the way, $\QQ$ is never really unconstrained since it is a variance-covariance matrix and the upper and lower triangles are shared.  However, because the shared values are only the symmetric values in the matrix, the derivation still works even though it's technically incorrect \citep{HendersonSearle1979}.  The constrained update equation for $\QQ$ shown in section \ref{sec:constrained.Q} explicitly deals with the shared lower and upper triangles.

\subsection{Update equation for $\aa$ (unconstrained)}\label{sec:unconstA}
Take the derivative of $\Psi$ with respect to $\aa$, where $\aa$ is a $n \times 1$ matrix.  Terms not involving $\aa$, equal 0 and drop out.  
\begin{equation}\label{eq:a.unconstrained1}
\begin{split}
&\partial\Psi/\partial\aa = - \frac{1}{2}\sum_{t=1}^T \bigg(- \partial(\E[\YY_t^\top\RR^{-1}\aa])/\partial\aa
- \partial(\E[\aa^\top\RR^{-1}\YY_t])/\partial\aa \\
&\quad + \partial(\E[(\ZZ\XX_t)^\top\RR^{-1}\aa])/\partial\aa 
+ \partial(\E[\aa^\top\RR^{-1}\ZZ\XX_t])/\partial\aa 
+ \partial(\E[\aa^\top\RR^{-1}\aa])/\partial\aa \bigg)
\end{split}
\end{equation}
The expectations around constants can be dropped\footnote{
because $\EXy(C)=C$, where $C$ is a constant.}.  Using relations \eqref{eq:derivaTc} and \eqref{eq:derivaTCa} and using $\RR^{-1} = (\RR^{-1})^\top$, we have then
\begin{equation}\label{eq:a.unconstrained2}
\begin{split}
&\partial\Psi/\partial\aa = - \frac{1}{2}\sum_{t=1}^T\bigg(-\E[\YY_t^\top\RR^{-1}]
-\E[\YY_t^\top\RR^{-1}] + \E[(\ZZ\XX_t)^\top\RR^{-1}] 
  + \E[(\ZZ\XX_t)^\top\RR^{-1}] + 2\aa^\top\RR^{-1} \bigg)
\end{split}
\end{equation}
Pull the parameters out of the expectations, use $(\aa\bb)^\top = \bb^\top\aa^\top$ and $\RR^{-1} = (\RR^{-1})^\top$ where needed, and remove the $-1/2$ to get
\begin{equation}\label{eq:a.unconstrained3}
\begin{split}
&\partial\Psi/\partial\aa = \sum_{t=1}^T\bigg(\E[\YY_t]^\top\RR^{-1}
 - \E[\XX_t]^\top\ZZ^\top\RR^{-1} - \aa^\top\RR^{-1} \bigg)
\end{split}
\end{equation}
Set the left side to zero (a $1 \times n$ matrix of zeros), take the transpose, and cancel out $\RR^{-1}$ by multiplying by $\RR$, giving
\begin{equation}\label{eq:a.unconstrained4}
\mathbf{0} = \sum_{t=1}^{T}\big(\E[\YY_t] - \ZZ\E[\XX_t] - \aa \big)
=\sum_{t=1}^{T}\big(\hatyt - \ZZ\hatxt - \aa\big)
\end{equation}

Solving for $\aa$ gives us the update equation for $\aa$: 
\begin{equation}\label{eq:a.update.unconstrained}
\aa_{j+1} = \frac{1}{T}\sum_{t=1}^{T}\big(\hatyt - \ZZ\hatxt \big)
\end{equation}

\subsection{The update equation for $\ZZ$ (unconstrained)}
Take the derivative of $\Psi$  with respect to $\ZZ$.  Terms not involving $\ZZ$, equal 0 and drop out. The expectations around terms involving only constants have been dropped. 
\begin{equation}\label{eq:Z.unconstrained1}
\begin{split}
&\partial\Psi/\partial\ZZ = \text{(note $\partial\ZZ$ is $m \times n$ while $\ZZ$ is $n \times m$)}\\
&\quad -\frac{1}{2} \sum_{t=1}^T\bigg(-\E[\partial(\YY_t^\top\RR^{-1}\ZZ\XX_t)/\partial\ZZ] 
 - \E[\partial((\ZZ\XX_t)^\top\RR^{-1}\YY_t)/\partial\ZZ] + \E[\partial((\ZZ\XX_t)^\top\RR^{-1}\ZZ\XX_t)/\partial\ZZ] \\
&\quad +  \E[\partial((\ZZ\XX_t)^\top\RR^{-1}\aa)/\partial\ZZ] 
+ \E[\partial(\aa^\top\RR^{-1}\ZZ\XX_t)/\partial\ZZ]\bigg)\\
&= -\frac{1}{2} \sum_{t=1}^T\bigg(-\E[\partial(\YY_t^\top\RR^{-1}\ZZ\XX_t)/\partial\ZZ] 
 -\E[\partial(\XX_t^\top\ZZ^\top\RR^{-1}\YY_t)/\partial\ZZ] 
+ \E[\partial(\XX_t^\top\ZZ^\top\RR^{-1}\ZZ\XX_t)/\partial\ZZ]  \\
&\quad + \E[\partial(\XX_t^\top\ZZ^\top\RR^{-1}\aa)/\partial\ZZ] 
+ \E[\partial(\aa^\top\RR^{-1}\ZZ\XX_t)/\partial\ZZ]\bigg)\\
\end{split}
\end{equation}
Using the matrix derivative relations (table \ref{tab:MatrixDerivatives}) and using $\RR^{-1} = (\RR^{-1})^\top$, we get 
\begin{equation}\label{eq:Z.unconstrained2}
\begin{split}
\partial\Psi/\partial\ZZ = -\frac{1}{2} \sum_{t=1}^T\bigg(-\E[ \XX_t\YY_t^\top\RR^{-1} ] - &\E[ \XX_t\YY_t^\top\RR^{-1} ]  \\
&+ 2 \E[\XX_t\XX_t^\top\ZZ^\top\RR^{-1}] + \E[ \XX_{t-1}\aa^\top\RR^{-1} ]  + \E[ \XX_t\aa^\top\RR^{-1} ] \bigg) 
\end{split}
\end{equation}
Pulling the parameters out of the expectations and getting rid of the $-1/2$, we have
\begin{equation}\label{eq:Z.unconstrained3}
\begin{split}
&\partial\Psi/\partial\ZZ =  \sum_{t=1}^T\bigg(\E[ \XX_t \YY_t^\top]\RR^{-1} 
- \E[ \XX_t\XX_t^\top ]\ZZ^\top\RR^{-1}  -\E[ \XX_t ]\aa^\top\RR^{-1} \bigg) \\
\end{split}
\end{equation}
Set the left side to zero (a $m \times n$ matrix of zeros), transpose it all, and cancel out $\RR^{-1}$ by multiplying by $\RR$ on the left, to give
\begin{equation}\label{eq:Z.unconstrained4}
\begin{split}
\mathbf{0}  =  \sum_{t=1}^T\big(\E[\YY_t\XX_t^\top] - \ZZ \E[\XX_t\XX_t^\top] - \aa\E[\XX_t^\top]\big) 
 =  \sum_{t=1}^T \big( \hatYXt  - \ZZ \hatPt - \aa\hatxt^\top \big)
\end{split}
\end{equation}
Solving for $\ZZ$ and noting that $\hatPt$ is invertible, gives us the new $\ZZ$: 
\begin{equation}\label{eq:Z.update.unconstrained}
\ZZ_{j+1}= \bigg( \sum_{t=1}^T \big(\hatYXt - \aa\hatxt^\top\big)\bigg) \bigg(\sum_{t=1}^T \hatPt\bigg)^{-1}
\end{equation}

\subsection{The update equation for $\RR$ (unconstrained)}
Take the derivative of $\Psi$ with respect to $\RR$.  Terms not involving $\RR$, equal 0 and drop out.  The expectations around terms involving constants have been removed. 
\begin{equation}\label{eq:R.unconstrained1}
\begin{split}
&\partial\Psi/\partial\RR = -\frac{1}{2} \sum_{t=1}^T\bigg(
\E[\partial(\YY_t^\top\RR^{-1}\YY_t)/\partial\RR]
-\E[\partial(\YY_t^\top\RR^{-1}\ZZ\XX_t)/\partial\RR] -\E[\partial((\ZZ\XX_t)^\top\RR^{-1}\YY_t)/\partial\RR]\\
&\quad   - \E[\partial(\YY_t^\top\RR^{-1}\aa)/\partial\RR] 
 - \E[\partial(\aa^\top\RR^{-1}\YY_t)/\partial\RR] 
+ \E[\partial((\ZZ\XX_t)^\top\RR^{-1}\ZZ\XX_t)/\partial\RR] \\
&\quad + \E[\partial((\ZZ\XX_t)^\top\RR^{-1}\aa)/\partial\RR] 
+ \E[\partial(\aa^\top\RR^{-1}\ZZ\XX_t)/\partial\RR] 
 + \partial(\aa^\top\RR^{-1}\aa)/\partial\RR
\bigg) - \partial\big(\frac{T}{2}\log |\RR| \big)/\partial\RR 
\end{split}
\end{equation}
We use relations \eqref{eq:derivInv} and \eqref{eq:derivlogDet} to do the differentiation. Notice that all the terms in the summation are of the form $\cc^\top\RR^{-1}\bb$, and thus after differentiation, we group all the $\cc^\top\bb$ inside one set of parentheses. Also there is a minus that comes from equation \eqref{eq:derivInv} and cancels out the minus in front of $-1/2$.
\begin{equation}\label{eq:R.unconstrained2}
\begin{split}
&\partial\Psi/\partial\RR = \frac{1}{2} \sum_{t=1}^T \RR^{-1} \bigg(  
 \E[\YY_t\YY_t^\top] -\E[\YY_t(\ZZ\XX_t)^\top] - \E[\ZZ\XX_t\YY_t^\top ] -  \E[\YY_t\aa^\top]  -  \E[\aa\YY_t^\top]\\
&\quad    + \E[ \ZZ\XX_t(\ZZ\XX_t)^\top ] + \E[\ZZ\XX_t\aa^\top] + \E[ \aa(\ZZ\XX_t)^\top ] 
 + \aa\aa^\top \bigg)\RR^{-1} - \frac{T}{2}\RR^{-1} 
\end{split}
\end{equation}
Pulling the parameters out of the expectations and using $(\ZZ\YY_t)^\top = \YY_t^\top\ZZ^\top$, we have
\begin{equation}\label{eq:R.unconstrained3}
\begin{split}
&\partial\Psi/\partial\RR = \frac{1}{2} \sum_{t=1}^T \RR^{-1} \bigg(  
 \E[\YY_t\YY_t^\top] -\E[\YY_t\XX_t^\top ]\ZZ^\top - \ZZ\E[\XX_t\YY_t^\top] 
 -  \E[\YY_t]\aa^\top - \aa\E[\YY_t^\top] \\
&\quad  + \ZZ\E[ \XX_t\XX_t^\top ]\ZZ^\top + \ZZ\E[\XX_t]\aa^\top + \aa\E[\XX_t^\top ]\ZZ^\top 
 + \aa\aa^\top \bigg)\RR^{-1} - \frac{T}{2}\RR^{-1} 
\end{split}
\end{equation}
We rewrite the partial derivative in terms of expectations:
\begin{equation}\label{eq:R.unconstrained4}
\begin{split}
&\partial\Psi/\partial\RR = \frac{1}{2} \sum_{t=1}^T \RR^{-1} \bigg( 
\hatOt - \hatYXt\ZZ^\top - \ZZ\hatYXt^\top 
 - \hatyt\aa^\top - \aa\hatyt^\top \\
&\quad + \ZZ\hatPt\ZZ^\top + \ZZ\hatxt\aa^\top + \aa\hatxt^\top\ZZ^\top + \aa\aa^\top \bigg)\RR^{-1} - \frac{T}{2}\RR^{-1} 
\end{split}
\end{equation}
Setting this to zero (a $n \times n$ matrix of zeros), we cancel out $\RR^{-1}$ by multiplying by $\RR$ twice, once on the left and once on the right, and get rid of the $1/2$. 
\begin{equation}\label{eq:R.unconstrained5}
\begin{split}
T\RR = \sum_{t=1}^T \bigg(  
\hatOt - \hatYXt\ZZ^\top - \ZZ\hatYXt^\top 
 - \hatyt\aa^\top - \aa\hatyt^\top 
 + \ZZ\hatPt\ZZ^\top + \ZZ\hatxt\aa^\top + \aa\hatxt^\top\ZZ^\top 
+ \aa\aa^\top \bigg)  
\end{split}
\end{equation}

We can then solve for $\RR$, giving us the new $\RR$ that maximizes $\Psi$, 
\begin{equation}\label{eq:R.update.unconstrained}
\begin{split}
\RR_{j+1} = \frac{1}{T}\sum_{t=1}^T \bigg(  
 \hatOt - \hatYXt\ZZ^\top - \ZZ\hatYXt^\top 
 - \hatyt\aa^\top - \aa\hatyt^\top 
 + \ZZ\hatPt\ZZ^\top + \ZZ\hatxt\aa^\top + \aa\hatxt^\top\ZZ^\top 
 + \aa\aa^\top \bigg)
\end{split}
\end{equation}
As with $\QQ$, this derivation immediately generalizes to a block diagonal matrix:
\begin{equation*}
\RR =
\begin{bmatrix}
\RR_1&0&0\\
0&\RR_2&0\\
0&0&\RR_3\\
\end{bmatrix}
\end{equation*}
In this case,
\begin{equation}\label{eq:R.update.blockdiag}
\begin{split}
\RR_{i,j+1} = \frac{1}{T}\sum_{t=1}^T \bigg(  
 \hatOt - \hatYXt\ZZ^\top - \ZZ\hatYXt^\top 
 - \hatyt\aa^\top - \aa\hatyt^\top 
 + \ZZ\hatPt\ZZ^\top + \ZZ\hatxt\aa^\top + \aa\hatxt^\top\ZZ^\top 
 + \aa\aa^\top \bigg)_i
\end{split}
\end{equation}
where the subscript $i$ means we take the elements in the matrix in the big parentheses that are analogous to $\RR_i$.  If $\RR_i$ is comprised of rows $a$ to $b$ and columns $c$ to $d$ of matrix $\RR$, then we take rows $a$ to $b$ and columns $c$ to $d$ of matrix subscripted by $i$ in equation \eqref{eq:R.update.blockdiag}.

\subsection{Update equation for $\xixi$ and $\LAM$ (unconstrained), stochastic initial state}
\citet{ShumwayStoffer2006} and \citet{GhahramaniHinton1996} imply in their discussion of the EM algorithm that both $\xixi$ and $\LAM$ can be estimated (though not simultaneously).  Harvey (1989), however, discusses that there are only two allowable cases: $\xx_0$ is treated as fixed ($\LAM=0$) and equal to the unknown parameter $\xixi$ or $\xx_0$ is treated as stochastic with a known mean $\xixi$ and variance $\LAM$.  For completeness, we show here the update equation in the case of $\xx_0$ stochastic with unknown mean $\xixi$ and variance $\LAM$ (a case that Harvey (1989) says is not consistent).

We proceed as before and solve for the new $\xixi$ by minimizing $\Psi$.
Take the derivative of $\Psi$  with respect to $\xixi$ .  Terms not involving $\xixi$, equal 0 and drop out.  
\begin{equation}\label{eq:pi.unconstrained1}
\begin{split}
\partial\Psi/\partial\xixi = - \frac{1}{2} \big(-  \partial(\E[\xixi^\top\LAM^{-1}\XX_0])/\partial\xixi 
- \partial(\E[\XX_0^\top\LAM^{-1}\xixi])/\partial\xixi 
 + \partial(\xixi^\top\LAM^{-1}\xixi)/\partial\xixi \big)
\end{split}
\end{equation}
Using relations \eqref{eq:derivaTc} and \eqref{eq:derivaTCa} and using $\LAM^{-1} = (\LAM^{-1})^\top$, we have
\begin{equation}\label{eq:pi.unconstrained2}
\partial\Psi/\partial\xixi = - \frac{1}{2} \big(- \E[ \XX_0^\top\LAM^{-1} ] 
- \E[ \XX_0^\top\LAM^{-1} ] + 2\xixi^\top\LAM^{-1} \big)
\end{equation}
Pulling the parameters out of the expectations, we get
\begin{equation}\label{eq:pi.unconstrained3}
\partial\Psi/\partial\xixi = - \frac{1}{2} \big(- 2\E[ \XX_0^\top ]\LAM^{-1} + 2\xixi^\top\LAM^{-1} \big)
\end{equation}
We then set the left side to zero, take the transpose, and cancel out $-1/2$ and $\LAM^{-1}$ (by noting that it is a variance-covariance matrix and is invertible).   
\begin{equation}\label{eq:pi.unconstrained4}
\mathbf{0} = \big(\LAM^{-1}\E[ \XX_0 ] + \LAM^{-1}\xixi \big)=(\widetilde{\mbox{$\mathbf x$}}_0 - \xixi)
\end{equation}

Thus,
\begin{equation}\label{eq:pi.update.unconstrained}
\xixi_{j+1} = \widetilde{\mbox{$\mathbf x$}}_0
\end{equation}
$\widetilde{\mbox{$\mathbf x$}}_0$ is the expected value of $\XX_0$ conditioned on the data from $t=1$ to $T$, which comes from the Kalman smoother recursions with initial conditions defined as $\E[\XX_0|\YY_0=\yy_0] \equiv \xixi$ and $\var(\XX_0 \XX_0^\top|\YY_0=\yy_0)\equiv \LAM$.
A similar set of steps gets us to the update equation for $\LAM$,
\begin{equation}\label{eq:V0.update.unconstrained}
\LAM_{j+1} = \widetilde{\VV}_0
\end{equation}
$\widetilde{\VV}_0$ is the variance of $\XX_0$ conditioned on the data from $t=1$ to $T$ and is an output from the Kalman smoother recursions.

If the initial state is defined as at $t=1$ instead of $t=0$, the update equation is derived in an identical fashion and the update equation is similar:
\begin{equation}\label{eq:pix1.update.unconstrained}
\xixi_{j+1} = \widetilde{\mbox{$\mathbf x$}}_1
\end{equation}
\begin{equation}
\LAM_{j+1} = \widetilde{\VV}_1
\end{equation}
These are output from the Kalman smoother recursions with initial conditions defined as $\E[\XX_1|\YY_0=\yy_0] \equiv \xixi$ and $\var(\XX_1 \XX_1^\top|\YY_0=\yy_0)\equiv \LAM$.  Notice that the recursions are initialized slightly differently; you will see the Kalman filter and smoother equations presented with both types of initializations depending on whether the author defines the initial state at $t=0$ or $t=1$.

\subsection{Update equation for $\xixi$ (unconstrained), fixed $\xx_0$}
For the case where $\xx_0$ is treated as fixed, i.e. as another parameter, then there is no $\LAM$, and we need to maximize $\partial\Psi/\partial\xixi$ using the slightly different $\Psi$ shown in equation \eqref{eq:logL.V0.is.0}.  Now $\xixi$ appears in the state equation part of the likelihood.
\begin{equation}\label{eq:pi.unconstrained.V0.is.0}
\begin{split}
&\partial\Psi/\partial\xixi = -\frac{1}{2} \bigg(-\E[\partial(\XX_1^\top\QQ^{-1}\BB\xixi)/\partial\xixi] 
 - \E[\partial((\BB\xixi)^\top\QQ^{-1}\XX_1)/\partial\xixi] + \E[\partial((\BB\xixi)^\top\QQ^{-1}(\BB\xixi))/\partial\xixi] \\
&\quad +  \E[\partial((\BB\xixi)^\top\QQ^{-1}\uu)/\partial\xixi] 
+ \E[\partial(\uu^\top\QQ^{-1}\BB\xixi)/\partial\xixi]\bigg)\\
&= -\frac{1}{2} \bigg(-\E[\partial(\XX_1^\top\QQ^{-1}\BB\xixi)/\partial\xixi] 
 - \E[\partial(\xixi^\top\BB^\top\QQ^{-1}\XX_1)/\partial\xixi] 
+ \E[\partial(\xixi^\top\BB^\top\QQ^{-1}(\BB\xixi))/\partial\xixi] \\
&\quad +  \E[\partial(\xixi^\top\BB^\top\QQ^{-1}\uu)/\partial\xixi] 
+ \E[\partial(\uu^\top\QQ^{-1}\BB\xixi)/\partial\xixi]\bigg)
\end{split}
\end{equation}
After pulling the constants out of the expectations, we use relations \eqref{eq:derivaTDb} and \eqref{eq:derivbDTCDd} to take the derivative:
\begin{equation}\label{eq:pi.unconstrained.V0.is.0.2}
\begin{split}
\partial\Psi/\partial\xixi = -\frac{1}{2} \bigg(-\E[\XX_1]^\top\QQ^{-1}\BB - \E[ \XX_1]^\top\QQ^{-1}\BB 
 + 2 \xixi^\top\BB^\top\QQ^{-1}\BB + \uu^\top\QQ^{-1}\BB  +  \uu^\top\QQ^{-1}\BB \bigg) \\
\end{split}
\end{equation}
This can be reduced to
\begin{equation}\label{eq:pi.unconstrained.V0.is.0.3}
\begin{split}
\partial\Psi/\partial\xixi = \E[\XX_1]^\top\QQ^{-1}\BB - \xixi^\top\BB^\top\QQ^{-1}\BB - \uu^\top\QQ^{-1}\BB 
\end{split}
\end{equation}
To solve for $\xixi$, set the left side to zero (an $m \times 1$ matrix of zeros), transpose the whole equation, and then cancel out $\BB^\top\QQ^{-1}\BB$ by multiplying by its inverse on the left, and solve for $\xixi$.  This step requires that this inverse exists.
\begin{equation}\label{eq:pi.unconstrained.V0.is.0.4}
\begin{split}
\xixi = (\BB^\top\QQ^{-1}\BB)^{-1}\BB^\top\QQ^{-1}(\E[\XX_1]  - \uu) 
\end{split}
\end{equation}
Thus, in terms of the Kalman filter/smoother output the new $\xixi$ for EM iteration $j+1$ is
\begin{equation}\label{eq:pi.unconstrained.V0.is.0.5}
\begin{split}
\xixi_{j+1} = (\BB^\top\QQ^{-1}\BB)^{-1}\BB^\top\QQ^{-1}(\widetilde{\mbox{$\mathbf x$}}_1  - \uu) 
\end{split}
\end{equation}
Note that using, $\widetilde{\mbox{$\mathbf x$}}_0$ output from the Kalman smoother would not work since $\LAM=0$.  As a result, $\xixi_{j+1} \equiv \xixi_j$ in the EM algorithm, and it is impossible to move away from your starting condition for $\xixi$.

This is conceptually similar to using a generalized least squares estimate of $\xixi$ to concentrate it out of the likelihood as discussed in Harvey (1989), section 3.4.4.  However, in the context of the EM algorithm, dealing with the fixed $\xx_0$ case requires nothing special; one simply takes care to use the likelihood for the case where $\xx_0$ is treated as an unknown parameter (equation \ref{eq:logL.V0.is.0}).  For the other parameters, the update equations are the same whether one uses the log-likelihood equation with $\xx_0$ treated as stochastic (equation \ref{eq:logL}) or fixed (equation \ref{eq:logL.V0.is.0}).

If your MARSS model is stationary\footnote{meaning the $\XX$'s have a stationary distribution} and your data appear stationary, however, equation \eqref{eq:pi.unconstrained.V0.is.0.4} probably is not what you want to use.  The estimate of $\xixi$ will be the maximum-likelihood value, but it will not be drawn from the stationary distribution; instead it could be some wildly different value that happens to give the maximum-likelihood.  If you are modeling the data as stationary, then you should probably assume that $\xixi$ is drawn from the stationary distribution of the $\XX$'s, which is some function of your model parameters.  This would mean that the model parameters would enter the part of the likelihood that involves $\xixi$ and $\LAM$. Since you probably don't want to do that (if might start to get circular), you might try an iterative process to get decent $\xixi$ and $\LAM$ or try fixing $\xixi$ and estimating $\LAM$ (above).  You can fix $\xixi$ at, say, zero, by making sure the model you fit has a stationary distribution with mean zero.  You might also need to demean your data (or estimate the $\aa$ term to account for non-zero mean data).  A second approach is to estimate $\xx_1$ as the initial state instead of $\xx_0$.

\subsection{Update equation for $\xixi$ (unconstrained), fixed $\xx_1$}\label{sec:xi.unconstrained.x1}
In some cases, the estimate of $\xx_0$ from $\xx_1$ using equation \ref{eq:pi.unconstrained.V0.is.0.5} will be highly sensitive to small changes in the parameters.  This is particularly the case for certain $\BB$ matrices, even if they are stationary.  The result is that your $\xixi$ estimate is wildly different from the data at $t=1$.  The estimates are correct given how you defined the model, just not realistic given the data.  In this case, you can specify $\xixi$ as being the value of $\xx$ at $t=1$ instead of $t=0$.  That way, the data at $t=1$ will constrain the estimated $\xixi$.  In this case, we treat $\xx_1$ as fixed but unknown parameter $\xixi$.  The likelihood is then:
\begin{equation}
\begin{split}
&\log\LL(\yy,\xx ; \Theta) = -\sum_1^T \frac{1}{2}(\yy_t - \ZZ \xx_t - \aa)^\top \RR^{-1} (\yy_t - \ZZ \xx_t - \aa) -\sum_1^T\frac{1}{2} \log |\RR|\\
&\quad  -\sum_2^T \frac{1}{2} (\xx_t - \BB \xx_{t-1} - \uu)^\top \QQ^{-1} (\xx_t - \BB \xx_{t-1} - \uu) - \sum_1^T\frac{1}{2}\log |\QQ|
\end{split}
\end{equation}

\begin{equation}\label{}
\begin{split}
&\partial\Psi/\partial\xixi = -\frac{1}{2} \bigg(-\E[\partial(\YY_1^\top\RR^{-1}\ZZ\xixi)/\partial\xixi] 
 - \E[\partial((\ZZ\xixi)^\top\RR^{-1}\YY_1)/\partial\xixi] + \E[\partial((\ZZ\xixi)^\top\RR^{-1}(\ZZ\xixi))/\partial\xixi] \\
&\quad +  \E[\partial((\ZZ\xixi)^\top\RR^{-1}\aa)/\partial\xixi] 
+ \E[\partial(\aa^\top\RR^{-1}\ZZ\xixi)/\partial\xixi]\bigg)\\
&\quad -\frac{1}{2} \bigg(-\E[\partial(\XX_2^\top\QQ^{-1}\BB\xixi)/\partial\xixi] 
 - \E[\partial((\BB\xixi)^\top\QQ^{-1}\XX_2)/\partial\xixi] + \E[\partial((\BB\xixi)^\top\QQ^{-1}(\BB\xixi))/\partial\xixi] \\
&\quad +  \E[\partial((\BB\xixi)^\top\QQ^{-1}\uu)/\partial\xixi] 
+ \E[\partial(\uu^\top\QQ^{-1}\BB\xixi)/\partial\xixi]\bigg)
\end{split}
\end{equation}
Note that the second summation starts at $t=2$ and $\xixi$ is $\xx_1$ instead of $\xx_0$.

After pulling the constants out of the expectations, we use relations \eqref{eq:derivaTDb} and \eqref{eq:derivbDTCDd} to take the derivative:
\begin{equation}\label{eq:pi.unconstrained.V0.is.0.t.1.1}
\begin{split}
&\partial\Psi/\partial\xixi = -\frac{1}{2} \bigg(-\E[\YY_1]^\top\RR^{-1}\ZZ - \E[ \YY_1]^\top\RR^{-1}\ZZ 
 + 2 \xixi^\top\ZZ^\top\RR^{-1}\ZZ + \aa^\top\RR^{-1}\ZZ  +  \aa^\top\RR^{-1}\ZZ \bigg) \\
&\quad-\frac{1}{2} \bigg(-\E[\XX_2]^\top\QQ^{-1}\BB - \E[ \XX_2]^\top\QQ^{-1}\BB 
 + 2 \xixi^\top\BB^\top\QQ^{-1}\BB + \uu^\top\QQ^{-1}\BB  +  \uu^\top\QQ^{-1}\BB \bigg) 
\end{split}
\end{equation}
This can be reduced to
\begin{equation}\label{}
\begin{split}
&\partial\Psi/\partial\xixi = \E[\YY_1]^\top\RR^{-1}\ZZ - \xixi^\top\ZZ^\top\RR^{-1}\ZZ - \aa^\top\RR^{-1}\ZZ 
 + \E[\XX_2]^\top\QQ^{-1}\BB - \xixi^\top\BB^\top\QQ^{-1}\BB - \uu^\top\QQ^{-1}\BB \\
&\quad =  - \xixi^\top(\ZZ^\top\RR^{-1}\ZZ + \BB^\top\QQ^{-1}\BB) + \E[\YY_1]^\top\RR^{-1}\ZZ - \aa^\top\RR^{-1}\ZZ 
+ \E[\XX_2]^\top\QQ^{-1}\BB - \uu^\top\QQ^{-1}\BB
\end{split}
\end{equation}
To solve for $\xixi$, set the left side to zero (an $m \times 1$ matrix of zeros), transpose the whole equation, and solve for $\xixi$.  
\begin{equation}\label{eq:pi.unconstrained.V0.is.0.t.1.2}
\begin{split}
\xixi = (\ZZ^\top\RR^{-1}\ZZ + \BB^\top\QQ^{-1}\BB)^{-1}(\ZZ^\top\RR^{-1}(\E[\YY_1]-\aa) +\BB^\top\QQ^{-1}(\E[\XX_2]  - \uu)) \\
\end{split}
\end{equation}
Thus, when $\xixi \equiv \xx_1$,  the new $\xixi$ for EM iteration $j+1$ is
\begin{equation}\label{eq:pi.unconstrained.V0.is.0.t.1.3}
\begin{split}
\xixi_{j+1} = (\ZZ^\top\RR^{-1}\ZZ + \BB^\top\QQ^{-1}\BB)^{-1}(\ZZ^\top\RR^{-1}(\widetilde{\mbox{$\mathbf y$}}_1-\aa) +\BB^\top\QQ^{-1}(\widetilde{\mbox{$\mathbf x$}}_2  - \uu))
\end{split}
\end{equation}

\section{The time-varying MARSS model with linear constraints}\label{sec:tvMARSS}
The first part of this report dealt with the case of a MARSS model (equation \ref{eq:MARSS}) where the parameters are time-constant and where all the elements in a parameter matrix are estimated with no constraints.   I will now describe the derivation of an EM algorithm to solve a much more general MARSS model (equation \ref{eq:MARSS.ex2}), which is a time-varying MARSS model where the MARSS parameter matrices are written as a linear equation $\ff+\DD\mm$.  This is a very general form of a MARSS model, of which many (most) multivariate autoregressive Gaussian models are a special case.  This general MARSS model includes as special cases, MARSS models with covariates (many VARSS models with exogeneous variables), multivariate AR lag-p models and multivariate moving average models, and MARSS models with linear constraints placed on the elements within the model parameters.  The objective is to derive one EM algorithm for the whole class, thus a uniform approach to fitting these models. 

The time-varying MARSS model is written:
\begin{subequations}\label{eq:MARSS.ex2}
\begin{gather}
\xx_t = \BB_t\xx_{t-1} + \uu_t + \HH_t\ww_t, \text{ where } \WW_t \sim \MVN(0,\QQ_t)\\
\yy_t = \ZZ_t\xx_t + \aa_t + \GG_t\vv_t, \text{ where } \VV_t \sim \MVN(0,\RR_t)\\
\xx_{t_0} = \xixi+\FF\ll, \text{ where } t_0=0 \text{ or } t_0=1\\
\LL \sim \MVN(0,\LAM)\\
\begin{bmatrix}\ww_t\\ \vv_t\end{bmatrix} \sim \MVN(0,\Sigma), \quad \Sigma=\begin{bmatrix}\QQ_t&0\\ 0&\RR_t\end{bmatrix}
\end{gather}
\end{subequations}
This looks quite similar to the previous non-time varying MARSS model, but now the model parameters, $\BB$, $\uu$, $\QQ$, $\ZZ$, $\aa$ and $\RR$, have a $t$ subscript and we have a multiplier matrix on the error terms $\vv_t$, $\ww_t$, $\ll$. The $\HH_t$ multiplier is $m \times s$, so we now have $s$ state errors instead of $m$.  The $\GG_t$ multiplier is $n \times k$, so we now have $k$ observation errors instead of $n$.  The $\FF$ multiplier is $m \times j$, so now we can have some initial states ($j$ of them) be stochastic and others be fixed.  I assume that appropriate constraints are put on $\GG$ and $\HH$ so that the resulting MARSS model is not under- or over-constrained\footnote{For example, if both $\GG$ and $\HH$ are column vectors, then the system is over-constrained and has no solution.}.
The notation/presentation here was influenced by SJ Koopman's work, esp. \citet{KoopmanOoms2011} and \citet{Koopman1993}, but in these works, $\QQ_t$ and $\RR_t$ equal $\II$ and the variance-covariance structures are instead specified only by $\HH_t$ and $\GG_t$.  I keep $\QQ_t$ and $\RR_t$ in my formulation as it seems more intuitive (to me) in the context of the EM algorithm and the required joint-likelihood function. 

We can rewrite this MARSS model using vec relationships (table \ref{tab:VecRelations}):
\begin{equation}\label{eq:MARSS.ex.vec}
\begin{gathered}
\xx_t = (\xx_{t-1}^\top \otimes \II_m)\vec(\BB_t) + \vec(\uu_t) + \HH_t\ww_t, \WW_t  \sim \MVN(0,\QQ_t)\\
\yy_t = (\xx_t^\top \otimes \II_n)\vec(\ZZ_t) + \vec(\aa_t) + \GG_t\vv_t, \VV_t \sim \MVN(0,\RR_t)\\
\xx_{t_0} = \xixi+\FF\ll, \LL \sim \MVN(0,\LAM)
\end{gathered}
\end{equation}
Each model parameter, $\BB_t$, $\uu_t$, $\QQ_t$, $\ZZ_t$, $\aa_t$, and $\RR_t$, is written as a time-varying linear model, $\ff_t+\DD_t\mm$, where $\ff$ and $\DD$ are fully-known (not estimated and no missing values) and $\mm$ is a column vector of the estimates elements of the parameter matrix:
\begin{equation}\label{eq:MARSS.ex.vec.2}
\begin{split}
\vec(\BB_t) &= \ff_{t,b} + \DD_{t,b}\bbeta\\
\vec(\uu_t) &= \ff_{t,u} + \DD_{t,u}\uupsilon\\
\vec(\QQ_t) &= \ff_{t,q} + \DD_{t,q}\qq\\
\vec(\ZZ_t) &= \ff_{t,z} + \DD_{t,z}\zzeta\\
\vec(\aa_t) &= \ff_{t,a} + \DD_{t,a}\aalpha\\
\vec(\RR_t) &= \ff_{t,r} + \DD_{t,r}\rr\\
\vec(\LAM)&= \ff_\lambda+\DD_\lambda\llambda \\
\vec(\xixi)&= \ff_\xi+\DD_\xi\pp
\end{split}
\end{equation}

The estimated parameters are now the column vectors, $\bbeta$, $\uupsilon$, $\qq$, $\zzeta$, $\aalpha$, $\rr$, $\pp$ and $\llambda$.  The time-varying aspect comes from the time-varying $\ff$ and $\DD$.  Note that variance-covariance matrices must be positive-definite and we cannot specify a form that cannot be estimated.  Fixing the diagonal terms and estimating the off-diagonals would not be allowed.  Thus the $\ff$ and $\DD$ terms for $\QQ$, $\RR$ and $\LAM$ are limited.  For the other parameters, the forms are fairly unrestricted, except that the $\DD$s need to be full rank so that we are not specifying an under-constrained model.  'Full rank' will imply that we are not trying to estimate confounded matrix elements; for example, trying to estimate $a_1$ and $a_2$ but only $a_1+a_2$ appear in the model.

The temporally variable MARSS model, equation \eqref{eq:MARSS.ex.vec} together with \eqref{eq:MARSS.ex.vec.2}, looks rather different than other temporally variable MARSS models, such as a VARSSX or MARSS with covariates model, in the literature.  But those models are special cases of this equation.  By deriving an EM algorithm for this more general (if unfamiliar) form, I then have an algorithm for many different types of time-varying MARSS models with linear constraints on the parameter elements.  Below I show some examples. 

\subsection{MARSS model with linear constraints}
We can use equation \eqref{eq:MARSS.ex.vec} to put linear constraints on the elements of the parameters, $\BB$, $\uu$, $\QQ$, $\ZZ$, $\aa$, $\RR$, $\xixi$ and $\LAM$. Here is an example of a simple MARSS model with linear constraints:
\begin{gather*}
\begin{bmatrix}x_1\\ x_2\end{bmatrix}_t
= \begin{bmatrix}a&0\\0&2a\end{bmatrix}
\begin{bmatrix}x_1\\x_2\end{bmatrix}_{t-1}
+ \begin{bmatrix}w_1\\ w_2\end{bmatrix}_t,\quad 
\begin{bmatrix}w_1\\ w_2\end{bmatrix}_t \sim \MVN\begin{pmatrix}\begin{bmatrix}0.1\\u+0.1\end{bmatrix},\begin{bmatrix}q_{11}&q_{12}\\q_{21}&q_{22}\end{bmatrix} \end{pmatrix}  \\
\\
\begin{bmatrix}y_1\\ y_2\\ y_3\end{bmatrix}_t
= \begin{bmatrix}c&3c+2d+1\\ c& d\\ c+e+2 &e\end{bmatrix}
\begin{bmatrix}x_1\\ x_2\end{bmatrix}_t
+ \begin{bmatrix}v_1\\ v_2\\ v_3\end{bmatrix}_t,\\
\begin{bmatrix}v_1\\ v_2\\ v_3\end{bmatrix}_t \sim \MVN\begin{pmatrix}\begin{bmatrix}a_1\\ a_2\\ 0\end{bmatrix},
 \begin{bmatrix}r&0&0\\0&2r&0\\0&0&4r\end{bmatrix} \end{pmatrix}  \\
\\
\begin{bmatrix}x_1\\ x_2\end{bmatrix}_0 \sim \MVN\begin{pmatrix}\begin{bmatrix}\pi\\ \pi\end{bmatrix},\begin{bmatrix}1&0\\ 0&1\end{bmatrix} \end{pmatrix}
\end{gather*}
Linear constraints mean that elements of a matrix may be fixed to a specific numerical value or specified as a linear combination of values (which can be shared within a matrix but not shared between matrices).

Let's say we have some parameter matrix $\MM$ (here $\MM$ could be any of the parameters in the MARSS model) where each matrix element is written as a linear model of some potentially shared values:
\begin{equation*}
\MM=
\begin{bmatrix}
a+2c+2&0.9&c\\
-1.2&a&0\\
0&3c+1&b
\end{bmatrix}
\end{equation*}
Thus each $i$-th element in $\MM$ can be written as $\beta_i+\beta_{a,i} a + \beta_{b,i} b + \beta_{c,i} c$, which is a linear combination of three estimated values $a$, $b$ and $c$. The matrix $\MM$ can be rewritten in terms of a $\beta_i$ part and the part involving the $\beta_{-,j}$'s:
\begin{equation*}
\MM=	
\begin{bmatrix}
2&0.9&0\\
-1.2&0&0\\
0&1&0
\end{bmatrix}
+
\begin{bmatrix}
a+2c&0&c\\
0&a&0\\
0&3c&b
\end{bmatrix}
=\MM_\text{fixed}+\MM_\text{free}
\end{equation*}
The vec function turns any matrix into a column vector by stacking the columns on top of each other.  Thus,
\begin{equation*}
\vec(\MM)=
\begin{bmatrix}
a+2c+2\\
-1.2\\
0\\
0.9\\
a\\
3c+1\\
c\\
0\\
b
\end{bmatrix}	
\end{equation*}
We can now write $\vec(\MM)$ as a linear combination of $\ff = \vec(\MM_\text{fixed})$ and $\DD\mm = \vec(\MM_\text{free})$.  $\mm$ is a $p \times 1$ column vector of the $p$ free values, in this case $p=3$ and the free values are $a, b, c$. $\DD$ is a design matrix that translates $\mm$ into $\vec(\MM_\text{free})$.  For example,
\begin{equation*}
\vec(\MM)=
\begin{bmatrix}
a+2c+2\\
-1.2\\
0\\
0.9\\
a\\
3c+1\\
c\\
0\\
b
\end{bmatrix}	
=
\begin{bmatrix}
0\\
-1.2\\
2\\
0.9\\
0\\
1\\
0\\
0\\
0
\end{bmatrix}
+
\begin{bmatrix}
1&2&0\\
0&0&0\\
0&0&0\\
0&0&0\\
1&0&0\\
0&0&3\\
0&0&1\\
0&0&0\\
0&1&0
\end{bmatrix}	
\begin{bmatrix}
a\\
b\\
c
\end{bmatrix}
= \ff + \DD\mm 
\end{equation*}
There are constraints on $\DD$.  Your $\DD$ matrix needs to describe a solvable linear set of equations.  Basically it needs to be full rank (rank $p$ where $p$ is the number of columns in $\DD$ or free values you are trying to estimate), so that you can estimate each of the $p$ free values.  For example, if $a+b$ always appeared together, then $a+b$ can be estimated but not $a$ and $b$ separately.  Note, if $\MM$ is fixed, then $\DD$ is undefined but that is fine because in this case, there will be no update equation needed; you just use the fixed value of $\MM$ in the algorithm.

\begin{table}
	\caption{Kronecker and vec relations.  Here $\AA$ is $n \times m$, $\BB$ is $m \times p$, $\CC$ is $p \times q$, and $\EE$ and $\DD$ are $p \times p$. $\aa$ is a $m \times 1$ column vector and $\bb$ is a $p \times 1$ column vector. The symbol $\otimes$ stands for the Kronecker product:  $\AA \otimes \CC$ is a $np \times mq$  matrix.	The identity matrix, $\II_n$, is a $n \times n$ diagonal matrix with ones on the diagonal.}
	\label{tab:VecRelations}
	\begin{center}
		\begin{tabular}{lr}
\hline
\\
\refstepcounter{equation}\label{eq:vec.a}
$\vec(\aa) = \vec(\aa^\top) = \aa$
&\multirow{3}{*}{(\theequation)} \\
The vec of a column vector (or its transpose) is itself. & \\
$\aa=(\aa^\top \otimes \II_1)$ & \\
\\
\refstepcounter{equation}\label{eq:vec.Aa}
$\vec(\AA\aa) = (\aa^\top \otimes \II_n)\vec(\AA) = \AA\aa$
&\multirow{2}{*}{(\theequation)} \\
$\vec(\AA\aa) = \AA\aa$ since $\AA\aa$ is itself an $m \times 1$ column vector. & \\
\\
\refstepcounter{equation}\label{eq:vec.AB}
$\vec(\AA\BB) = (\II_p \otimes \AA)\vec(\BB) = (\BB^\top \otimes \II_n)\vec(\AA)$
& (\theequation) \\
\\
\refstepcounter{equation}\label{eq:vec.ABC}
$\vec(\AA\BB\CC) = (\CC^\top \otimes \AA)\vec(\BB)$
& (\theequation) \\
\\
\refstepcounter{equation}\label{eq:kron.prod}
$(\AA \otimes \BB)(\CC \otimes \DD) = (\AA\CC \otimes \BB\DD)$
& (\theequation) \\
\\
\refstepcounter{equation}\label{eq:kron.column.vec}
$(\aa \otimes \II_p)\CC = (\aa \otimes \CC)$ &\multirow{4}{*}{(\theequation)} \\
$\CC(\aa^\top \otimes \II_q) = (\aa^\top \otimes \CC)$ &\\
$\EE(\aa^\top \otimes \DD)=\EE\DD(\aa^\top \otimes \II_p)=(\aa^\top \otimes \EE\DD)$ &\\
\\
\refstepcounter{equation}\label{eq:kron.column.quad.vec}
$(\aa \otimes \II_p)\CC(\bb^\top \otimes \II_q) = (\aa\bb^\top \otimes \CC)$ &
(\theequation) \\
\\
\refstepcounter{equation}\label{eq:kron.column.column.vec}
$(\aa \otimes \aa)=\vec(\aa\aa^\top)$ 
&\multirow{2}{*}{(\theequation)} \\
$(\aa^\top \otimes \aa^\top)=(\aa \otimes \aa)^\top=(\vec(\aa\aa^\top))^\top$ &\\
\\
\refstepcounter{equation}\label{eq:kron.trans}
$(\AA^\top \otimes \BB^\top)=(\AA \otimes \BB)^\top$ &
(\theequation) \\
\\\hline
\end{tabular}
\end{center}
\end{table}

\subsection{A MARSS model with exogenous variables}
The following is a commonly seen MARSS model with covariates $\gg_t$ and $\hh_t$ appearing as additive elements:
\begin{equation*}
\begin{split}
\xx_t &= \BB\xx_{t-1} + \CC\gg_t + \ww_t\\
\yy_t &= \ZZ\xx_t + \FF\hh_t + \vv_t 
\end{split}
\end{equation*}
We would typically want to estimate $\CC$ or $\FF$ which are the influence of our covariates on our responses, $\xx$ or $\yy$.  Let's say there are $p$ covariates in $\hh_t$ and $q$ covariates in $\gg_t$.  Then we can write the above in vec form:
\begin{equation}\label{eq:MARSS.simple.ex}
\begin{split}
\xx_t &= (\xx_{t-1}^\top \otimes \II_m)\vec(\BB) + (\hh_t^\top \otimes \II_p)\vec(\CC) + \ww_t\\
\yy_t &= (\xx_t^\top \otimes \II_n)\vec(\ZZ) + (\gg_t^\top \otimes \II_q)\vec(\DD) + \vv_t 
\end{split}
\end{equation}
Let's say we put no constraints $\BB$, $\ZZ$, $\QQ$, $\RR$, $\xixi$, or $\LAM$.  Then in the form of equation \eqref{eq:MARSS.ex.vec},
\begin{equation*}
\begin{split}
\xx_t &= (\xx_{t-1}^\top \otimes \II_m)\vec(\BB_t) + \vec(\uu_t) + \ww_t\\
\yy_t &= (\xx_t^\top \otimes \II_n)\vec(\ZZ_t) + \vec(\aa_t) + \vv_t,
\end{split}
\end{equation*}
with the parameters defined as follows:
\begin{equation*}
\begin{split}
\vec(\BB_t) &= \ff_{t,b} + \DD_{t,b}\bbeta;
\, \ff_{t,b} = 0;\, \DD_{t,b}=1;\, \bbeta=\vec(\BB)\\
\vec(\uu_t) &= \ff_{t,u} + \DD_{t,u}\uupsilon;
\, \ff_{t,u} = 0;\, \DD_{t,u}=(\hh_t^\top \otimes \II_p);\, \uupsilon=\vec(\CC)\\
\vec(\QQ_t) &= \ff_{t,q} + \DD_{t,q}\qq;
\, \ff_{t,q}= 0;\,  \DD_{t,q}= \DD_q \\
\vec(\ZZ_t) &= \ff_{t,z} + \DD_{t,z}\zzeta;
\, \ff_{t,z} = 0;\, \DD_{t,z}=1;\, \zzeta=\vec(\ZZ)\\
\vec(\aa_t) &= \ff_{t,a} + \DD_{t,a}\aalpha;
\, \ff_{t,a} = 0;\, \DD_{t,a}=(\gg_t^\top \otimes \II_q);\, \aalpha=\vec(\FF)\\
\vec(\RR_t) &= \ff_{t,r} + \DD_{t,r}\rr;
\, \ff_{t,r}= 0;\,  \DD_{t,r}= \DD_r \\
\vec(\LAM)&= \ff_\lambda+\DD_\lambda\llambda;
\, \ff_\lambda= 0 \\
\vec(\xixi)&= \xixi=\ff_\xi+\DD_\xi\pp;
\, \ff_\xi = 0; \, \DD_\xi = 1
\end{split}
\end{equation*}
Note that variance-covariance matrices are never unconstrained really so we use $\DD_q$, $\DD_r$ and $\DD_\lambda$ to specify the symmetry within the matrix.

The transformation of the simple MARSS with covariates (equation \ref{eq:MARSS.simple.ex}) into the form of equation \eqref{eq:MARSS.ex.vec} may seem a little painful, but the advantage is that a single  EM algorithm can be used for a large class of models.  Presumably, the transformation of the equation will be hidden from users by a wrapper function that does the reformulation before passing the model to the general EM algorithm.  In the MARSS R package, this reformultion is done in the \verb@MARSS.marxss@ function.

\subsection{A general MARSS model with exogenous variables}
Let's imagine now a very general MARSS model with various `inputs'.  `
input' here just means that it is some fully known matrix rather than something we are estimating.  It could be a sequence of 0s and 1s if for example we were fitting a before/after sort of model.  Below the letters with a $t$ subscript are the inputs, except $\xx$, $\yy$, $\ww$ and $\vv$.
\begin{equation}\label{eq:MARSS.general.ex}
\begin{split}
\xx_t &= \Bb_t\BB\Ba_t\xx_{t-1} + \Ub_t\UU\Ua_t + \Qb_t\ww_t\\
\yy_t &= \Zb_t\ZZ\Za_t\xx_t + \Ab_t\AA\Aa_t + \Rb_t\vv_t 
\end{split}
\end{equation}
In vec form, this is:
\begin{equation}\label{eq:MARSS.ex3}
\begin{split}
\xx_t &= (\xx_{t-1}^\top \otimes \II_m)(\Ba_t^\top \otimes \Bb_t)\vec(\BB) + (\Ua_t^\top \otimes \Ub_t)\vec(\UU)
+ \Qb_t\ww_t\\
&= (\xx_{t-1}^\top \otimes \II_m)(\Ba_t^\top \otimes \Bb_t)(\ff_b+\DD_b\bbeta) + (\Ua_t^\top \otimes \Ub_t)(\ff_u + \DD_u\uupsilon) + \Qb_t\ww_t\\
\WW_t & \sim \MVN(0,\Qb_t\QQ\Qb_t^\top)\\
\\ 
\yy_t &= (\xx_t^\top \otimes \II_n)(\Za_t^\top \otimes \Zb_t)\vec(\ZZ) + (\Aa_t^\top \otimes \Ab_t)\vec(\AA) + \Rb_t\vv_t \\
&= (\xx_t^\top \otimes \II_n)\mathbb{Z}_t(\ff_z+\DD_z\zzeta) + \mathbb{A}_t(\ff_a+\DD_a\aalpha) + \Rb_t\vv_t \\
\VV_t &\sim \MVN(0,\Rb_t\RR\Rb_t^\top)\\
\\
\XX_{t_0} &\sim \MVN(\ff_\xi+\DD_\xi\pp,\FF\LAM\FF^\top), \text{ where } \vec(\LAM)=\ff_\lambda+\DD_\lambda\llambda
\end{split}
\end{equation}
We could write down a likelihood function for this model but written this way, the model presumes that $\Rb_t\RR\Rb_t^\top$, $\Qb_t\QQ\Qb_t^\top$, and $\FF\LAM\FF^\top$ are valid variance-covariance matrices.  I will actually write this model differently below because I don't want to make that assumption.

We define the $\ff$ and $\DD$ parameters as follows.
\begin{equation*}
\begin{split}
\vec(\BB_t) &= \ff_{t,b} + \DD_{t,b}\bbeta = (\Ba_t^\top \otimes \Bb_t)\ff_b + (\Ba_t^\top \otimes \Bb_t)\DD_b\bbeta\\
\vec(\uu_t) &= \ff_{t,u} + \DD_{t,u}\uupsilon = (\Ua_t^\top \otimes \Ub_t)\ff_u + (\Ua_t^\top \otimes \Ub_t)\DD_u\uupsilon\\
\vec(\QQ_t) &= \ff_{t,q} + \DD_{t,q}\qq = (\Qb_t \otimes \Qb_t)\ff_q + (\Qb_t \otimes \Qb_t)\DD_q\qq \\
\vec(\ZZ_t) &= \ff_{t,z} + \DD_{t,z}\zzeta = (\Za_t^\top \otimes \Zb_t)\ff_z + (\Za_t^\top \otimes \Zb_t)\DD_z\zzeta\\
\vec(\aa_t) &= \ff_{t,a} + \DD_{t,a}\aalpha = (\Aa_t^\top \otimes \Ab_t)\ff_a + (\Aa_t^\top \otimes \Ab_t)\DD_a\aalpha\\
\vec(\RR_t) &= \ff_{t,r} + \DD_{t,r}\rr = (\Rb_t \otimes \Rb_t)\ff_q + (\Rb_t \otimes \Rb_t)\DD_r\rr\\
\vec(\LAM)&= \ff_\lambda+\DD_\lambda\llambda = 0 + \DD_\lambda\llambda\\
\vec(\xixi)&= \xixi=\ff_\xi+\DD_\xi\pp = 0+1\pp
\end{split}
\end{equation*}
Here, for example $\ff_b$ and $\DD_b$ indicate the linear constraints on $\BB$ and $\ff_{t,b}$ is $(\Ba_t^\top \otimes \Bb_t)\ff_b$ and $\DD_{t,b}$ is $(\Ba_t^\top \otimes \Bb_t)\DD_b$.  The elements of $\BB$ that are being estimated are $\bbeta$ arranged as a column vector.

As usual, this reformulation looks cumbersome, but would be hidden from the user presumably.

\subsection{The expected log-likelihood function}
As mentioned above, we do not necessarily want to assume that $\GG_t\RR_t\GG_t^\top$, $\HH_t\QQ_t\HH_t^\top$, and $\FF\LAM\FF^\top$ are valid variance-covariance matrices.  This would rule out many MARSS models that we would like to fit.  For example, if $\QQ=\sigma^2$ and $\HH=\begin{bmatrix}1\\ 1\\ 1\end{bmatrix}$,  $\HH\QQ\HH^\top$ would be an invalid variance-variance matrix.  However, this is a valid MARSS model.

Instead I will define  $\Phi_t=(\HH_t^\top\HH_t)^{-1}\HH_t^\top$, $\Xi_t=(\GG_t^\top\GG_t)^{-1}\GG_t^\top$, and $\Pi = (\FF^\top\FF)^{-1}\FF^\top$. I then require that the inverses of $\GG_t^\top\GG_t$, $\HH_t^\top\HH_t$, and $\FF^\top\FF$ exist and that  $\ff_{t,q}+\DD_{t,q}\qq$, $\ff_{t,r}+\DD_{t,r}\rr$, and $\ff_\lambda+\DD_\lambda\llambda$ specify valid variance-covariance matrices. These are much less stringent restrictions.

For the purpose of writing down the expected log-likelihood, our MARSS model is now written
\begin{equation}\label{eq:MARSS.ex.reformed}
\begin{gathered}
\Phi_t\xx_t = \Phi_t( \xx_{t-1}^\top \otimes \II_m)\vec(\BB_t) + \Phi_t\vec(\uu_t) + \ww_t, \quad
\text{ where } \WW_t \sim \mathrm{MVN}(0,\QQ_t)\\
\Xi_t\yy_t = \Xi_t( \xx_t^\top \otimes \II_n)\vec(\ZZ_t) + \Xi_t\vec(\aa_t) + \vv_t,\quad \text{ where } \VV_t \sim \mathrm{MVN}(0,\RR_t)\\
\Pi\xx_{t_0}=\Pi\xixi+\ll, \quad \text{ where } \LL \sim \MVN(0,\LAM)
\end{gathered}
\end{equation}
As mentioned before, this relies on $\GG$ and $\HH$ having forms that do not lead to over- or under-constrained linear systems.

To derive the EM update equations, we need the expected log-likelihood function for the time-varying MARSS model.  Using equation \eqref{eq:MARSS.ex.reformed}, we get
\begin{equation}\label{eq:logL.vec.general}
\begin{split}
&\EXy[\log\LL(\YY,\XX ; \Theta)] = -\frac{1}{2}\EXy\bigg(
 \sum_1^T(\YY_t - ( \XX_t^\top \otimes \II_m)\vec(\ZZ_t) - \vec(\aa_t) )^\top\Xi_t^\top \RR_t^{-1}\Xi_t\\
&\quad (\YY_t-(\XX_t^\top \otimes \II_m)\vec(\ZZ_t) - \vec(\aa_t))+\sum_1^T \log |\RR_t|\\
&\quad +\sum_{t_0+1}^T(\XX_t-(\XX_{t-1}^\top \otimes \II_m)\vec(\BB_t) - \vec(\uu_t) )^\top \Phi_t^\top\QQ_t^{-1}\Phi_t \\
&\quad ( \XX_t-(\XX_{t-1}^\top \otimes \II_m)\vec(\BB_t) - \vec(\uu_t) )+\sum_{t_0+1}^T\log |\QQ_t|\\
&\quad +(\XX_{t_0}-\vec(\xixi))^\top \Pi^\top\LAM^{-1}\Pi (\XX_{t_0}-\vec(\xixi)) + \log |\LAM| + \log 2\pi \bigg)
\end{split}
\end{equation}
If any $\GG_t$, $\HH_t$ or $\FF$ is all zero, then the line in the likelihood with $\RR_t$, $\QQ_t$ or $\LAM$, respectively, does not appear.  If any $\xx_{t_0}$ are fixed, meaning all zero row in $\FF$, that $\XX_{t_0}\equiv\xixi$ anywhere it appears in the likelihood.  The way I have written the general equation, some $\xx_{t_0}$ might be fixed and others stochastic.

The vec of the model parameters are defined as follows:
\begin{equation*}
\begin{split}
\vec(\BB_t)&=\ff_{t,b}+\DD_{t,b}\bbeta\\
\vec(\uu_t) &= \ff_{t,u}+\DD_{t,u}\uupsilon\\
\vec(\ZZ_t)&=\ff_{t,z}+\DD_{t,z}\zzeta\\
\vec(\aa_t) &= \ff_{t,a}+\DD_{t,a}\aalpha\\
\vec(\QQ_t)&=\ff_{t,q}+\DD_{t,q}\qq\\
\vec(RR_t)&=\ff_{t,r}+\DD_{t,r}\rr\\
\vec(\xixi)&=\ff_\xi+\DD_\xi\pp\\
\vec(\LAM)&=\ff_\lambda+\DD_\lambda\llambda\\
\Phi_t&=(\GG_t^\top\GG_t)^{-1}\GG_t^\top\\
\Xi_t&=(\HH_t^\top\HH_t)^{-1}\HH_t^\top\\
\Pi&=(\FF^\top\FF)^{-1}\FF^\top
\end{split}
\end{equation*}

\section{The constrained update equations}\label{sec:constrained}
The derivation proceeds by taking the partial derivative of equation \ref{eq:logL.vec.general} with respect to the estimated terms, the $\zzeta$, $\aalpha$, etc, setting the derivative to zero, and solving for those estimated terms.  Conceptually, the algebraic steps in the derivation are similar to those in  the unconstrained derivation.  

\subsection{The general $\uu$ update equations}
We take the derivative of $\Psi$ (equation \ref{eq:logL.vec.general}) with respect to $\uupsilon$.
\begin{equation}\label{eq:u.general1}
\begin{split}
\partial\Psi/\partial\pmb{\upsilon}
 &= - \frac{1}{2}\sum_{t=1}^T \bigg(-\partial(\E[\XX_t^\top\Qm_t\DD_{t,u}\uupsilon])/\partial\pmb{\upsilon}
 - \partial(\E[\uupsilon^\top\DD_{t,u}^\top\Qm_t\XX_t])/\partial\pmb{\upsilon} \\
&+ \partial(\E[(( \XX_{t-1}^\top \otimes \II_m)\vec(\BB_t))^\top\Qm_t\DD_{t,u}\uupsilon] )/\partial\pmb{\upsilon} 
 + \partial(\E[\uupsilon^\top\DD_{t,u}^\top\Qm_t( \XX_{t-1}^\top \otimes \II_m)\vec(\BB_t)])/\partial\pmb{\upsilon}\\
&+ \partial(\uupsilon^\top\DD_{t,u}^\top\Qm_t\DD_{t,u}\uupsilon)/\partial\pmb{\upsilon} 
+ \partial(\E[\ff_{t,u}^\top\Qm_t\DD_{t,u}\uupsilon])/\partial\pmb{\upsilon}
 + \partial(\E[\uupsilon^\top\DD_{t,u}^\top\Qm_t\ff_{t,u}])/\partial\pmb{\upsilon}
\bigg)
\end{split}
\end{equation}
where $\Qm_t = \Phi_t^\top\QQ_t^{-1}\Phi_t$.

Since $\uupsilon$ is to the far left or right in each term, the derivative is simple using the derivative terms in table \ref{sec:MatrixDerivatives}.
$\partial\Psi/\partial\pmb{\upsilon}$ becomes:
\begin{equation}\label{eq:u.general1b}
\begin{split}
\partial\Psi/\partial\pmb{\upsilon}
&= - \frac{1}{2}\sum_{t=1}^T \bigg(  
-2\E[\XX_t^\top\Qm_t\DD_{t,u}] +  2\E[(( \XX_{t-1}^\top \otimes \II_m)\vec(\BB_t))^\top \Qm_t\DD_{t,u}] \\
&\quad + 2(\uupsilon^\top\DD_{t,u}^\top\Qm_t\DD_{t,u}) + 2\E[ \ff_{t,u}^\top\Qm_t\DD_{t,u} ] \bigg)
\end{split}\end{equation}
Set the left side to zero and transpose the whole equation. 
\begin{equation}\label{eq:u.general4}
\begin{split}
\mathbf{0}= \sum_{t=1}^T \bigg( \DD_{t,u}^\top\Qm_t\E[\XX_t] - \DD_{t,u}^\top\Qm_t(\E[\XX_{t-1}]^\top \otimes \II_m)\vec(\BB_t) 
- \DD_{t,u}^\top\Qm_t\DD_{t,u}\uupsilon
 - \DD_{t,u}^\top\Qm_t\ff_{t,u} \bigg)
\end{split}
\end{equation}
Thus,
\begin{equation}\label{eq:u.general5}
\big(\sum_{t=1}^T \DD_{t,u}^\top\Qm_t\DD_{t,u} \big)\uupsilon
 = \sum_{t=1}^T \DD_{t,u}^\top\Qm_t\big(\E[ \XX_t ]- (\E[\XX_{t-1}]^\top \otimes \II_m)\vec(\BB_t) - \ff_{t,u} \big)
\end{equation}
We solve for $\uupsilon$, and the new $\uupsilon$ for the $j+1$ iteration of the EM algorithm is
\begin{equation}\label{eq:u.general.update1}
\begin{split}
\uupsilon_{j+1} &= \bigg(\sum_{t=1}^T \DD_{t,u}^\top\Qm_t\DD_{t,u} \bigg)^{-1}
  \sum_{t=1}^T \DD_{t,u}^\top\Qm_t\big(\hatxt- (\hatxtm^\top \otimes \II_m)\vec(\BB_t) - \ff_{t,u} \big)
\end{split}
\end{equation}

The update equation requires that $\sum_{t=1}^T \DD_{t,u}^\top\Qm_t\DD_{t,u}$ is invertible. It generally will be if $\Phi_t\QQ_t\Phi_t^\top$ is a proper variance-covariance matrix (positive semi-definite) and $\DD_{t,u}$ is full rank.  If $\GG_t$ has all-zero rows then $\Phi_t\QQ_t\Phi_t^\top$ has zeros on the diagonal and we have a partially deterministic model.  In this case, $\Qm_t$ will have all-zero row/columns and $\DD_{t,u}^\top\Qm_t\DD_{t,u}$ will not be invertible unless the corresponding row of $\DD_{t,u}$ is zero.  This means that if one of the $\xx$ rows is fully deterministic then the corresponding row of $\uu$ would need to be fixed.  We can get around this, however.  See section \ref{sec:degenerate} on the modifications to the update equation when some of the $\xx$'s are fully deterministic.

\subsection{The general $\aa$ update equation}\label{sec:constA}
The derivation of the update equation for $\aalpha$ with fixed and shared values is completely analogous to the derivation for $\uupsilon$. We take the derivative of $\Psi$ with respect to $\aalpha$ and arrive at the analogous:
\begin{equation}\label{eq:general.a.update}
\begin{split}
\aalpha_{j+1} &= \big(\sum_{t=1}^T \DD_{t,a}^\top\Rm_t\DD_{t,a} \big)^{-1}
  \sum_{t=1}^T \DD_{t,a}^\top\Rm_t\big(\hatyt- (\hatxt^\top \otimes \II_n)\vec(\ZZ_t) - \ff_{t,a} \big) \\
  &= \big(\sum_{t=1}^T \DD_{t,a}^\top\Rm_t\DD_{t,a} \big)^{-1}
  \sum_{t=1}^T \DD_{t,a}^\top\Rm_t\big(\hatyt- \ZZ_t\hatxt - \ff_{t,a} \big)
\end{split}
\end{equation}
$\sum_{t=1}^T \DD_{t,a}^\top\Rm_t\DD_{t,a}$ must be invertible.

\subsection{The general $\xixi$ update equation, stochastic initial state}
When $\xx_0$ is treated as stochastic with an unknown mean and known variance, the derivation of the update equation for $\xixi$ with fixed and shared values is as follows. Take the derivative of $\Psi$ (using equation \ref{eq:logL.vec.general}) with respect to $\pp$:
\begin{equation}\label{eq:pi.general1}
\partial\Psi/\partial\pp =  
\big(\widetilde{\mbox{$\mathbf x$}}_0^\top \LAMm - \xixi^\top\LAMm \big)
\end{equation}
Replace $\xixi$ with $\ff_\xi + \DD_\xi\pp$, set the left side to zero and transpose:
\begin{equation}\label{eq:pi.general2}
\mathbf{0} = 
\DD_\xi^\top\big(\LAMm\widetilde{\mbox{$\mathbf x$}}_0 - \LAMm\ff_\xi + \LAMm\DD_\xi\pp \big) 
\end{equation}
Thus,
\begin{equation}\label{eq:pi.update.general1}
\pp_{j+1} = \big(\DD_\xi^\top\LAMm\DD_\xi \big)^{-1}
\DD_\xi^\top\LAMm(\widetilde{\mbox{$\mathbf x$}}_0 - \ff_\xi)
\end{equation}
and the new $\xixi$ is then,
\begin{equation}\label{eq:pi.update.general2}
\xixi_{j+1} = \ff_\xi + \DD_\xi\pp_{j+1},
\end{equation}
When the initial state is defined as at $t=1$, replace $\widetilde{\mbox{$\mathbf x$}}_0$ with $\widetilde{\mbox{$\mathbf x$}}_1$ in equation \ref{eq:pi.update.general1}.

\subsection{The general $\xixi$ update equation, fixed $\xx_0$}\label{sec:xi.constrained.x0}
For the case, $\xx_0$ is treated as fixed, i.e. as another parameter, and $\LAM$ does not appear in the equation. It will be easier to work with $\Psi$ written as follows:
\begin{equation}\label{eq:logL.vec.general.V0.is.0b}
\begin{split}
&\EXy[\log\LL(\YY,\XX ; \Theta)] = -\frac{1}{2}\EXy\bigg(
 \sum_1^T(\YY_t - \ZZ_t\XX_t - \aa_t)^\top \Rm_t(\YY_t - \ZZ_t\XX_t - \aa_t)
 +\sum_1^T \log |\RR_t|\\
&\quad +\sum_1^T(\XX_t - \BB_t\XX_{t-1} - \uu_t)^\top \Qm_t (\XX_t - \BB_t\XX_{t-1} - \uu_t)
 +\sum_1^T\log |\QQ_t| + \log 2\pi\bigg)\\
&\quad \xx_0 \equiv \ff_\xi+\DD_\xi\pp  
\end{split}
\end{equation}
This is the same as equation \eqref{eq:logL.vec.general} except not written in vec form and $\LAM$ does not appear.  Take the derivative of $\Psi$ using equation \eqref{eq:logL.vec.general.V0.is.0b}. Terms not involving $\pp$ will drop out:
\begin{equation}\label{eq:pi.constrained.V0.is.0}
\begin{split}
&\partial\Psi/\partial\pp = -\frac{1}{2} \bigg( -\E[\partial(\mathbb{P}_1^\top\Qm_1\BB_1\DD_\xi\pp)/\partial\pp]  -\E[\partial(\pp^\top(\BB_1\DD_\xi)^\top\Qm_1\mathbb{P}_1)/\partial\pp] \\
&\quad +\E[\partial(\pp^\top(\BB_1\DD_\xi)^\top\Qm_1\BB_1\DD_\xi\pp)/\partial\pp]  
\bigg)
\end{split}
\end{equation}
where
\begin{equation}
\mathbb{P}_1=\XX_1 - \BB_1\ff_\xi  - \uu_1
\end{equation}

After pulling the constants out of the expectations and taking the derivative, we arrive at:
\begin{equation}
\begin{split}
\partial\Psi/\partial\pp = -\frac{1}{2} \bigg( - 2\E[\mathbb{P}_1]^\top\Qm_1\BB_1\DD_\xi
+ 2\pp^\top(\BB_1\DD_\xi)^\top\Qm_1\BB_1\DD_\xi  
\bigg)
\end{split}
\end{equation}
Set the left side to zero, and solve for $\pp$.
\begin{equation}\label{eq:pi.constrained.V0.is.0.4}
\pp = (\DD_\xi^\top\BB_1^\top\Qm_1\BB_1\DD_\xi)^{-1}\DD_\xi^\top\BB_1^\top\Qm_1(\hatxone - \BB_1\ff_\xi  - \uu_1)
\end{equation}

This equation requires that the inverse right of the $=$ exists and it might not if $\BB_t$ or $\Qm_1$ has any all zero rows/columns.  In that case, defining $\xixi \equiv \xx_1$ might work (section \ref{sec:general.x1.update}) or the problematic rows of $\xixi$ could be fixed.
The new $\xixi$ is then,
\begin{equation}
\xixi_{j+1} = \ff_\xi + \DD_\xi\pp_{j+1},
\end{equation}

\subsection{The general $\xixi$ update equation, fixed $\xx_1$}\label{sec:general.x1.update}
When $\xx_1$ is treated as fixed, i.e. as another parameter, and $\LAM$ does not appear, the expected log likelihood, $\Psi$, is written as follows:
\begin{equation}\label{eq:logL.vec.general.V0.is.0.x1}
\begin{split}
&\EXy[\log\LL(\YY,\XX ; \Theta)] = -\frac{1}{2}\EXy\bigg(
 \sum_1^T(\YY_t - \ZZ_t\XX_t - \aa_t)^\top \Rm_t (\YY_t - \ZZ_t\XX_t - \aa_t) +\sum_1^T \log |\RR_t|\\
&\quad +\sum_2^T(\XX_t - \BB_t\XX_{t-1} - \uu_t)^\top \Qm_t (\XX_t - \BB_t\XX_{t-1} - \uu_t)
 +\sum_2^T\log |\QQ_t| + \log 2\pi\bigg)\\
&\quad \xx_1 \equiv \ff_\xi+\DD_\xi\pp  
\end{split}
\end{equation}
Take the derivative of $\Psi$ using equation \eqref{eq:logL.vec.general.V0.is.0.x1}:
\begin{equation}
\begin{split}
&\partial\Psi/\partial\pp = -\frac{1}{2} \bigg(
-\E[\partial(\mathbb{O}_1^\top\Rm_1\ZZ_1\DD_\xi\pp)/\partial\pp]  -\E[\partial((\ZZ_1\DD_\xi\pp)^\top\Rm_1\mathbb{O}_1)/\partial\pp] \\
&\quad +\E[\partial((\ZZ_1\DD_\xi\pp)^\top\Rm_1\ZZ_1\DD_\xi\pp)/\partial\pp]
 -\E[\partial(\mathbb{P}_2^\top\Qm_2\BB_2\DD_\xi\pp)/\partial\pp]  -\E[\partial((\BB_2\DD_\xi\pp)^\top\Qm_2\mathbb{P}_2)/\partial\pp] \\
&\quad +\E[\partial((\BB_2\DD_\xi\pp)^\top\Qm_2\BB_2\DD_\xi\pp)/\partial\pp]  
\bigg)
\end{split}
\end{equation}
where
\begin{equation}
\begin{split}
\mathbb{P}_2&=\XX_2 - \BB_2\ff_\xi  - \uu_2\\
\mathbb{O}_1&=\YY_1 - \ZZ_1\ff_\xi  - \aa_1\\
\end{split}
\end{equation}
 
In terms of the Kalman smoother output the new $\xixi$ for EM iteration $j+1$ when $\xixi \equiv \xx_1$ is
\begin{equation}\label{eq:pix1.unconstrained.V0.is.0.t.1.3}
\begin{split}
\pp_{j+1} &= ((\ZZ_1\DD_\xi)^\top\Rm_1\ZZ_1\DD_\xi + 
(\BB_2\DD_\xi)^\top\Qm_2\BB_2\DD_\xi)^{-1}
( (\ZZ_1\DD_\xi)^\top\Rm_1\widetilde{\mathbb{O}}_1 + (\BB_2\DD_\xi)^\top\Qm_2\widetilde{\mathbb{P}}_2)
\end{split}
\end{equation}
where
\begin{equation}
\begin{split}
\widetilde{\mathbb{P}}_2 &=\widetilde{\mbox{$\mathbf x$}}_2 - \BB_2\ff_\xi  - \uu_2\\
\widetilde{\mathbb{O}}_1 &=\widetilde{\mbox{$\mathbf y$}}_1 - \ZZ_1\ff_\xi  - \aa_1
\end{split}
\end{equation}
The new $\xixi$ is
\begin{equation}
\xixi_{j+1} = \ff_\xi + \DD_\xi\pp_{j+1},
\end{equation}

\subsection{The general $\BB$ update equation}
Take the derivative of $\Psi$ with respect to $\bbeta$; terms in $\Psi$  do not involve $\bbeta$ will equal 0 and drop out.  
\begin{equation}\label{eq:B.general1}
\begin{split}
\partial\Psi/\partial\bbeta
 &= - \frac{1}{2}\sum_{t=1}^T \bigg( -\partial(\E[\XX_t^\top\Qm_t\Bm_t\DD_{t,b}\bbeta])/\partial\bbeta
 - \partial(\E[(\Bm_t\DD_{t,b}\bbeta)^\top\Qm_t\XX_t])/\partial\bbeta \\
&+ \partial(\E[(\Bm_t\DD_{t,b}\bbeta)^\top\Qm_t\Bm_t\DD_{t,b}\bbeta])/\partial\bbeta
 + \partial(\E[\uu_t^\top\Qm_t\Bm_t\DD_{t,b}\bbeta])/\partial\bbeta
+ \partial((\Bm_t\DD_{t,b}\bbeta)^\top\Qm_t\uu_t)/\partial\bbeta \\
&+ \partial(\E[(\Bm_t\ff_{t,b})^\top\Qm_t\Bm_t\DD_{t,b}\bbeta])/\partial\bbeta
 + \partial(\E[(\Bm_t\DD_{t,b}\bbeta)^\top\Qm_t\Bm_t\ff_{t,b}])/\partial\bbeta
\bigg)\end{split}
\end{equation}
where
\begin{equation}
\Bm_t = (\XX_{t-1}^\top \otimes \II_m)
\end{equation}Since $\bbeta$ is to the far left or right in each term, the derivative is simple using the derivative terms in table \ref{sec:MatrixDerivatives}.
$\partial\Psi/\partial\bbeta$ becomes:
\begin{equation}\label{eq:B.general1b}
\begin{split}
\partial\Psi/\partial\pmb{\upsilon}
&= - \frac{1}{2}\sum_{t=1}^T \bigg(  
-2\E[\XX_t^\top\Qm_t\Bm_t\DD_{t,b}] + 2(\beta^\top\DD_{t,b}^\top\Bm_t^\top\Qm_t\Bm_t\DD_{t,b})  \\
&+ 2\E[\uu_t^\top\Qm_t\Bm_t\DD_{t,b}] + 2\E[(\Bm_t\ff_{t,b})^\top\Qm_t\Bm_t\DD_{t,b}] \bigg)
\end{split}\end{equation}
Note that $\XX$ appears in $\Bm_t$ but not in other terms. We need to keep track of where $\XX$ appears so the we keep the expectation brackets around any terms involving $\XX$.
\begin{equation}
\begin{split}
\partial\Psi/\partial\bbeta
= \sum_{t=1}^T \bigg(  
\E[\XX_t^\top\Qm_t\Bm_t]\DD_{t,b} - \uu_t^\top\Qm_t\E[\Bm_t]\DD_{t,b} 
- \bbeta^\top\DD_{t,b}^\top\E[\Bm_t^\top\Qm_t\Bm_t]\DD_{t,b} - \ff_{t,b}^\top\E[\Bm_t^\top\Qm_t\Bm_t]\DD_{t,b} \bigg)
\end{split}
\end{equation}
Set the left side to zero and transpose the whole equation. 
\begin{equation}
\mathbf{0}
 = \sum_{t=1}^T \bigg(
\DD_{t,b}^\top\E[\Bm_t^\top\Qm_t\XX_t]- \DD_{t,b}^\top\E[\Bm_t]^\top\Qm_t\uu_t
 - \DD_{t,b}^\top\E[\Bm_t^\top\Qm_t\Bm_t]\ff_{t,b}
 - \DD_{t,b}^\top\E[\Bm_t^\top\Qm_t\Bm_t]\DD_{t,b}\bbeta \bigg)
\end{equation}
Thus,
\begin{equation}
\big(\sum_{t=1}^T \DD_{t,b}^\top\E[\Bm_t^\top\Qm_t\Bm_t]\DD_{t,b} \big)\bbeta
 = \sum_{t=1}^T \DD_{t,b}^\top\big(
 \E[\Bm_t^\top\Qm_t\XX_t]
 - \E[\Bm_t]^\top\Qm_t\uu_t
  - \E[\Bm_t^\top\Qm_t\Bm_t]\ff_{t,b} \big)
\end{equation}
Now we need to deal with the expectations.
\begin{equation}
\begin{split}
\E[\Bm_t^\top\Qm_t\Bm_t]
&=\E[(\XX_{t-1}^\top \otimes \II_m)^\top\Qm_t(\XX_{t-1}^\top \otimes \II_m)]\\
&=\E[(\XX_{t-1} \otimes \II_m)\Qm_t(\XX_{t-1}^\top \otimes \II_m)]\\
&=\E[\XX_{t-1}\XX_{t-1}^\top \otimes \Qm_t]\\
&=\E[\XX_{t-1}\XX_{t-1}^\top] \otimes \Qm_t\\
&=\hatPtm \otimes \Qm_t
\end{split}
\end{equation}

\begin{equation}
\begin{split}
\E[\Bm_t^\top\Qm_t\XX_t]
&=\E[(\XX_{t-1}^\top \otimes \II_m)^\top\Qm_t\XX_t]\\
&=\E[(\XX_{t-1} \otimes \II_m)\Qm_t\XX_t]\\
&=\E[(\XX_{t-1} \otimes \Qm_t)\XX_t]\\
&=\E[\vec(\Qm_t\XX_t\XX_{t-1}^\top)]\\
&=\vec(\Qm_t\hatPttm)
\end{split}
\end{equation}

\begin{equation}
\begin{split}
\E[\Bm_t]^\top\Qm_t\uu_t&=(\E[\XX_{t-1}] \otimes \II_m)\Qm_t\uu_t \\
&= (\hatxtm \otimes \Qm_t)\uu_t \\
&= \vec(\Qm_t\uu_t\hatxtm^\top)
\end{split}
\end{equation}

Thus,
\begin{equation}
\begin{split}
\big(\sum_{t=1}^T &\DD_{t,b}^\top(\hatPtm  \otimes \Qm_t)\DD_{t,b} \big)\bbeta 
 =  \sum_{t=1}^T \DD_{t,b}^\top \big(\vec(\Qm_t\hatPttm) - (\hatPtm \otimes \Qm_t)\ff_{t,b} - \vec(\Qm_t\uu_t\hatxtm^\top)  \big)
\end{split}
\end{equation}
Then $\bbeta$ for the $j+1$ iteration of the EM algorithm is then:
\begin{equation}\label{eq:B.general4}
\begin{split}
\bbeta = \bigg(\sum_{t=1}^T \DD_{t,b}^\top(\hatPtm  \otimes \Qm_t)\DD_{t,b} \bigg)^{-1}
\times \sum_{t=1}^T \DD_{t,b}^\top \big(\vec(\Qm_t\hatPttm) - (\hatPtm \otimes \Qm_t)\ff_{t,b} - \vec(\Qm_t\uu_t\hatxtm^\top)  \big)
\end{split}
\end{equation}

This requires that $\DD_{t,b}^\top(\hatPtm  \otimes \Qm_t)\DD_{t,b}$ is invertible, and as usual we will run into trouble if $\Phi_t\QQ_t\Phi_t^\top$ has zeros on the diagonal. See section \ref{sec:degenerate}.

\subsection{The general $\ZZ$ update equation}\label{sec:constZ}
The derivation of the update equation for $\zzeta$ with fixed and shared values is analogous to the derivation for $\bbeta$.  The update equation for $\zzeta$ is
\begin{equation}\label{eq:general.Z.update}
\begin{split}
\zzeta_{j+1} = 
\big(\sum_{t=1}^T \DD_{t,z}^\top(\hatPt  \otimes \Rm_t)\DD_{t,z} \big)\bbeta 
\times  \sum_{t=1}^T \DD_{t,z}^\top \big(\vec(\Rm_t\hatYXt) - (\hatPt \otimes \Rm_t)\ff_{t,z} - \vec(\Rm_t\aa_t\hatxt^\top)  \big)
\end{split}
\end{equation}

This requires that $\DD_{t,z}^\top(\hatPt \otimes \Rm_t)\DD_{t,z}$ is invertible. If $\Xi_t\RR_t\Xi_t^\top$ has zeros on the diagonal, this will not be the case. See section \ref{sec:degenerate}. 

\subsection{The general $\QQ$ update equation}\label{sec:constrained.Q}
A general analytical solution for $\QQ$ is problematic because the inverse of $\QQ_t$ appears in the likelihood and  $\QQ_t^{-1}$ cannot always be rewritten as a function of $\vec(\QQ_t)$. However, in a few important special---yet quite broad--- cases, an analytical solution can be derived.  The most general of these special cases is a block-symmetric matrix with optional independent fixed blocks (subsection \ref{sec:Q.general}).  Indeed, all other cases (diagonal, block-diagonal, unconstrained, equal variance-covariance) except one (a replicated block-diagonal) are special cases of the blocked matrix with optional independent fixed blocks. 

Unlike the other parameters, I need to put constraints on $\ff$ and $\DD$.  I constrain $\DD$ to be a design matrix.  It has only 1s and 0s, and the rows sums are either 1 or 0. Thus terms like $q_1+q_2$ are not allowed.  A non-zero value in $\ff$ is only allowed if the corresponding row in $\DD$ is all zero.  Thus elements like $f_1+q_1$ are not allowed in $\QQ$.  These constraints, especially the constraint that $\DD$ only has 0s and 1s, might be loosened, but with the addition of $\GG_t$, we still have a very wide class of $\QQ$ matrices.

The general update equation for $\QQ$ with these constraints is
\begin{equation}\label{eq:Q.general}
\begin{split}
\qq_{j+1} &= \big(\sum_{t=1}^T(\DD_{t,q}^\top\DD_{t,q})\big)^{-1} \sum_{t=1}^T\DD_{t,q}^\top\vec(\SS_t)\\
\text{where }\SS_t&=\Phi_t\big(\hatPt - \hatPttm \BB_t^\top - \BB_t\hatPtmt 
- \hatxt\uu_t^\top - \uu_t\hatxt^\top + \\
&\quad \BB_t\hatPtm\BB_t^\top + \BB_t\hatxtm\uu_t^\top + \uu_t\hatxtm^\top\BB_t^\top + \uu_t\uu_t^\top \big)\Phi_t^\top\\
\QQ_t&=\ff_{t,q}+\DD_{t,q}\qq \\
\text{where}\\
\Phi_t = (\GG_t^\top\GG_t)^{-1}\GG_t^\top
\end{split}
\end{equation}

The vec of $\QQ_t$ is written in the form of $\vec(\QQ_t) = \ff_{t,q} + \DD_{t,q} \pmb{q}$, where $\ff_{t,q}$ is a $p^2 \times 1$ column vector of the fixed values including zero, $\DD_{t,q}$ is the $p^2 \times s$ design matrix, and $\pmb{q}$ is a column vector of the $s$ free values in $\QQ_t$.  This requires that $(\DD_{t,q}^\top\DD_{t,q})$ be invertible, which in a valid model must be true; if is not true you have specified an invalid variance-covariance structure since the implied variance-covariance matrix will not be full-rank and not invertible and thus an invalid variance-covariance matrix.

Below I show how the $\QQ$ update equation arises by working through a few of the special cases.  In these derivations the $q$ subscript is left off the $\DD$ and $\ff$ matrices.

\subsubsection{Special case: diagonal $\QQ$ matrix (with shared or unique parameters)}
Let $\QQ$ be a non-time varying diagonal matrix with fixed and shared values such that it takes a form like so:
\begin{equation*}
\QQ=
\begin{bmatrix}
q_1&0&0&0&0\\
0&f_1&0&0&0\\
0&0&q_2&0&0\\
0&0&0&f_2&0\\
0&0&0&0&q_2
\end{bmatrix}
\end{equation*}
Here, $f$'s are fixed values (constants) and $q$'s are free parameters elements.  The $f$ and $q$ do not occur together; i.e. there are no terms like $f_1+q_1$.

The vec of $\QQ^{-1}$ can be written then as $\vec(\QQ^{-1}) = \ff^{*}_q + \DD_q \pmb{q^{*}}$, where $\ff^{*}$ is like $\ff_q$ but with the corresponding $i$-th non-zero fixed values replaced by $1/f_i$ and $\pmb{q^{*}}$ is a column vector of 1 over the $q_i$ values.  For the example above,
\begin{equation*}
\pmb{q^{*}} =
\begin{bmatrix}
1/q_1 \\ 1/q_2
\end{bmatrix}
\end{equation*}

Take the partial derivative of $\Psi$ with respect to $\pmb{q^{*}}$.  We can do this because $\QQ^{-1}$ is diagonal and thus each element of $\pmb{q^{*}}$ is independent of the other elements; otherwise we would not necessarily be able to vary one element of $\pmb{q^{*}}$ while holding the other elements constant.
\begin{equation}
\begin{split}
&\partial\Psi/\partial\pmb{q^{*}} = -\frac{1}{2} \sum_{t=1}^T\partial\bigg(
\E[\XX_t^\top\Phi_t^\top\QQ^{-1}\Phi_t\XX_t]
-\E[\XX_t^\top\Phi_t^\top\QQ^{-1}\Phi_t\BB_t\XX_{t-1}] \\
&\quad -\E[(\BB_t\XX_{t-1})^\top\Phi_t^\top\QQ^{-1}\Phi_t\XX_t] 
 - \E[\XX_t^\top\Phi_t^\top\QQ^{-1}\Phi_t\uu_t] \\
&\quad - \E[\uu_t^\top\Phi_t^\top\QQ^{-1}\Phi_t\XX_t] 
+ \E[(\BB_t\XX_{t-1})^\top\Phi_t^\top\QQ^{-1}\Phi_t\BB_t\XX_{t-1}] \\
&\quad + \E[(\BB_t\XX_{t-1})^\top\Phi_t^\top\QQ^{-1}\Phi_t\uu_t] 
+ \E[\uu_t^\top\Phi_t^\top\QQ^{-1}\Phi_t\BB_t\XX_{t-1}] + \uu_t^\top\Phi_t^\top\QQ^{-1}\Phi_t\uu_t \bigg)/\partial\pmb{q^{*}}\\
& - \partial\big(\frac{T}{2}\log |\QQ| \big)/\partial\pmb{q^{*}} \\
\end{split}
\end{equation}
Using the same vec operations as in the derivations for $\BB$ and $\ZZ$, pull $\QQ^{-1}$ out from the middle and replace the expectations with the Kalman smoother output.\footnote{Another, more common, way to do this is to use a ``trace trick'', $\trace(\aa^\top\AA\bb)=\trace(\AA\bb\aa^\top)$, to pull $\QQ^{-1}$ out.}
\begin{equation}\label{eq:Q.gendiag2}
\begin{split}
&\partial\Psi/\partial\pmb{q^{*}} = -\frac{1}{2} \sum_{t=1}^T\partial\bigg(
\E[\XX_t^\top \otimes \XX_t^\top ]
-\E[\XX_t^\top \otimes (\BB_t\XX_{t-1})^\top] -\E[(\BB_t\XX_{t-1})^\top \otimes \XX_t^\top] \\
&\quad  - \E[\XX_t^\top \otimes \uu_t^\top] - \E[\uu_t^\top \otimes \XX_t^\top]
+ \E[(\BB_t\XX_{t-1})^\top \otimes (\BB_t\XX_{t-1})^\top] \\
&\quad + \E[(\BB_t\XX_{t-1})^\top \otimes \uu_t^\top]
+ \E[\uu_t^\top \otimes (\BB\XX_{t-1})^\top] + (\uu_t^\top \otimes \uu_t^\top) \bigg)(\Phi_t \otimes \Phi_t)^\top\vec(\QQ^{-1})/\partial\pmb{q^{*}}\\
& - \partial\bigg(\frac{T}{2}\log |\QQ| \bigg)/\partial\pmb{q^{*}} \\
&\quad = -\frac{1}{2} \sum_{t=1}^T\vec(\SS_t)^\top\partial\big(\vec(\QQ^{-1})\big)/\partial\pmb{q^{*}}
+ \partial\big(\frac{T}{2}\log|\QQ^{-1}| \big)/\partial\pmb{q^{*}} \\
&\text{where }\\
&\SS_t=\Phi_t\big(\hatPt - \hatPttm \BB_t^\top - \BB\hatPtmt 
- \hatxt\uu_t^\top - \uu_t \hatxt^\top + \\
&\quad \BB_t\hatPtm\BB_t^\top + \BB_t\hatxtm\uu_t^\top + \uu_t\hatxtm^\top\BB_t^\top + \uu_t\uu_t^\top \big)\Phi_t^\top
\end{split}
\end{equation}
This reduction used
$$(\Phi_t \otimes \Phi_t)(\XX \otimes \XX)=\vec(\XX\XX^\top)=\vec(\Phi_t\vec(\XX\XX^\top)\Phi_t^\top).$$
I also  replaced $\log|\QQ|$ with $-\log|\QQ^{-1}|$; the determinant of a diagonal matrix is the product of its diagonal elements.  Thus,
\begin{equation}\label{eq:Q.gendiag3}
\begin{split}
&\partial\Psi/\partial\pmb{q^{*}}= 
 -\bigg(\frac{1}{2} \sum_{t=1}^T \vec(\SS_t)^\top (\ff^{*} + \DD_q\pmb{q^{*}})  \\
&\quad - \frac{1}{2}\sum_{t=1}^T(\log(f^{*}_1) + \log(f^{*}_2) ... k\log(q^{*}_1) + l\log(q^{*}_2)...)\bigg)/\partial\pmb{q^{*}}\\
\end{split}
\end{equation}
where $k$ is the number of times $q_1$ appears on the diagonal of $\QQ$ and $l$ is the number of times $q_2$ appears, etc.
Taking the derivatives,
\begin{equation}\label{eq:Q.gendiag4}
\begin{split}
&\partial\Psi/\partial\pmb{q^{*}} = 
= \frac{1}{2} \sum_{t=1}^T\DD_q^\top \vec(\SS_t) - \frac{1}{2}\sum_{t=1}^T(\log(f^{*}_1) + ... k\log(q^{*}_1) + l\log(q^{*}_2)...)/\partial\pmb{q^{*}}\\
&\quad = \frac{1}{2} \sum_{t=1}^T\DD_q^\top \vec(\SS_t) - \frac{1}{2}\sum_{t=1}^T\DD_q^\top\DD_q\pmb{q}
\end{split}
\end{equation}
$\DD_q^\top\DD_q$ is a $s \times s$ matrix with $k$, $l$, etc. along the diagonal and thus is invertible; as usual, $s$ is the number of free elements in $\QQ$.  Set the left side to zero (a $1 \times s$ matrix of zeros) and solve for $\pmb{q}$.  This gives us the update equation for $\qq$ and $\QQ$:
\begin{equation}\label{eq:Q.diag.update}
\begin{split}
\qq_{j+1} &= \big(\sum_{t=1}^T\DD_q^\top\DD_q\big)^{-1}\sum_{t=1}^T\DD_q^\top\vec(\SS_t)\\
\vec(\QQ)_{j+1} &= \ff + \DD_q\pmb{q}_{j+1}
\end{split}
\end{equation}
Since in this example, $\DD_q$ is time-constant, this reduces to 
\begin{equation*}
\pmb{q}_{j+1} = \frac{1}{T}(\DD_q^\top\DD_q)^{-1}\DD_q^\top\sum_{t=1}^T\vec(\SS_t)
\end{equation*}
$\SS_t$ is defined in equation \eqref{eq:Q.gendiag2}.

\subsubsection{Special case: $\QQ$ with one variance and one covariance}
\begin{equation*}
\QQ=
\begin{bmatrix}
\alpha&\beta&\beta&\beta\\
\beta&\alpha&\beta&\beta\\
\beta&\beta&\alpha&\beta\\
\beta&\beta&\beta&\alpha
\end{bmatrix}\quad\quad
\QQ^{-1}=
\begin{bmatrix}
f(\alpha,\beta)&g(\alpha,\beta)&g(\alpha,\beta)&g(\alpha,\beta)\\
g(\alpha,\beta)&f(\alpha,\beta)&g(\alpha,\beta)&g(\alpha,\beta)\\
g(\alpha,\beta)&g(\alpha,\beta)&f(\alpha,\beta)&g(\alpha,\beta)\\
g(\alpha,\beta)&g(\alpha,\beta)&g(\alpha,\beta)&f(\alpha,\beta)
\end{bmatrix}
\end{equation*}
This is a matrix with a single shared variance parameter on the diagonal and a single shared covariance on the off-diagonals.  The derivation is the same as for the diagonal case, until the step involving the differentiation of $\log|\QQ^{-1}|$:
\begin{equation}\label{eq:Q.eqvarcov1}
\begin{split}
&\partial\Psi/\partial\pmb{q^{*}} = 
 \partial\bigg(-\frac{1}{2} \sum_{t=1}^T\big(\vec(\SS_t)^\top\big)\vec(\QQ^{-1})
+ \frac{T}{2}\log |\QQ^{-1}|\bigg)/\partial\pmb{q^{*}} \\
\end{split}
\end{equation}
It does not make sense to take the partial derivative of $\log |\QQ^{-1}|$ with respect to $\vec(\QQ^{-1})$ because many elements of $\QQ^{-1}$ are shared so it is not possible to fix one element while varying another.  Instead, we can take the partial derivative of $\log |\QQ^{-1}|$ with respect to $g(\alpha,\beta)$ which is $\sum_{\{i,j\}\in \text{set}_g}\partial\log|\QQ^{-1}|/\partial\pmb{q^{*}}_{i,j}$.  Set $g$ is those $i,j$ values where $\pmb{q^{*}}=g(\alpha,\beta)$.  Because $g()$ and $f()$ are different functions of both $\alpha$ and $\beta$, we can hold one constant while taking the partial derivative with respect to the other (well, presuming there exists some combination of $\alpha$ and $\beta$ that would allow that).  But if we have fixed values on the off-diagonal, this would not be possible.  In this case (see below), we cannot hold $g()$ constant while varying $f()$ because both are only functions of $\alpha$:
\begin{equation*}
\QQ=
\begin{bmatrix}
\alpha&f&f&f\\
f&\alpha&f&f\\
f&f&\alpha&f\\
f&f&f&\alpha
\end{bmatrix}\quad\quad
\QQ^{-1}=
\begin{bmatrix}
f(\alpha)&g(\alpha)&g(\alpha)&g(\alpha)\\
g(\alpha)&f(\alpha)&g(\alpha)&g(\alpha)\\
g(\alpha)&g(\alpha)&f(\alpha)&g(\alpha)\\
g(\alpha)&g(\alpha)&g(\alpha)&f(\alpha)
\end{bmatrix}
\end{equation*}

Taking the partial derivative of $\log |\QQ^{-1}|$ with respect to $\pmb{q^{*}}=\big[\begin{smallmatrix}f(\alpha,\beta)\\g(\alpha,\beta)\end{smallmatrix}\big]$, we arrive at the same equation as for the diagonal matrix:
\begin{equation}\label{eq:Q.eqvarcov2}
\begin{split}
&\partial\Psi/\partial\pmb{q^{*}} = 
\frac{1}{2} \sum_{t=1}^T\DD^\top \vec(\SS_t) - \frac{1}{2}\sum_{t=1}^T(\DD^\top\DD)\pmb{q}
\end{split}
\end{equation}
where here $\DD^\top\DD$ is a $2 \times 2$ diagonal matrix with the number of times $f(\alpha,\beta)$ appears in element $(1,1)$ and the number of times $g(\alpha,\beta)$ appears in element $(2,2)$ of $\DD$; $s=2$ here since there are only 2 free parameters in $\QQ$.

Setting to zero and solving for $\pmb{q^{*}}$ leads to the exact same update equation as for the diagonal $\QQ$, namely equation \eqref{eq:Q.diag.update} in which $\ff_q = 0$ since there are no fixed values.

\subsubsection{Special case: a block-diagonal matrices with replicated blocks}
\label{sec:Q.block.diagonal}
Because these operations extend directly to block-diagonal matrices, all results for individual matrix types can be extended to a block-diagonal matrix with those types:
\begin{equation*}
\QQ=
\begin{bmatrix}
\mathbb{B}_1&0&0\\
0&\mathbb{B}_2&0\\
0&0&\mathbb{B}_3\\
\end{bmatrix}
\end{equation*}
where $\mathbb{B}_i$ is a matrix from any of the allowed matrix types, such as unconstrained, diagonal (with fixed or shared elements), or equal variance-covariance.   Blocks can also be shared:
\begin{equation*}\QQ=
\begin{bmatrix}
\mathbb{B}_1&0&0\\
0&\mathbb{B}_2&0\\
0&0&\mathbb{B}_2\\
\end{bmatrix}
\end{equation*}
but the entire block must be identical $(\mathbb{B}_2 \equiv \mathbb{B}_3)$; one cannot simply share individual elements in different blocks.  Either all the elements in two (or 3, or 4...) blocks are shared or none are shared. 

This is ok:
\begin{equation*}
\begin{bmatrix}
c&d&d&0&0&0\\
d&c&d&0&0&0\\
d&d&c&0&0&0\\
0&0&0&c&d&d\\
0&0&0&d&c&d\\
0&0&0&d&d&c\\
\end{bmatrix}
\end{equation*}
This is not ok:
\begin{equation*}
\begin{bmatrix}
c&d&d&0&0\\
d&c&d&0&0\\
d&d&c&0&0\\
0&0&0&c&d\\
0&0&0&d&c
\end{bmatrix}
\text{ nor }
\begin{bmatrix}
c&d&d&0&0&0\\
d&c&d&0&0&0\\
d&d&c&0&0&0\\
0&0&0&c&e&e\\
0&0&0&e&c&e\\
0&0&0&e&e&c\\
\end{bmatrix}\end{equation*}
The first is bad because the blocks are not identical; they need the same dimensions as well as the same values.  The second is bad because again the blocks are not identical; all values must be the same.

\subsubsection{Special case: a symmetric blocked matrix}
\label{sec:Q.symmetric blocked}
The same derivation translates immediately to blocked symmetric $\QQ$ matrices with the following form:
\begin{equation*}
\QQ=
\begin{bmatrix}
\mathbb{E}_1&\mathbb{C}_{1,2}&\mathbb{C}_{1,3}\\
\mathbb{C}_{1,2}&\mathbb{E}_2&\mathbb{C}_{2,3}\\
\mathbb{C}_{1,3}&\mathbb{C}_{2,3}&\mathbb{E}_3\\
\end{bmatrix}
\end{equation*}
where the $\mathbb{E}$ are as above matrices with one value on the diagonal and another on the off-diagonals (no zeros!). The $\mathbb{C}$ matrices have only one free value or are all zero.  Some $\mathbb{C}$ matrices can be zero while are others are non-zero, but a individual $\mathbb{C}$ matrix cannot have a combination of free values and zero values; they have to be one or the other. Also the whole matrix must stay block symmetric. Additionally, there can be shared $\mathbb{E}$ or $\mathbb{C}$ matrices but the whole matrix needs to stay block-symmetric.  Here are the forms that $\mathbb{E}$ and $\mathbb{C}$ can take:
\begin{equation*}
\mathbb{E}_i=
\begin{bmatrix}
\alpha&\beta&\beta&\beta\\
\beta&\alpha&\beta&\beta\\
\beta&\beta&\alpha&\beta\\
\beta&\beta&\beta&\alpha
\end{bmatrix}
\quad\quad
\mathbb{C}_i=
\begin{bmatrix}
\chi&\chi&\chi&\chi\\
\chi&\chi&\chi&\chi\\
\chi&\chi&\chi&\chi\\
\chi&\chi&\chi&\chi
\end{bmatrix}
\text{ or }
\begin{bmatrix}
0&0&0&0\\
0&0&0&0\\
0&0&0&0\\
0&0&0&0
\end{bmatrix}
\end{equation*}
The following are block-symmetric:
\begin{equation*}
\begin{bmatrix}
\mathbb{E}_1&\mathbb{C}_{1,2}&\mathbb{C}_{1,3}\\
\mathbb{C}_{1,2}&\mathbb{E}_2&\mathbb{C}_{2,3}\\
\mathbb{C}_{1,3}&\mathbb{C}_{2,3}&\mathbb{E}_3\\
\end{bmatrix}
\text{ and }
\begin{bmatrix}
\mathbb{E}&\mathbb{C}&\mathbb{C}\\
\mathbb{C}&\mathbb{E}&\mathbb{C}\\
\mathbb{C}&\mathbb{C}&\mathbb{E}\\
\end{bmatrix}
\end{equation*}
\begin{equation*}
\text{ and }
\begin{bmatrix}
\mathbb{E}_1&\mathbb{C}_1&\mathbb{C}_{1,2}\\
\mathbb{C}_1&\mathbb{E}_1&\mathbb{C}_{1,2}\\
\mathbb{C}_{1,2}&\mathbb{C}_{1,2}&\mathbb{E}_2\\
\end{bmatrix}
\end{equation*}
The following are NOT block-symmetric:
\begin{equation*}
\begin{bmatrix}
\mathbb{E}_1&\mathbb{C}_{1,2}&0\\
\mathbb{C}_{1,2}&\mathbb{E}_2&\mathbb{C}_{2,3}\\
0&\mathbb{C}_{2,3}&\mathbb{E}_3
\end{bmatrix}
\text{ and }
\begin{bmatrix}
\mathbb{E}_1&0&\mathbb{C}_1\\
0&\mathbb{E}_1&\mathbb{C}_2\\
\mathbb{C}_1&\mathbb{C}_2&\mathbb{E}_2
\end{bmatrix}
\text{ and }
\begin{bmatrix}
\mathbb{E}_1&0&\mathbb{C}_{1,2}\\
0&\mathbb{E}_1&\mathbb{C}_{1,2}\\
\mathbb{C}_{1,2}&\mathbb{C}_{1,2}&\mathbb{E}_2\\
\end{bmatrix}
\end{equation*}
\begin{equation*}
\text{ and }
\begin{bmatrix}
\mathbb{U}_1&\mathbb{C}_{1,2}&\mathbb{C}_{1,3}\\
\mathbb{C}_{1,2}&\mathbb{E}_2&\mathbb{C}_{2,3}\\
\mathbb{C}_{1,3}&\mathbb{C}_{2,3}&\mathbb{E}_3
\end{bmatrix}
\text{ and }
\begin{bmatrix}
\mathbb{D}_1&\mathbb{C}_{1,2}&\mathbb{C}_{1,3}\\
\mathbb{C}_{1,2}&\mathbb{E}_2&\mathbb{C}_{2,3}\\
\mathbb{C}_{1,3}&\mathbb{C}_{2,3}&\mathbb{E}_3\end{bmatrix}
\end{equation*}
In the first row, the matrices have fixed values (zeros) and free values (covariances) on the same off-diagonal row and column.  That is not allowed.  If there is a zero on a row or column, all other terms on the off-diagonal row and column must be also zero.  In the second row, the matrix is not block-symmetric since the upper corner is an unconstrained block ($\mathbb{U}_1$) in the left matrix and diagonal block ($\mathbb{D}_1$) in the right matrix  instead of a equal variance-covariance matrix ($\mathbb{E}$).

\subsubsection{The general case: a block-diagonal matrix with general blocks}
\label{sec:Q.general}
In it's most general form, $\QQ$ is allowed to have a block-diagonal form where the blocks, here called $\mathbb{G}$ are any of the previous allowed cases.  No shared values across $\mathbb{G}$'s; shared values are allowed within $\mathbb{G}$'s.
\begin{equation*}
\QQ=
\begin{bmatrix}
\mathbb{G}_1&0&0\\
0&\mathbb{G}_2&0\\
0&0&\mathbb{G}_3\\
\end{bmatrix}
\end{equation*}
The $\mathbb{G}$'s must be one of the special cases listed above: unconstrained, diagonal (with fixed or shared values), equal variance-covariance, block diagonal (with shared or unshared blocks), and block-symmetric (with shared or unshared blocks).  Fixed blocks are allowed, but then the covariances with the free blocks must be zero:
\begin{equation*}
\QQ=
\begin{bmatrix}
\mathbb{F}&0&0&0\\
0&\mathbb{G}_1&0&0\\
0&0&\mathbb{G}_2&0\\
0&0&0&\mathbb{G}_3
\end{bmatrix}
\end{equation*}
Fixed blocks must have only fixed values (zero is a fixed value) but the fixed values can be different from each other.  The free blocks must have only free values (zero is not a free value).  

\subsection{The general $\RR$ update equation}
The $\RR$ update equation for blocked symmetric matrices with optional independent fixed blocks is completely analogous to the $\QQ$ equation.  Thus if $\RR$ has the form
\begin{equation*}
\RR=
\begin{bmatrix}
\mathbb{F}&0&0&0\\
0&\mathbb{G}_1&0&0\\
0&0&\mathbb{G}_2&0\\
0&0&0&\mathbb{G}_3
\end{bmatrix}
\end{equation*}
Again the $\mathbb{G}$'s must be one of the special cases listed above: unconstrained, diagonal (with fixed or shared values), equal variance-covariance, block diagonal (with shared or unshared blocks), and block-symmetric (with shared or unshared blocks).  Fixed blocks are allowed, but then the covariances with the free blocks must be zero.  Elements like $f_i+r_j$ and $r_i+r_j$ are not allowed in $\RR$. Only elements of the form $f_i$ and $r_i$ are allowed.  If an element has a fixed component, it must be completely fixed.  Each element in $\RR$ can have only one of the elements in $\rr$, but multiple elements in $\RR$ can have the same $\rr$ element.

The update equation is
\begin{equation}\label{eq:R.general}
\begin{split}
&\rr_{j+1}  =   
\bigg(\sum_{t=1}^T\DD_{t,r}^\top \DD_{t,r}\bigg)^{-1} \vec\bigg(\sum_{t=1}^T \DD_{t,r}^\top\RR_{t,{j+1}}  
 \bigg)\\
&\quad\quad\quad\quad\vec(\RR)_{t,j+1} = \ff_{t,r} + \DD_{t,r} \rr_{j+1}
\end{split}
\end{equation}
The $\RR_{t,j+1}$ used at time step $t$ in equation \eqref{eq:R.general} is the term that appears in the summation in the unconstrained update equation with no missing values (equation \ref{eq:R.update.unconstrained}):
\begin{equation}
\begin{split}
\RR_{t,j+1}=\Xi_t\bigg(  
 \hatOt - \hatYXt\ZZ_t^\top - \ZZ_t\hatYXt^\top 
 - \hatyt\aa_t^\top - \aa_t\hatyt^\top 
 + \ZZ_t\hatPt\ZZ_t^\top + \ZZ_t\hatxt\aa_t^\top + \aa_t\hatxt^\top\ZZ_t^\top 
 + \aa_t\aa_t^\top \bigg)\Xi_t^\top
\end{split}
\end{equation}
where $\Xi_t = (\HH_t^\top\HH_t)^{-1}\HH_t^\top$.

\section{Computing the expectations in the update equations}\label{sec:compexpectations}
For the update equations, we need to compute the expectations of $\XX_t$ and $\YY_t$ and their products conditioned on 1) the observed data $\YY(1)=\yy(1)$ and 2) the parameters at time $t$, $\Theta_j$.  This section shows how to compute these expectations.  Throughout the section, I will normally leave off the conditional $\YY(1)=\yy(1),\Theta_j$ when specifying an expectation. Thus any $\E[ ]$ appearing without its conditional is conditioned on $\YY(1)=\yy(1),\Theta_j$.  However if there are additional or different conditions those will be shown.  Also all expectations are over the joint distribution of $XY$ unless explicitly specified otherwise.

Before commencing, we need some notation for the observed and unobserved elements of the data.
The $n \times 1$ vector $\yy_t$ denotes the potential observations at time $t$. If some elements of $\yy_t$ are missing, that means some elements are equal to NA (or some other missing values marker):
\begin{equation}
\yy_t=\begin{bmatrix}
y_1\\
NA\\
y_3\\
y_4\\
NA\\
y_6
\end{bmatrix}
\end{equation}
We denote the non-missing observations as $\yy_t(1)$ and the missing observations as $\yy_t(2)$.  Similar to $\yy_t$, $\YY_t$ denotes all the $\YY$ random variables at time $t$.  The $\YY_t$'s with an observation are  $\YY_t(1)$ and those without an observation are denoted $\YY_t(2)$. 

Let $\OMG_t^{(1)}$ be the matrix that extracts only $\YY_t(1)$ from $\YY_t$ and $\OMG_t(2)$ be the matrix that extracts only $\YY_t(2)$.  For the example above,
\begin{equation}
\begin{split}
&\YY_t(1)=\OMG_t^{(1)} \YY_t,\quad \OMG_t^{(1)} = 
\begin{bmatrix}
1&0&0&0&0&0\\
0&0&1&0&0&0\\
0&0&0&1&0&0\\
0&0&0&0&0&1\\
\end{bmatrix}\\
&\YY_t(2)=\OMG_t^{(2)} \YY_t,\quad \OMG_t^{(2)} = 
\begin{bmatrix}
0&1&0&0&0&0\\
0&0&0&0&1&0
\end{bmatrix}
\end{split}
\end{equation}

We will define another set of matrices that zeros out the missing or non-missing values. Let $\IIm_t^{(1)}$ denote a diagonal matrix that zeros out the $\YY_t(2)$ in $\YY_t$ and $\IIm_t^{(2)}$ denote a matrix that zeros out the $\YY_t(1)$ in $\YY_t$.  For the example above, 
\begin{equation}
\begin{split}
\IIm_t^{(1)} &= (\OMG_t^{(1)})^\top\OMG_t^{(1)} =
\begin{bmatrix}
1&0&0&0&0&0\\
0&0&0&0&0&0\\
0&0&1&0&0&0\\
0&0&0&1&0&0\\
0&0&0&0&0&0\\
0&0&0&0&0&1\\
\end{bmatrix}\quad\text{and}\\
\IIm_t^{(2)} &= (\OMG_t^{(2)})^\top\OMG_t^{(2)} =
\begin{bmatrix}
0&0&0&0&0&0\\
0&1&0&0&0&0\\
0&0&0&0&0&0\\
0&0&0&0&0&0\\
0&0&0&0&1&0\\
0&0&0&0&0&0\\
\end{bmatrix}.
\end{split}
\end{equation}

\subsection{Expectations involving only $\XX_t$}\label{sec:kalman.smoother}
The Kalman smoother provides the expectations involving only $\XX_t$ conditioned on all the data from time 1 to $T$.  
\begin{subequations}\label{eq:kfsoutput}
\begin{align}
&\hatxt = \E[\XX_t]\label{eq:xt}\\
&\widetilde{\VV}_t = \var[\XX_t]\\
&\widetilde{\VV}_{t,t-1} = \cov[\XX_t,\XX_{t-1}]\\
&\text{From $\hatxt$, $\widetilde{\VV}_t$, and $\widetilde{\VV}_{t,t-1}$, we compute}\nonumber\\
&\hatPt=\E[\XX_t\XX_t^\top]= \widetilde{\VV}_t + \hatxt\hatxt^\top\label{eq:Pt}\\
&\hatPttm=\E[\XX_{t}\XX_{t-1}^\top]=\widetilde{\VV}_{t,t-1} +\hatxt\hatxtm^\top\label{eq:Ptt1}
\end{align}
\end{subequations}
The $\hatPt$ and $\hatPttm$ equations arise from the computational formula for variance (equation \ref{eq:comp.formula.variance}). Note the smoother is different than the Kalman filter as the filter does not provide the expectations of $\XX_t$ conditioned on all the data (time 1 to $T$) but only on the data up to time $t$.

The first part of the Kalman smoother algorithm is the Kalman filter which gives the expectation at time $t$ conditioned on the data up to time $t$.  The following the filter as shown in \citep[sec. 6.2, p. 331]{ShumwayStoffer2006}, although the notation is a little different.  
\begin{subequations}\label{eq:kffilter}
\begin{align}
&\xx_t^{t-1}=\BB_t\xx_{t-1}^{t-1}+\uu_t\label{eq:xtt1}\\
&\VV_t^{t-1}=\BB_t\VV_{t-1}^{t-1}\BB_t^\top + \GG_t\QQ_t\GG_t^\top\label{eq:Vtt1}\\
&\xx_t^t=\xx_t^{t-1}+\KK_t(\yy_t-\ZZ_t\xx_t^{t-1}-\aa_t)\label{eq:xtt}\\
&\VV_t^t=(\II_m-\KK_t\ZZ_t)\VV_t^{t-1}\label{eq:Vtt}\\
&\KK_t=\VV_t^{t-1}\ZZ_t^\top(\ZZ_t\VV_t^{t-1}\ZZ_t^\top+\HH_t\RR_t\HH_t^\top)^{-1}\label{eq:Kt}
\end{align}
\end{subequations}

The Kalman smoother and lag-1 covariance smoother compute the expectations conditioned on all the data, 1 to $T$:
\begin{subequations}\label{eq:kfsmoother}
\begin{align}
&\xx_{t-1}^T=\xx_{t-1}^{t-1}+\JJ_{t-1}(\xx_t^T-\xx_t^{t-1})\label{eq:xt1T}\\
&\VV_{t-1}^T=\VV_{t-1}^{t-1}+\JJ_{t-1}(\VV_t^T-\VV_t^{t-1})\JJ_t^\top\label{eq:Vt1T}\\
&\JJ_{t-1}=\VV_{t-1}^{t-1}\BB_t^\top(\VV_t^{t-1})^{-1}\label{eq:Jt}\\
\\
&\VV_{T,T-1}^T=(\II-\KK_T\ZZ_T)\BB_T\VV_{T-1}^{T-1}\label{eq:VTT1T}\\
&\VV_{t-1,t-2}^T=\VV_{t-1}^{t-1}\JJ_{t-2}^\top + \JJ_{t-1}((\VV_{t,t-1}^T-\BB_t\VV_{t-1}^{t-1}))\JJ_{t-2}^\top\label{eq:Vtt1T}
\end{align}
\end{subequations}

The classic Kalman smoother is an algorithm to compute these expectations conditioned on no missing values in $\yy$. However, the algorithm  can be easily modified to give the expected values of $\XX$ conditioned on the incomplete data, $\YY(1)=\yy(1)$ \citep[sec. 6.4, eqn 6.78, p. 348]{ShumwayStoffer2006}.  
In this case, the usual filter and smoother equations are used with the following modifications to the parameters and data used in the equations.  If the $i$-th element of $\yy_t$ is missing, zero out the $i$-th rows in $\yy_t$, $\aa$ and $\ZZ$.  Thus if the 2nd and 5th elements of $\yy_t$ are missing,
\begin{equation}\label{eq:yaZ.miss}
\yy_t^*=\begin{bmatrix}
y_1\\
0\\
y_3\\
y_4\\
0\\
y_6\\
\end{bmatrix}, \quad
\aa_t^*=\begin{bmatrix}
a_1\\
0\\
a_3\\
a_4\\
0\\
a_6\\
\end{bmatrix}, \quad
\ZZ_t^*=\begin{bmatrix}
z_{1,1}&z_{1,2}&...\\
0&0&...\\
z_{3,1}&z_{3,2}&...\\
z_{4,1}&z_{4,2}&...\\
0&0&...\\
z_{6,1}&z_{6,2}&...\\
\end{bmatrix}
\end{equation}

The $\RR_t$ parameter used in the filter equations is also modified.  We need to zero out the covariances between the non-missing, $\yy_t(1)$, and missing, $\yy_t(2)$, data.  For the example above, if
\begin{equation}
\RR_t = \Rb_t\RR\Rb_t^\top = \begin{bmatrix}
r_{1,1}&r_{1,2}&r_{1,3}&r_{1,4}&r_{1,5}&r_{1,6}\\
r_{2,1}&r_{2,2}&r_{2,3}&r_{2,4}&r_{2,5}&r_{2,6}\\
r_{3,1}&r_{3,2}&r_{3,3}&r_{3,4}&r_{3,5}&r_{3,6}\\
r_{4,1}&r_{4,2}&r_{4,3}&r_{4,4}&r_{4,5}&r_{4,6}\\
r_{5,1}&r_{5,2}&r_{5,3}&r_{5,4}&r_{5,5}&r_{5,6}\\
r_{6,1}&r_{6,2}&r_{6,3}&r_{6,4}&r_{6,5}&r_{6,6}\\
\end{bmatrix}
\end{equation}
then the $\RR_t$ we use at time $t$, will have zero covariances between the non-missing elements 1,3,4,6 and the missing elements 2,5:
\begin{equation}
\RR_t^*
 = \begin{bmatrix}
r_{1,1}&0&r_{1,3}&r_{1,4}&0&r_{1,6}\\
0&r_{2,2}&0&0&r_{2,5}&0\\
r_{3,1}&0&r_{3,3}&r_{3,4}&0&r_{3,6}\\
r_{4,1}&0&r_{4,3}&r_{4,4}&0&r_{4,6}\\
0&r_{5,2}&0&0&r_{5,5}&0\\
r_{6,1}&0&r_{6,3}&r_{6,4}&0&r_{6,6}\\
\end{bmatrix}
\end{equation}

Thus, the data and parameters used in the filter and smoother equations are
\begin{equation}
\begin{split}
\yy_t^*&=\IIm_t^{(1)}\yy_t\\
\aa_t^*&=\IIm_t^{(1)}\aa_t\\
\ZZ_t^*&=\IIm_t^{(1)}\ZZ_t\\
\RR_t^* &= \IIm_t^{(1)}\RR_t\IIm_t^{(1)} + \IIm_t^{(2)}\RR_t\II_t^{(2)}
\end{split}
\end{equation}
$\aa_t^*$, $\ZZ_t^*$ and $\RR_t^*$ only are used in the Kalman filter and smoother.  They are not used in the EM update equations.  However when coding the algorithm, it is convenient to replace the NAs (or whatever the missing values placeholder is) in $\yy_t$ with zero so that there is not a problem with NAs appearing in the computations.

\subsection{Expectations involving $\YY_t$}\label{sec:exp.Y}
First, replace the missing values in $\yy_t$ with zeros\footnote{The only reason is so that in your computer code, if you use NA or NaN as the missing value marker, NA-NA=0 and 0*NA=0 rather than NA.} and then the expectations are given by the following equations.  The derivations for these equations are given in the subsections to follow.
\begin{subequations}\label{eq:Yt.exp}
\begin{align}
\hatyt &= \E[\YY_t]=\yy_t-\IR_t(\yy_t-\ZZ_t\hatxt-\aa_t)\label{eq:hatyt}\\
\hatOt &= \E[\YY_t\YY_t^\top]=\IIm_t^{(2)}(\IR_t\HH_t\RR_t\HH_t^\top + \IR_t\ZZ_t\hatVt\ZZ_t^\top\IR_t^\top)\IIm_t^{(2)}  +  \hatyt\hatyt^\top \label{eq:hatOt}\\
\hatYXt&= \E[\YY_t\XX_t^\top] = \IR_t\ZZ_t\hatVt + \hatyt\hatxt^\top \label{eq:hatyxttm}\\
\hatYXttm&= \E[\YY_t\XX_{t-1}^\top] = \IR_t\ZZ_t\hatVttm + \hatyt\hatxtm^\top \label{eq:hatyxt}\\
\text{where }\IR_t &= \II-\HH_t\RR_t\HH_t^\top(\OMG_t^{(1)})^\top(\OMG_t^{(1)}\HH_t\RR_t\HH_t^\top(\OMG_t^{(1)})^\top)^{-1}\OMG_t^{(1)}\label{eq.IRt}\\
\text{and }\IIm_t^{(2)}&=(\OMG_t^{(2)})^\top\OMG_t^{(2)}
\label{eq:IRt}
\end{align}
\end{subequations}
If $\yy_t$ is all missing, $\OMG_t^{(1)}$ is a $0 \times n$ matrix, and we define $(\OMG_t^{(1)})^\top(\OMG_t^{(1)}\RR(\OMG_t^{(1)})^\top)^{-1}\OMG_t^{(1)}$ to be a $n \times n$ matrix of zeros.  If $\RR_t$ is diagonal, then $\RR_t(\OMG_t^{(1)})^\top(\OMG_t^{(1)}\RR_t(\OMG_t^{(1)})^\top)^{-1}\OMG_t^{(1)}=\IIm_t^{(1)}$ and $\IR_t=\IIm_t^{(2)}$.  This will mean that in $\hatyt$ the $\yy_t(2)$ are given by $\ZZ_t\hatxt+\aa_t$, as expected when $\yy_t(1)$ and $\yy_t(2)$ are independent.

If there are zeros on the diagonal of $\RR_t$ (section \ref{sec:degenerate}), the definition of $\IR_t$ is changed slightly from that shown in equation \ref{eq:Yt.exp}. Let $\mho_t^{(r)}$ be the matrix that extracts the elements of $\yy_t$ where $\yy_t(i)$ is not missing AND $\HH_t\RR_t(i,i)\HH_t^\top$ is not zero. Then
\begin{equation}
\IR_t = \II-\HH_t\RR_t\HH_t^\top(\mho_t^{(r)})^\top
(\mho_t^{(r)}\HH_t\RR_t\HH_t^\top(\mho_t^{(r)})^\top)^{-1}\mho_t^{(r)}
\label{eq:IRt.degen}
\end{equation}

\subsection{Derivation of the expected value of $\YY_t$}
In the MARSS equation, the observation errors are denoted $\HH_t\vv_t$.  $\vv_t$ is a specific realization from a random variable $\VV_t$ that is distributed multivariate normal with mean 0 and variance $\RR_t$.  $\VV_t$ is not to be confused with $\widetilde{\VV}_t$ in equation \ref{eq:kfsoutput}, which is unrelated\footnote{I apologize for the confusing notation, but $\widetilde{\VV}_t$ and $\vv_t$ are somewhat standard in the MARSS literature and it is standard to use a capital letter to refer to a random variable.  Thus $\VV_t$ would be the standard way to refer to the random variable associated with $\vv_t$.} to $\VV_t$. If there are no missing values, then we condition on $\YY_t=\yy_t$ and
\begin{equation}
\begin{split}
\E[\YY_t|\YY(1)=\yy(1)] = \E[\YY_t|\YY_t=\yy_t] = \yy_t\\
\end{split}
\end{equation}
If there are no observed values, then 
\begin{equation}
\begin{split}
\E[\YY_t|\YY(1)=\yy(1)]=\E[\YY_t] = \E[\ZZ_t\XX_t+\aa_t+\VV_t] = \ZZ_t\hatxt+\aa_t
\end{split}
\end{equation}
If only some of the $\YY_t$ are observed, then we use the conditional probability for a multivariate normal distribution (here shown for a bivariate case):
\begin{equation}\label{eq:cond.multi.normal}
\text{If, }\begin{bmatrix}
Y_1\\
Y_2\end{bmatrix}
\sim 
\MVN\biggl( \begin{bmatrix}
\mu_1\\
\mu_2
\end{bmatrix}, \begin{bmatrix}
\Sigma_{11}&\Sigma_{12}\\
\Sigma_{21}&\Sigma_{22}\end{bmatrix}\biggr)
\end{equation}
Then, 
\begin{equation}
\begin{split}
(Y_1|Y_1=y_1) &=y_1,\quad\text{and}\\
(Y_2|Y_1=y_1) &\sim \MVN(\bar{\mu},\bar{\Sigma}),\quad\text{where}\\
\bar{\mu}&= \mu_2+\Sigma_{21}\Sigma_{11}^{-1}(y_1-\mu_1)\\
\bar{\Sigma} &= \Sigma_{22}-\Sigma_{21}\Sigma_{11}^{-1}\Sigma_{12} 
\end{split}
\end{equation}

From this property, we can write down the distribution of $\YY_t$ conditioned on $\YY_t(1)=\yy_t(1)$ and $\XX_t=\xx_t$:
\begin{equation}
\begin{split}
\begin{bmatrix}
\YY_t(1)|\XX_t=\xx_t\\
\YY_t(2)|\XX_t=\xx_t
\end{bmatrix}
&\sim \\
&\MVN\biggl( \begin{bmatrix}
\OMG_t^{(1)}(\ZZ_t\xx_t+\aa_t)\\
\OMG_t^{(2)}(\ZZ_t\xx_t+\aa_t)
\end{bmatrix}, \begin{bmatrix}
(\HH_t\RR_t\HH_t^\top)_{11}&(\HH_t\RR_t\HH_t^\top)_{12}\\
(\HH_t\RR_t\HH_t^\top)_{21}&(\HH_t\RR_t\HH_t^\top)_{22}
\end{bmatrix}\biggr)
\end{split}
\end{equation}
Thus, 
\begin{equation}\label{eq:varY}
\begin{split}
(\YY_t(1)&|\YY_t(1)=\yy_t(1),\XX_t=\xx_t) = \OMG_t^{(1)}\yy_t\quad\text{and}\\
(\YY_t(2)&|\YY_t(1)=\yy_t(1),\XX_t=\xx_t) \sim \MVN(\ddot{\mu},\ddot{\Sigma})\quad\text{where}\\
\ddot{\mu}&= \OMG_t^{(2)}(\ZZ_t\xx_t+\aa_t)+\ddot{\RR}_{t,21}(\ddot{\RR}_{t,11})^{-1}\OMG_t^{(1)}(\yy_t-\ZZ_t\xx_t-\aa_t)\\
\ddot{\Sigma}&= \ddot{\RR}_{t,22} - \ddot{\RR}_{t,21}(\ddot{\RR}_{t,11})^{-1}\ddot{\RR}_{t,12} \\
\ddot{\RR}_t&= \HH_t\RR_t\HH_t^\top 
\end{split}
\end{equation}

Note that since we are conditioning on $\XX_t=\xx_t$, we can replace $\YY$ by $\YY_t$ in the conditional:
$$\E[\YY_t|\YY(1)=\yy(1),\XX_t=\xx_t]=\E[\YY_t|\YY_t(1)=\yy_t(1),\XX_t=\xx_t].$$
From this and the distributions in equation \eqref{eq:varY}, we can write down $\hatyt=\E[\YY_t|\YY(1)=\yy(1),\Theta_j]$:
\begin{equation}
\begin{split}
\hatyt&=\E_{XY}[\YY_t|\YY(1)=\yy(1)]\\
&=\int_{\xx_t}\int_{\yy_t}\yy_t f(\yy_t|\yy_t(1),\xx_t)d\yy_t f(\xx_t)d\xx_t \\
&=\E_X[\E_{Y|x}[\YY_t|\YY_t(1)=\yy_t(1),\XX_t=\xx_t]]\\
&=\E_X[\yy_t-\IR_t(\yy_t-\ZZ_t\XX_t-\aa_t)]\\
&=\yy_t-\IR_t(\yy_t-\ZZ_t\hatxt-\aa_t)\\
\text{where }\IR_t &= \II-\ddot{\RR}_t(\OMG_t^{(1)})^\top(\ddot{\RR}_{t,11})^{-1}\OMG_t^{(1)}
\end{split}
\end{equation}
$(\OMG_t^{(1)})^\top(\ddot{\RR}_{t,11})^{-1}\OMG_t^{(1)}$ is a $n \times n$ matrix with 0s in the non-(11) positions. If the $k$-th element of $\yy_t$ is observed, then $k$-th row and column of $\IR_t$ will be zero.
Thus if there are no missing values at time $t$, $\IR_t=\II-\II=0$. If there are no observed values at time $t$, $\IR_t$ will reduce to $\II$. 

\subsection{Derivation of the expected value of $\YY_t\YY_t^\top$}
The following outlines a\footnote{The following derivations are painfully ugly, but appear to work.  There are surely more elegant ways to do this; at least, there must be more elegant notations.} derivation.  If there are no missing values, then we condition on $\YY_t=\yy_t$ and
\begin{equation}
\begin{split}
\E[\YY_t &\YY_t^\top|\YY(1)=\yy(1)] = \E[\YY_t \YY_t^\top|\YY_t=\yy_t]\\
&=\yy_t\yy_t^\top.
\end{split}
\end{equation}
If there are no observed values at time $t$, then 
\begin{equation}
\begin{split}
\E[\YY_t &\YY_t^\top]\\ 
&=\var[\ZZ_t\XX_t+\aa_t+\HH_t\VV_t]+\E[\ZZ_t\XX_t+\aa_t+\HH_t\VV_t]\E[\ZZ_t\XX_t+\aa_t+\HH_t\VV_t]^\top\\
&=\var[\VV_t]+\var[\ZZ_t\XX_t]+(\E[\ZZ_t\XX_t+\aa_t]+\E[\HH_t\VV_t])(\E[\ZZ_t\XX_t+\aa_t]+\E[\HH_t\VV_t])^\top\\
&=\ddot{\RR}_t+\ZZ_t\hatVt\ZZ_t^\top + (\ZZ_t\hatxt+\aa_t)(\ZZ_t\hatxt+\aa_t)^\top
\end{split}
\end{equation}

When only some of the $\YY_t$ are observed, we use again the conditional probability of a multivariate normal (equation \ref{eq:cond.multi.normal}).  From this property, we know that 
\begin{equation}
\begin{split}
&\var_{Y|x}[\YY_t(2)\YY_t(2)^\top|\YY_t(1)=\yy_t(1),\XX_t=\xx_t]=\ddot{\RR}_{t,22} - \ddot{\RR}_{t,21}(\ddot{\RR}_{t,11})^{-1}\ddot{\RR}_{t,12},\\
&\var_{Y|x}[\YY_t(1)|\YY_t(1)=\yy_t(1),\XX_t=\xx_t]=0\\
\text{and }&\cov_{Y|x}[\YY_t(1),\YY_t(2)|\YY_t(1)=\yy_t(1),\XX_t=\xx_t]=0\\
\\
\text{Thus }&\var_{Y|x}[\YY_t|\YY_t(1)=\yy_t(1),\XX_t=\xx_t]\\
&=(\OMG_t^{(2)})^\top(\ddot{\RR}_{t,22} - \ddot{\RR}_{t,21}(\ddot{\RR}_{t,11})^{-1}\ddot{\RR}_{t,12})\OMG_t^{(2)}\\
&=(\OMG_t^{(2)})^\top(\OMG_t^{(2)}\ddot{\RR}_t(\OMG_t^{(2)})^\top - \OMG_t^{(2)}\ddot{\RR}_t(\OMG_t^{(1)})^\top(\ddot{\RR}_{t,11})^{-1}\OMG_t^{(1)}\ddot{\RR}_t(\OMG_t^{(2)})^\top)\OMG_t^{(2)}\\
&=\IIm_t^{(2)}(\ddot{\RR}_t-\ddot{\RR}_t(\OMG_t^{(1)})^\top(\ddot{\RR}_{t,11})^{-1}\OMG_t^{(1)}\ddot{\RR}_t)\IIm_t^{(2)}\\
&=\IIm_t^{(2)}\IR_t\ddot{\RR}_t\IIm_t^{(2)}
\end{split}
\end{equation}
The $\IIm_t^{(2)}$ bracketing both sides is zero-ing out the rows and columns corresponding to the $\yy_t(1)$ values.

Now we can compute the $\E_{XY}[\YY_t\YY_t^\top|\YY(1)=\yy(1)]$.  The subscripts are added to the $\E$ to emphasize that we are breaking the multivariate expectation into an inner and outer expectation.
\begin{equation}
\begin{split}
\hatOt&=\E_{XY}[\YY_t\YY_t^\top|\YY(1)=\yy(1)]=\E_X[\E_{Y|x}[\YY_t\YY_t^\top|\YY_t(1)=\yy_t(1),\XX_t=\xx_t]]\\
&=\E_X\bigl[\var_{Y|x}[\YY_t|\YY_t(1)=\yy_t(1),\XX_t=\xx_t] \\
&\quad + \E_{Y|x}[\YY_t|\YY_t(1)=\yy_t(1),\XX_t=\xx_t]
\E_{Y|x}[\YY_t|\YY_t(1)=\yy_t(1),\XX_t=\xx_t]^\top\bigr]\\
&=\E_X[\IIm_t^{(2)}\IR_t\ddot{\RR}_t\IIm_t^{(2)}] +\E_X[(\yy_t-\IR_t(\yy_t-\ZZ_t\XX_t-\aa_t))(\yy_t-\IR_t(\yy_t-\ZZ_t\XX_t-\aa_t))^\top]\\
&=\IIm_t^{(2)}\IR_t\ddot{\RR}_t\IIm_t^{(2)} + \var_X\bigl[\yy_t-\IR_t(\yy_t-\ZZ_t\XX_t-\aa_t)\bigr]\\ &\quad +\E_X[\yy_t-\IR_t(\yy_t-\ZZ_t\XX_t-\aa_t)]\E_X[\yy_t-\IR_t(\yy_t-\ZZ_t\XX_t-\aa_t)]^\top\\
&=\IIm_t^{(2)}\IR_t\ddot{\RR}_t\IIm_t^{(2)}  + \IIm_t^{(2)}\IR_t\ZZ_t\hatVt\ZZ_t^\top\IR_t^\top\IIm_t^{(2)} + \hatyt\hatyt^\top  \\
\end{split}
\end{equation}
Thus,
\begin{equation}
\hatOt=\IIm_t^{(2)}(\IR_t\ddot{\RR}_t + \IR_t\ZZ_t\hatVt\ZZ_t^\top\IR_t^\top)\IIm_t^{(2)}  +  \hatyt\hatyt^\top  \\
\end{equation}

\subsection{Derivation of the expected value of $\YY_t\XX_t^\top$}
If there are no missing values, then we condition on $\YY_t=\yy_t$ and
\begin{equation}
\begin{split}
\E[\YY_t \XX_t^\top|\YY(1)=\yy(1)] = \yy_t\E[\XX_t^\top]=\yy_t\hatxt^\top
\end{split}
\end{equation}
If there are no observed values at time $t$, then 
\begin{equation}
\begin{split}
\E[\YY_t &\XX_t^\top|\YY(1)=\yy(1)]\\ 
&=\E[(\ZZ_t\XX_t+\aa_t+\VV_t)\XX_t^\top]\\
&=\E[\ZZ_t\XX_t\XX_t^\top+\aa_t\XX_t^\top+\VV_t\XX_t^\top]\\
&=\ZZ_t\hatPt+\aa_t\hatxt^\top+\cov[\VV_t,\XX_t]+\E[\VV_t]\E[\XX_t]^\top\\
&=\ZZ_t\hatPt+\aa_t\hatxt^\top
\end{split}
\end{equation}
Note that $\VV_t$ and $\XX_t$ are independent (equation \ref{eq:MARSS}). $\E[\VV_t]=0$ and $\cov[\VV_t,\XX_t]=0$.

Now we can compute the $\E_{XY}[\YY_t\XX_t^\top|\YY_(1)=\yy(1)]$.  
\begin{equation}
\begin{split}
\hatYXt&=\E_{XY}[\YY_t\XX_t^\top|\YY(1)=\yy(1)]\\
&=\cov[\YY_t,\XX_t|\YY_t(1)=\yy_t(1)]+\E_{XY}[\YY_t|\YY(1)=\yy(1)]\E_{XY}[\XX_t^\top|\YY(1)=\yy(1)]^\top\\
&=\cov[\yy_t-\IR_t(\yy_t-\ZZ_t\XX_t-\aa_t)+\VV^*_t,\XX_t]+\hatyt\hatxt^\top \\
&=\cov[\yy_t,\XX_t]-\cov[\IR_t\yy_t,\XX_t]+\cov[\IR_t\ZZ_t\XX_t,\XX_t] +\cov[\IR_t\aa_t,\XX_t]\\
&\quad + \cov[\VV^*_t,\XX_t]+\hatyt\hatxt^\top \\
&=0 - 0 + \IR_t\ZZ_t\hatVt + 0 + 0 + \hatyt\hatxt^\top \\
&= \IR_t\ZZ_t\hatVt + \hatyt\hatxt^\top
\end{split}
\end{equation}
This uses the computational formula for covariance: $\E[\YY\XX^\top]=\cov[\YY,\XX]+\E[\YY]\E[\XX]^\top$. $\VV^*_t$ is a random variable with mean 0 and variance $\ddot{\RR}_{t,22} - \ddot{\RR}_{t,21}(\ddot{\RR}_{t,11})^{-1}\ddot{\RR}_{t,12}$ from equation \eqref{eq:varY}.  $\VV^*_t$ and $\XX_t$ are independent of each other, thus $\cov[\VV^*_t,\XX_t^\top]=0$.

\subsection{Derivation of the expected value of $\YY_t\XX_{t-1}^\top$}
The derivation of $\E[\YY_t\XX_{t-1}^\top]$ is similar to the derivation of $\E[\YY_t\XX_{t-1}^\top]$:
\begin{equation}
\begin{split}
\hatYXt&=\E_{XY}[\YY_t\XX_{t-1}^\top|\YY(1)=\yy(1)]\\
&=\cov[\YY_t,\XX_{t-1}|\YY_t(1)=\yy_t(1)] + \E_{XY}[\YY_t|\YY(1)=\yy(1)]\E_{XY}[\XX_{t-1}^\top|\YY(1)=\yy(1)]^\top\\
&=\cov[\yy_t-\IR_t(\yy_t-\ZZ_t\XX_t-\aa_t)+\VV^*_t,\XX_{t-1}]+\hatyt\hatxtm^\top \\
&=\cov[\yy_t,\XX_{t-1}]-\cov[\IR_t\yy_t,\XX_{t-1}]+\cov[\IR_t\ZZ_t\XX_t,\XX_{t-1}] \\
&\quad +\cov[\IR_t\aa_t,\XX_{t-1}] + \cov[\VV^*_t,\XX_{t-1}]+\hatyt\hatxtm^\top \\
&=0 - 0 + \IR_t\ZZ_t\hatVttm + 0 + 0 + \hatyt\hatxtm^\top \\
&= \IR_t\ZZ_t\hatVttm + \hatyt\hatxtm^\top
\end{split}
\end{equation}

\section{Degenerate variance models}\label{sec:degenerate}
It is possible that the model has deterministic and probabilistic elements; mathematically this means that either $\GG_t$, $\HH_t$ or $\FF$ have all zero rows, and this means that some of the observation or state processes are deterministic.  Such models often arise when a MAR-p is put into MARSS-1 form.  Assuming the model is solvable (one solution and not over-determined), we can modify the Kalman smoother and EM algorithm to handle models with deterministic elements.    

The motivation behind the degenerate variance modification is that we want to use one set of EM update equations for all models in the MARSS class---regardless of whether they are partially or fully degenerate.  The difficulties arise in getting the $\uu$ and $\xixi$ update equations.  If we were to fix these or make $\xixi$ stochastic (a fixed mean and fixed variance), most of the trouble in this section could be avoided.  However, fixing $\xixi$ or making it stochastic is putting a prior on it and placing a prior on the variance-covariance structure of $\xixi$ that conflicts logically with the model is often both unavoidable (since the correct variance-covariance structure depends on the parameters you are trying to estimate) and disasterous to one's estimation although the problem is often difficult to detect especially with long time series.  Many papers have commented on this subtle problem.  So, we want to be able to estimate $\xixi$ so we do not have to specify $\LAM$ (because we remove it from the model).  Note that in a univariate $\xx$ model (one state), $\LAM$ is just a variance so we do not run into this trouble.  The problems arise when $\xx$ is multivariate (>1 state) and then we have to deal with the variance-covariance structure of the initial states. 

\subsection{Rewriting the state and observation models for degenerate variance systems}
Let's start with an example: 
\begin{equation}
\RR_t=\begin{bmatrix}
1&.2\\
.2&1
\end{bmatrix}\\
\text{  and  }
\HH_t=\begin{bmatrix}
1&0\\
0&0\\
0&1
\end{bmatrix}
\end{equation}
Let $\OMG_{t,r}^+$ be a $p \times n$  matrix that extracts the $p$ non-zero rows from $\HH_t$. The diagonal matrix $(\OMG_{t,r}^+)^\top\OMG_{t,r}^+ \equiv \II_{t,r}^+$ is a diagonal matrix that can zero out the $\HH_t$ zero rows in any $n$ row matrix.   

\begin{equation}
\begin{split}
\OMG_{t,r}^+ &= 
\begin{bmatrix}
1&0&0\\
0&0&1
\end{bmatrix}\quad\quad
\II_{t,r}^+ = (\OMG_{t,r}^+)^\top\OMG_{t,r}^+ = 
\begin{bmatrix}
1&0&0\\
0&0&0\\
0&0&1
\end{bmatrix}\\
\yy_t^+&=\OMG_{t,r}^+ \yy_t\\
\end{split}
\end{equation}

Let $\OMG_{t,r}^{(0)}$ be a $(n-p) \times n$  matrix that extracts the $n-p$ zero rows from $\HH_t$. For the example above, 
\begin{equation}
\begin{split}
\OMG_{t,r}^{(0)} &= 
\begin{bmatrix}
0&1&0\\
\end{bmatrix}\quad\quad
\II_{t,r}^{(0)} = (\OMG_{t,r}^{(0)})^\top\OMG_{t,r}^{(0)} = 
\begin{bmatrix}
0&0&0\\
0&1&0\\
0&0&0
\end{bmatrix}\\
\yy_t^{(0)}&=\OMG_{t,r}^{(0)} \yy_t\\
\end{split}
\end{equation}
Similarly,  $\OMG_{t,q}^+$ extracts the non-zero rows from $\GG_t$ and  $\OMG_{t,q}^{(0)}$ extracts the zero rows.  $\II_{t,q}^+$ and $\II_{t,q}^{(0)}$ are defined similarly.

Using these definitions, we can rewrite the state process part of the MARSS model by separating out the deterministic parts:
\begin{equation}\label{eq:degen.model.x1}
\begin{split}
\xx_t^{(0)}&=\OMG_{t,q}^{(0)}\xx_t=\OMG_{t,q}^{(0)}(\BB_t\xx_{t-1} + \uu_t)\\
\xx_t^+&=\OMG_{t,q}^+\xx_t=\OMG_{t,q}^+(\BB_t\xx_{t-1} + \uu_t + \GG_t\ww_t)\\
\ww_t^+ &\sim \MVN(0,\QQ_t)\\
\xx_0& \sim \MVN(\xixi,\LAM)
\end{split}
\end{equation}
Similarly, we can rewrite the observation process part of the MARSS model by separating out the parts with no observation error:
\begin{equation}\label{eq:degen.model.y1}
\begin{split}
\yy_t^{(0)}&=\OMG_{t,r}^{(0)}\yy_t=\OMG_{t,r}^{(0)}(\ZZ_t\xx_t + \aa_t) \\
&=\OMG_{t,r}^{(0)}(\ZZ_t\II_{t,q}^+\xx_t + \ZZ_t\II_{t,q}^{(0)}\xx_t + \aa_t)\\
\yy_t^+&=\OMG_{t,r}^+\yy_t=\OMG_{t,r}^+(\ZZ_t\xx_t + \aa_t + \HH_t\vv_t)\\
&=\OMG_{t,r}^+(\ZZ_t\II_{t,q}^+\xx_t + \ZZ_t\II_{t,q}^{(0)}\xx_t + \aa_t+\HH_t\vv_t)\\
\vv_t^+ &\sim \MVN(0,\RR_t)
\end{split}
\end{equation}
I am treating $\LAM$ as fully stochastic for this example, but in general $\FF$ might have 0 rows.

In order for this to be solvable using an EM algorithm with the Kalman filter, we require that no estimated $\BB$ or $\uu$ elements appear in the equation for $\yy_t^{(0)}$.  Since the $\yy_t^{(0)}$ do not appear in the likelihood function (since $\HH_t^{(0)}=0$), $\yy_t^{(0)}$ would not affect the estimate for the parameters appearing in the $\yy_t^{(0)}$ equation.  This translates to the following constraints, $(1_{1 \times m} \otimes \OMG_{t,r}^{(0)}\ZZ_t\II_{t,q}^{(0)})\DD_{t,b}$ is all zeros and $\OMG_{t,r}^{(0)}\ZZ_t\II_{t,q}^{(0)}\DD_u$ is all zeros.
Also notice that $\OMG_{t,r}^{(0)}\ZZ_t$ and $\OMG_{t,r}^{(0)}\aa_t$ appear in the $\yy^{(0)}$ equation and not in the $\yy^+$ equation.  This means that $\OMG_{t,r}^{(0)}\ZZ_t$ and $\OMG_{t,r}^{(0)}\aa_t$ must be only fixed terms.

In summary, the degenerate model becomes
\begin{equation}\label{eq:degen.model2}
\begin{split}
\xx_t^{(0)}&=\BB_t^{(0)}\xx_{t-1} + \uu_t^{(0)}\\
\xx_t^+&=\BB_t^+\xx_{t-1} + \uu_t^+ + \GG_t^+\ww_t\\
\ww_t &\sim \MVN(0,\QQ_t)\\
\xx_0& \sim \MVN(\xixi,\LAM)
\\
\yy_t^{(0)}&=\ZZ^{(0)}\II_q^+\xx_t + \ZZ^{(0)}\II_q^{(0)}\xx_t + \aa^{(0)}_t\\
\yy_t^+&=\ZZ_t^+\xx_t + \aa_t^+ \HH_t^+\vv_t\\
&=\ZZ_t^+\II_q^+\xx_t + \ZZ_t^+\II_q^{(0)}\xx_t + \aa_t^+ + \HH_t^+\vv_t\\
\vv_t &\sim \MVN(0,\RR)
\end{split}
\end{equation}
where $\BB_t^{(0)}=\OMG_{t,q}^{(0)}\BB_t$ and $\BB_t^+=\OMG_{t,q}^+\BB_t$ so that $\BB_t^{(0)}$ are the rows of $\BB_t$ corresponding to the zero rows of $\GG_t$ and $\BB_t^+$ are the rows of $\BB_t$ corresponding to non-zero rows of  $\GG_t$.  The other parameters are similarly defined: $\uu_t^{(0)}=\OMG_{t,q}^{(0)}\uu_t$  and $\uu_t^+=\OMG_{t,q}^+\uu_t$, $\ZZ_t^{(0)}=\OMG_{t,r}^{(0)}\ZZ_t$  and $\ZZ_t^+=\OMG_{t,r}^+\ZZ_t$, and $\aa_t^{(0)}=\OMG_{t,r}^{(0)}\aa_t$  and $\aa_t^+=\OMG_{t,r}^+\aa_t$.

\subsection{Identifying the fully deterministic $\xx$  rows}\label{sec:ident.xds}
To derive EM update equations, we need to take the derivative of the expected log-likelihood holding everything but the parameter of interest constant.  If there are deterministic $\xx_t$ rows, then we cannot hold these constant and do this partial differentiation with respect to the state parameters.  We need to identify these $\xx_t$ rows and remove them from the likelihood function by rewriting them in terms of only the state parameters. For this derivation, I am going to make the simplifying assumption that the locations of the 0 rows in $\GG_t$ and $\HH_t$ are time-invariant.  This is not strictly necessary, but simplifies the algebra greatly.
 
For the  deterministic $\xx_t$ rows, the process equation is $\xx_t = \BB_t\xx_{t-1}+\uu_t$, with the $\ww_t$ term left off. When we do the partial differentiation step in deriving the EM update equation for $\uu$, $\BB$ or $\xixi$, we will need to take a partial derivative  while holding $\xx_t$ and $\xx_{t-1}$ constant.   We cannot hold the deterministic rows of $\xx_t$ and $\xx_{t-1}$ constant while changing the corresponding rows of $\uu_t$ and $\BB_t$ (or $\xixi$ if $t=0$ or $t=1$).  If a row of $\xx_t$ is fully deterministic, then that $x_{i,t}$ must change when row $i$ of $\uu_t$ or $\BB_t$ is changed.  Thus we  cannot do the partial differentiation step required in the EM update equation derivation.  

So we need to identify the fully deterministic $\xx_t$ and treat them differently in our likelihood so we can derive the update equation.  I will use the terms 'deterministic', 'indirectly stochastic' and 'directly stochastic' when referring to the $\xx_t$ rows.  Deterministic means that that $\xx_t$ row (denoted $\xx_t^d$) has no state error terms appearing in it (no $w$ terms) and can be written as a function of only the state parameters.  Indirectly stochastic (denoted $\xx_t^{is}$) means that the corresponding row of $\GG_t$ is all zero (an $\xx_t^{(0)}$ row), but the $\xx_t$ row has a state error term ($w$) which it picked up through $\BB$ in one of the prior $\BB_t\xx_t$ steps. Directly stochastic (the $\xx_t^+$) means that the corresponding row of $\GG_t$ is non-zero and thus these row pick up at state error term ($w_t$) at each time step. The stochastic $\xx_t$ are denoted $\xx_t^s$ whether they are indirectly or directly stochastic.

How do you determine the $d$, or deterministic, set of $\xx_t$ rows?  These are the rows with no $w$ terms, from time $t$ or from prior $t$, in them at time $t$.  Note that the location of the $d$ rows is time-dependent, a row may be deterministic at time $t$ but pick up a $w$ at time $t+1$ and thus be indirectly stochastic thereafter.  I am requiring that once a row becomes indirectly stochastic, it remains that way; rows are not allowed to flip back and forth between deterministic (no $w$ terms in them) and indirectly stochastic (containing a $w$ term).

I will work through an example and then show a general algorithm to keep track of the deterministic rows at time $t$.

Let $\xx_0=\xixi$ (so $\FF$ is all zero and $\xx_0$ is not stochastic).  Define $\II_t^{ds}$, $\II_t^{is}$, and $\II_t^d$ as diagonal indicator matrices with a 1 at $\II(i,i)$ if row $i$ is directly stochastic, indirectly stochastic, or deterministic respectively.  $\II_t^s+\II_t^{is}+\II_t^d=\II_m$.  Let our state equation be $\XX_t=\BB_t\XX_{t-1}+\GG_t\ww_t$.
Let
\begin{equation}
\BB=\begin{bmatrix}
1&1&0&0\\
1&0&0&0\\
0&1&0&0\\
0&0&0&1
\end{bmatrix}
\end{equation}
At $t=0$
\begin{equation}
\XX_0=\begin{bmatrix}
\pi_1\\
\pi_2\\
\pi_3\\
\pi_4
\end{bmatrix}
\end{equation}
\begin{equation}
\II_0^d=\begin{bmatrix}
1&0&0&0\\
0&1&0&0\\
0&0&1&0\\
0&0&0&1
\end{bmatrix}
\quad
\II_0^s=\II_0^{is}=\begin{bmatrix}
0&0&0&0\\
0&0&0&0\\
0&0&0&0\\
0&0&0&0
\end{bmatrix}
\end{equation}
At $t=1$
\begin{equation}
\XX_1=\begin{bmatrix}
\pi_1+\pi_2+w_1\\
\pi_1\\
\pi_2\\
\pi_4
\end{bmatrix}
\end{equation}
\begin{equation}
\II_0^d=\begin{bmatrix}
0&0&0&0\\
0&1&0&0\\
0&0&1&0\\
0&0&0&1
\end{bmatrix}
\quad
\II_0^s=\begin{bmatrix}
1&0&0&0\\
0&0&0&0\\
0&0&0&0\\
0&0&0&0
\end{bmatrix}
\quad
\II_0^{is}=\begin{bmatrix}
0&0&0&0\\
0&0&0&0\\
0&0&0&0\\
0&0&0&0
\end{bmatrix}
\end{equation}
At $t=2$
\begin{equation}
\XX_2=\begin{bmatrix}
\dots+w_2\\
\pi_1+\pi_2+w_1\\
\pi_1\\
\pi_4
\end{bmatrix}
\end{equation}
\begin{equation}
\II_0^d=\begin{bmatrix}
0&0&0&0\\
0&0&0&0\\
0&0&1&0\\
0&0&0&1
\end{bmatrix}
\quad
\II_0^s=\begin{bmatrix}
1&0&0&0\\
0&0&0&0\\
0&0&0&0\\
0&0&0&0
\end{bmatrix}
\quad
\II_0^{is}=\begin{bmatrix}
0&0&0&0\\
0&1&0&0\\
0&0&0&0\\
0&0&0&0
\end{bmatrix}
\end{equation}
By $t=3$, the system stabilizes
\begin{equation}
\XX_3=\begin{bmatrix}
\dots+w_1+w_2+w_3\\
\dots+w_1+w_2\\
\pi_1+\pi_2+w_1\\
\pi_4
\end{bmatrix}
\end{equation}
\begin{equation}
\II_0^d=\begin{bmatrix}
0&0&0&0\\
0&0&0&0\\
0&0&0&0\\
0&0&0&1
\end{bmatrix}
\quad
\II_0^s=\begin{bmatrix}
1&0&0&0\\
0&0&0&0\\
0&0&0&0\\
0&0&0&0
\end{bmatrix}
\quad
\II_0^{is}=\begin{bmatrix}
0&0&0&0\\
0&1&0&0\\
0&0&1&0\\
0&0&0&0
\end{bmatrix}
\end{equation}
After time $t=3$ the location of the deterministic and indirectly stochastic rows is stabilized and no longer changes.  

In general, it can take up to $m$ time steps for the location of the deterministic rows to stabilize.  This is because $\BB_t$ is like an adjacency matrix, and I require that the location of the 0s in $\BB_1\BB_2\dots\BB_t$ is time invariant. If we replace all non-zero elements in $\BB_t$ with 1, then we have an adjacency matrix, let's call it $\MM$.  If there is a path in $\MM$ from $x_{j,t}$ to an $x_{s,t}$ , then row $j$ will eventually be indirectly stochastic. Graph theory tells us that it takes at most $m$ steps for a $m \times m$ adjacency matrix to show full connectivity. This means that if element $j,i$ is 0 in $M^m$ then row $j$ is not connected to row $i$ by any path and thus will remain unconnected for $M^{t>m}$; note element $i,j$ can be 0 while $j,i$ is not.

This means that $\BB_1\BB_2\dots\BB_t$, $t>m$, can be rearranged to look something like so where $ds$ are directly stochastic, $is$ are indirectly stochastic, and $d$ are fully deterministic:
\begin{equation}\label{eq:block.B}
\BB_1\BB_2\dots\BB_t=\begin{bmatrix}
ds&ds&ds&ds&ds\\
ds&ds&ds&ds&ds\\
is&is&is&is&is\\
0&0&0&d&d\\
0&0&0&d&d\\
\end{bmatrix}
\end{equation}
The $ds$'s, $is$'s and $d$'s are not all equal nor are they necessarily all non-zero; I am just showing the blocks.  The $d$ rows will always be deterministic while the $is$ rows will only be deterministic for $t<m$ time steps; the number of time steps depends on the form of $\BB$.

Since my $\BB_t$ matrices are small, I use a  inefficient strategy in my code to construct the indicator matrices $\II_d^t$.  I define $\MM$ as $\BB_t$ with the non-zero $\BB$ replaced with 1; I require that the location of the non-zero elements in $\BB_t$ are time-invariant so there is only one $\MM$.  Within the product $\MM^t$, those rows where only 0s appear in the 'stochastic' columns (non-zero $\GG_t$ rows) are the fully deterministic $\xx_{t+1}$ rows. Note, $t+1$ so one time step ahead.  There are much faster algorithms for finding paths, but my $\MM$ tend to be small.  Also, unfortunately, using $\BB_1\BB_2\dots\BB_t$, needed for the $\xx_t^d$ function, in place of $\MM^t$ is not robust. Let's say $\BB=\bigl[ \begin{smallmatrix}-1&-1\\ 1&1\end{smallmatrix} \bigr]$ and $\GG=\bigl[ \begin{smallmatrix}1\\ 0\end{smallmatrix} \bigr]$.
Then $\BB^2$ is a matrix of all zeros even though the correct $\II_2^d$ is $\bigl[ \begin{smallmatrix}
0&0\\ 0&0
\end{smallmatrix} \bigr]$ not 
$\bigl[ \begin{smallmatrix}
0&0\\ 0&1
\end{smallmatrix} \bigr]$.

\subsubsection{Redefining the $\xx_t^d$ elements in the likelihood}
By definition, all the $\BB_t$ elements in the $ds$ and $is$ columns of the $d$ rows of $\BB_t$ are 0 (see equation \ref{eq:block.B}).  This is due to the constraint that I have imposed that locations of 0s in $\BB_t$ are time-invariant and the location of the zero rows in $\GG_t$ also time-invariant: $\II_q^+$ and $\II_q^{(0)}$ are time-constant. 

Consider this $\BB$ and $\GG$, which would arise in a MARSS version of an AR-3 model:
\begin{equation}
\BB=\begin{bmatrix}
b_1&b_2&b_3\\
1&0&0\\
0&1&0\end{bmatrix}
\quad
\GG=\begin{bmatrix}
1\\
0\\
0\end{bmatrix}
\end{equation}
Using $\xx_0=\xixi$:
\begin{equation}
\xx_0=\begin{bmatrix}
\pi_1\\
\pi_2\\
\pi_3
\end{bmatrix}
\quad
\xx_1=\begin{bmatrix}
\dots+w_1\\
\pi_1\\
\pi_2
\end{bmatrix}
\quad
\xx_2=\begin{bmatrix}
\dots+w_2\\
\dots+w_1\\
\pi_1\end{bmatrix}
\quad
\xx_3=\begin{bmatrix}
\dots+w_3\\
\dots+w_2\\
\dots+w_1
\end{bmatrix}
\end{equation}
The $\dots$ just represent 'some values'.  The key part is the $w$ appearing which is the stochasticity.  At $t=1$, rows 2 and 3 are deterministic.  At $t=2$, row 3 is deterministic, and at $t=3$, no rows are deterministic.

We can rewrite the equation for the deterministic rows in $\xx_t$ as follows.  Note that by definition, all the non-$d$ columns in the $d$-rows of $\BB_t$ are zero. $\xx_t^d$ is $\xx_t$ with the $d$ rows zeroed out, so $\xx_t^d = \II_{q,t}^d\xx_t$.
\begin{equation} \label{eq:xt.det.sum}
\begin{split}
\xx_1^{d}&=\BB_1^d \xx_0 + \uu_1^d  \\
				 &=\BB_1^d \xx_0 + \ff_{u,1}^d + \DD_{u,1}^d\uupsilon \\
				 &=\II_1^d(\BB_1 \xx_0 + \ff_{u,1} + \DD_{u,1}\uupsilon) \\
\xx_2^{d}&=\BB_2^d \xx_1 + \uu_2^{d}  \\
         &=\BB_2^d(\II_1^d(\BB_1 \xx_0 + \ff_{u,1} + \DD_{u,1}\uupsilon)) + \ff_{u,2}^{d} + \DD_{u,2}^{d}\uupsilon \\
         &=\II_2^d\BB_2\II_2^d(\II_1^d(\BB_1 \xx_0 + \ff_{u,1} + \DD_{u,1}\uupsilon)) + \II_2^d\ff_{u,2} + \II_2^d\DD_{u,2}\uupsilon \\
         &=\II_2^d\BB_2\BB_1 \xx_0 + \II_2^d(\BB_2\ff_{1,u} + \ff_{2,u}) + \II_2^d(\BB_2\DD_{u,1} +\DD_{u,2})\uupsilon \\
         &=\II_2^d(\BB_2\BB_1 \xx_0 + \BB_2\ff_{1,u} + \ff_{2,u} + (\BB_2\DD_{u,1} +\DD_{u,2})\uupsilon) \\
         \dots \\
\end{split}
\end{equation}
The messy part is keeping track of which rows are deterministic because this will potentially change up to time $t=m$.

We can rewrite the function for $\xx_t^d$, where $t_0$ is the $t$ at which the initial state is defined.  It is either $t=0$ or $t=1$.
\begin{equation} \label{eq:xt.det.sum2}
\begin{split}
\xx_t^d &= \II_t^d(\BB^*_t\xx_{t_0} + \ff^*_t +\DD^*_t\uupsilon) \\
\text{where}&\\
\BB^*_{t_0} &= \II_m \\
\BB^*_t &=\BB_t\BB^*_{t-1} \\ 
\\
\ff^*_{t_0} &= 0 \\
\ff^*_t &=\BB_t\ff^*_{t-1} + \ff_{t,u}\\ 
\\
\DD^*_{t_0} &= 0 \\
\DD^*_t &= \BB_t\DD^*_{t-1}  + \DD_{t,u}\\ 
\\
\II_{t_0}^d&=\II_m \\
\text{diag}(\II_{t_0+\tau}^d)&=\text{apply}(\OMG_q^{(0)}\MM^\tau\OMG_q^+ == 0, 1, \text{all}) 
\end{split}
\end{equation}
The bottom line is written in R: $\II_{t_0+\tau}^d$ is a diagonal matrix with a 1 at $(i,i)$ where row $i$ of $\GG$ is all 0 and all $ds$ and $is$ columns in row $i$ of $\MM^t$ are equal to zero.

In the expected log-likelihood, the term $\E[\XX_t^d]=\E[\XX_t^d|\YY=\yy]$, meaning the expected value of $\XX_t^d$ conditioned on the data, appears. Thus in the expected log-likelihood the function will be written:
\begin{equation}
\begin{split}
\XX_t^d &= \II_t^d(\BB^*_t\XX_{t_0} + \ff^*_t +\DD^*_t\uupsilon) \\
\E[\XX_t^d] &= \II_t^d(\BB^*_t\E[\XX_{t_0}] + \ff^*_t +\DD^*_t\uupsilon)
\end{split}
\end{equation}

When the $j$-th row of $\FF$ is all zero, meaning the $j$-th row of $\xx_0$ is fixed to be $\xi_j$, then $\E[X_{t_0,j}]\equiv\xi_j$.  This is the case where we treat $x_{t_0,j}$ as fixed and we either estimate or specify its value.  If $\xx_{t_0}$ is wholly treated as fixed, then $\E[\XX_{t_0}]\equiv\xixi$ and $\LAM$ does not appear in the model at all.  In the general case, where some $x_{t_0,j}$ are treated as fixed and some as stochastic, we can write $\E[\XX_t^d]$ appearing in the expected log-likelihood as:
\begin{equation}\label{eq:EXtd}
\begin{split}
\E[\XX_{t_0}]=(\II_m-\II_\lambda^{(0)})\E[\XX_{t_0}]+\II_\lambda^{(0)}\xixi
\end{split}
\end{equation}
$\II_\lambda^{(0)}$ is a diagonal indicator matrix with 1 at $(j,j)$ if row $j$ of $\FF$ is all zero.
  
If $\BB^{d,d}$ and $\uu^d$ are time-constant, we could use the matrix geometric series: 
\begin{equation}\label{eq:xt.geo}
\begin{split}
\xx_t^{d}=&(\BB^{d,d})^t \xx_0^{d} + \sum_{i=0}^{t-1}(\BB^{d,d})^i \uu^{d} = 
(\BB^{d,d})^t \xx_0^{d} + (\II - \BB^{d,d})^{-1}(\II-(\BB^{d,d})^t)\uu^{d}, \quad\text{if }\BB^{d,d} \neq \II\\
&\xx_0^d + \uu^d,\quad\text{if }\BB^{d,d} = \II
\end{split}
\end{equation}
where $\BB^{d,d}$ is the block of $d$'s in equation \ref{eq:block.B}. 

\subsubsection{Dealing with the $\xx_t^{is}$ elements in the likelihood and associated parameter rows}
Although $\ww_t^{is}=0$, these terms are connected to the stochastic $\xx$'s in earlier time steps though $\BB$, thus all $\xx_t^{is}$ are possible for a given $\uu_t$, $\BB_t$ or $\xixi$. However, all $\xx_t^{is}$ are not possible conditioned on $\xx_{t-1}$, so we are back in the position that we cannot both change $\xx_t$ and change $\uu_t$.

Recall that for the partial differentiation step in the EM algorithm, we need to be able to hold the $E[\XX_t]$ appearing in the likelihood constant.  We can deal with the deterministic $\xx_t$ because they are not stochastic and do not have 'expected values'.  They can be removed from the likelihood by rewriting $\xx_t^d$ in terms of the model parameters.  We cannot do that for $\xx_t^{is}$ because these $x$ are stochastic. There is no equation for them; all $\xx^{is}$ are possible but some are more likely than others.  We also cannot replace $\xx_t^{is}$ with $\BB_t^{is}E[\XX_{t-1}]+\uu_t^{is}$ to force $\BB_t^{is}$ and $\uu^{is}$ to appear in the $\yy$ part of the likelihood.  The reason is that $\E[\XX_t]$ and $\E[\XX_{t-1}]$ both appear in the likelihood and we cannot hold both constant (as we must for the partial differentiation) and at the same time change $\BB_t^{is}$ or $\uu_t^{is}$ as we are doing when we differentiate with respect to $\BB_t^{is}$ or $\uu)_t^{is}$.  We cannot do that because  $\xx_t^{is}$ is constrained to equal $\BB_t^{is}\xx_{t-1}+\uu_t^{is}$.

This effectively means that we cannot estimate $\BB_t^{is}$ and $\uu_t^{is}$ because we cannot rewrite $\xx_t^{is}$ in terms of only the model parameters.  This is specific to the EM algorithm because it is an iterative algorithm where the expected $\XX_t$ are computed with fixed parameters and then the $\E[\XX_t]$ are held fixed at their expected values while the parameters are updated.  In my $\BB$ update equation, I assume that $\BB_t^{(0)}$ is fixed for all $t$.  Thus I circumvent the problem altogether for $\BB$.  For $\uu$, I assume that only the $\uu^{is}$ elements are fixed.

\subsection{Expected log-likelihood for degenerate models}

The basic idea is to replace $\II_q^d\E[\XX_t]$ with a deterministic function involving only  the state parameters (and $\E[\XX_{t_0}]$ if $\XX_{t_0}$ is stochastic) . These appear in the $\yy$ part of the likelihood in $\ZZ_t\XX_t$ when the $d$ columns of $\ZZ_t$ have non-zero values.  They appear in the $\xx$ part of the likelihood in $\BB_t\XX_{t-1}$ when the $d$ columns of $\BB_t$ have non-zero values.  They do not appear in $\XX_t$ in the $\xx$ part of the likelihood because $\Qm_t$ has all the non-$s$ columns and rows zeroed out (non-$s$ includes both $d$ and $is$) and the element to the left of $\Qm_t$ is a row vector and to the right, it is a column vector.  Thus any $\xx_t^d$ in $\XX_t$ are being zeroed out by $\Qm_t$.

The first step is to pull out the $\II_t^{d}\XX_t$:
\begin{equation}
\begin{split}
\Psi^+ &= \E[\log\LL(\YY^+,\XX^+ ; \Theta)] = \E[-\frac{1}{2}\sum_1^T \\
& (\YY_t - \ZZ_t(\II_m-\II_t^d)\XX_t - \ZZ_t\II_t^d\XX_t - \aa_t)^\top \Rm_t\\ 
&(\YY_t - \ZZ_t(\II_m-\II_t^d)\XX_t - \ZZ_t\II_t^d\XX_t - \aa_t) -\frac{1}{2}\sum_1^T\log |\RR_t|\\
& -\frac{1}{2}\sum_{t_0+1}^T (\XX_t - \BB_t ((\II_m - \II_{t-1}^d)\XX_{t-1} + \II_{t-1}^d\XX_{t-1}) - \uu_t)^\top \Qm_t \\
&(\XX_t - \BB_t ((\II_m - \II_{t-1}^d)\XX_{t-1} + \II_{t-1}^d\XX_{t-1}) - \uu_t) - \frac{1}{2}\sum_{t_0+1}^T\log |\QQ_t| \\
&  - \frac{1}{2}(\XX_{t_0} - \xixi)^\top \LAMm(\XX_{t_0} - \xixi) - \frac{1}{2}\log |\LAM| -\frac{n}{2}\log 2\pi 
\end{split}
\end{equation}
See section \ref{sec:ident.xds} for the definition of  $\II_t^d$.

Next we replace $\II_q^d\XX_t$ with equation \eqref{eq:xt.det.sum2}.  $\XX_{t_0}$ will appear in this function instead of $\xx_{t_0}$.  I rewrite $\uu_t$ as $\ff_{u,t}+\DD_{u,t}\uupsilon$.  This gives us the expected log-likelihood:
\begin{equation}\label{eq:degen.logL.x0}
\begin{split}
\Psi^+ &= \E[\log\LL(\YY^+,\XX^+ ; \Theta)] = \E[-\frac{1}{2}\sum_1^T \\
& (\YY_t - \ZZ_t(\II_m-\II_t^d)\XX_t - \ZZ_t\II_t^d(\BB^*_t\XX_{t_0} + \ff^*_t +\DD^*_t\uupsilon) - \aa_t)^\top \Rm_t\\ 
&(\YY_t - \ZZ_t(\II_m-\II_t^d)\XX_t - \ZZ_t\II_t^d(\BB^*_t\XX_{t_0} + \ff^*_t +\DD^*_t\uupsilon) - \aa_t) -\frac{1}{2}\sum_1^T\log |\RR_t|\\
& -\frac{1}{2}\sum_{t_0+1}^T (\XX_t - \BB_t ((\II_m - \II_{t-1}^d)\XX_{t-1} + \II_{t-1}^d(\BB^*_{t-1}\XX_{t_0} + \ff^*_{t-1} +\DD^*_{t-1}\uupsilon)) - \ff_{u,t} - \DD_{u,t}\uupsilon)^\top \Qm_t \\
&(\XX_t - \BB_t ((\II_m - \II_{t-1}^d)\XX_{t-1} + \II_{t-1}^d(\BB^*_{t-1}\XX_{t_0} + \ff^*_{t-1} +\DD^*_{t-1}+\uupsilon)) - \ff_{u,t} - \DD_{u,t}\uupsilon) \\
&- \frac{1}{2}\sum_{t_0}^T\log |\QQ_t|   - \frac{1}{2}(\XX_{t_0} - \xixi)^\top \LAMm(\XX_{t_0} - \xixi) - \frac{1}{2}\log |\LAM| -\frac{n}{2}\log 2\pi 
\end{split}
\end{equation}
where $\BB^*$, $\ff^*$ and $\DD^*$ are defined in equation \eqref{eq:xt.det.sum2}.  $\Rm_t = \Xi_t^\top\RR_t^{-1}\Xi_t$ and $\Qm_t = \Phi_t^\top\QQ_t^{-1}\Phi_t$, $\LAMm = \Pi^\top\LAM^{-1}\Pi$.  When $\xx_{t_0}$ is treated as fixed, $\LAMm=0$ and the last line will drop out altogether, however in general some rows of $\xx_{t_0}$ could be fixed and others stochastic.  

We can see directly in equation \eqref{eq:degen.logL.x0} where $\uupsilon$ appears in the expected log-likelihood.  Where $\pp$ appears is less obvious because it depends on $\FF$, which specifies which rows of $\xx_{t_0}$ are fixed.  From equation \eqref{eq:EXtd}, $$\E[\XX_{t_0}]=(\II_m-\II_l^{(0)})\E[\XX_{t_0}]+\II_l^{(0)}\xixi$$
and $\xixi=\ff_\xi+\DD_\xi\pp$.  Thus where $\pp$ appears in the expected log-likelihood depends on the location of zero rows in $\FF$ (and thus the zero rows in the indicator matrix $\II_l^{(0)}$).  Recall that $\E[\XX_{t_0}]$ appearing in the expected log-likelihood function is conditioned on the data so $\E[\XX_{t_0}]$ in $\Psi$ is not equal to $\xixi$ if $\xx_{t_0}$ is stochastic. 

The case where $\xx_{t_0}$ is stochastic is a little odd because conditioned on $\XX_{t_0}=\xx_{t_0}$, $\xx_t^d$ is deterministic even though $\XX_0$ is a random variable in the model.  Thus in the model, $\xx_t^d$ is a random variable through $\XX_{t_0}$.  But when we do the partial differentiation step for the EM algorithm, we hold $\XX$ at its expected value thus we are holding $\XX_{t_0}$ at a specific value.  We cannot do that and change $\uu$ at the same time because once we fix $\XX_{t_0}$ the  $\xx_t^d$ are deterministic functions of $\uu$.

\subsection{Logical constraints to ensure a consistent system of equations}
We need to ensure that the model remains internally consistent when $\RR$ or $\QQ$ goes to zero and that we do not have an over- or under-constrained system.

As an example of a solvable versus unsolvable model, consider the following.
\begin{equation}
\HH_t\RR_t=\begin{bmatrix}
0&0\\
1&0\\
0&1\\
0&0
\end{bmatrix}
\begin{bmatrix}
a&0\\
0&b\\
\end{bmatrix} = \begin{bmatrix}
0&0&0&0\\
0&a&0&0\\
0&0&b&0\\
0&0&0&0\\\end{bmatrix},
\end{equation}
then following are bad versus ok $\ZZ$ matrices.
\begin{equation}
\ZZ_{\text{bad}}=\begin{bmatrix}
c&d&0\\
z(2,1)&z(2,2)&z(2,3)\\
z(3,1)&z(3,1)&z(3,1)\\
c&d&0
\end{bmatrix},\quad
\ZZ_{\text{ok}}=\begin{bmatrix}
c&0&0\\
z(2,1)&z(2,2)&z(2,3)\\
z(3,1)&z(3,1)&z(3,1)\\
c&d\neq0&0
\end{bmatrix}
\end{equation}
Because $y_t(1)$ and $y_t(4)$ have zero observation variance, the first $\ZZ$ reduces to this for $x_t(1)$ and $x_t(2)$:
\begin{equation}
\begin{bmatrix}
y_t(1)\\
y_t(4)
\end{bmatrix}=
\begin{bmatrix}
c x_t(1) + d x_t(2)\\
c x_t(1) + d x_t(2)
\end{bmatrix}
\end{equation}
and since $y_t(1)\neq y_t(4)$, potentially, that is not solvable.  The second $\ZZ$ reduces to
\begin{equation}
\begin{bmatrix}
y_t(1)\\
y_t(4)
\end{bmatrix}=
\begin{bmatrix}
c x_t(1)\\
c x_t(1) + d x_t(4)
\end{bmatrix}
\end{equation}and that is solvable for any $y_t(1)$ and $y_t(4)$ combination.  Notice that in the latter case, $x_t(1)$ and $x_t(2)$ are fully specified by $y_t(1)$ and $y_t(4)$. 

\subsubsection{Constraint 1: $\ZZ$ does not lead to an over-determined observation process}
We need to ensure that a $\xx_t$ exists for all $\yy^{(0)}_t$ such that:
$$\E[\yy^{(0)}_t]=\ZZ^{(0)}\E[\xx_t]+\aa^{(0)}.$$  If $\ZZ^{(0)}$ is invertible, such a $\xx_t$ certainly exists.  But we do not require that only one $\xx_t$ exists, simply that at least one exists.  Thus the system can be under-constrained but not over-constrained.  One way to test for this is to use the singular value decomposition (SVD) of $\ZZ^{(0)}$.  If the number of singular values of $\ZZ^{(0)}$ is less than the number of columns in $\ZZ$, which is the number of $\xx$ rows, then $\ZZ^{(0)}$ specifies an over-constrained system ($y=Zx$\footnote{This is the classic problem of solving the system of linear equations, which is standardly written $Ax=b$.}) Using the R language, you would test if the length of \verb@svd(Z)$d@ is less than than \verb@dim(Z)[2]@.   If $\ZZ^{(0)}$ specifies and under-determined system, some of the singular values would be equal to 0 (within machine tolerance).  It is possible that $\ZZ^{(0)}$ could specify both an over- and under-determined system at the same time.  That is, the number of singular values could be less than the number of columns in $\ZZ^{(0)}$ and some of the singular values could be 0.

Doesn't a $\ZZ$ with more rows than columns automatically specify a over-determined system? No.  Considered this $\ZZ$
\begin{equation}
\begin{bmatrix}
1&0\\
0&1\\
0&0
\end{bmatrix}
\end{equation}
This $\ZZ$ is fine, although obviously the last row of $\yy$ will not hold any information about the $\xx$.  But it could have information about $\RR$ and $\aa$, which might be shared with the other $\yy$, so we don't want to prevent the user from specifying a $\ZZ$ like this.

\subsubsection{Constraint 2: the state processes are not over-constrained. }
We also need to be concerned with the state process being over-constrained when both $\QQ=0$ and $\RR=0$ because we can have a situation where the constraint imposed by the observation process is at odds with the constraint imposed by the state process.  Here is an example:

\begin{equation}
\begin{split}
\yy_t=\begin{bmatrix}
1&0\\
0&1
\end{bmatrix}
\begin{bmatrix}
x_1\\x_2
\end{bmatrix}_t\\
\begin{bmatrix}
x_1\\x_2
\end{bmatrix}_t=
\begin{bmatrix}
1&0\\
0&0
\end{bmatrix}
\begin{bmatrix}
x_1\\x_2
\end{bmatrix}_{t-1}
+
\begin{bmatrix}
w_1\\0
\end{bmatrix}_{t-1}
\end{split}
\end{equation}

In this case, some of the $x$'s are deterministic, $\QQ=0$ and not linked through $\BB$ to a stochastic $x$, and the corresponding $y$ are also deterministic.  These cases will show up as errors in the Kalman filter/smoother because in the Kalman gain equation (equation \ref{eq:Kt}), the term $\ZZ_t\VV_t^{t-1}\ZZ_t^\top$ will appear when $\RR=0$.  We need to make sure that 0 rows in $\BB_t$, $\ZZ_t$ and $\QQ_t$ do not line up in such a way that 0 rows/cols do not appear in $\ZZ_t\VV_t^{t-1}\ZZ_t^\top$ at the same place as 0 rows/cols in $\RR$.  In MARSS, this is checked by doing a pre-run of the Kalman smoother to see if it throws an error in the Kalman gain step. 

\section{EM algorithm modifications for degenerate models}

The $\RR$, $\QQ$, $\ZZ$, and $\aa$ update equations are largely unchanged.  The real difficulties arise for the $\uu$ and $\xixi$ update equations when $\uu^{(0)}$ or $\xixi^{(0)}$ are estimated.  For $\BB$, I do not have a degenerate update equation, so I need to assume that $\BB^{(0)}$ elements are fixed (not estimated).

\subsection{$\RR$ and $\QQ$ update equations}
The constrained update equations for  $\QQ$ and $\RR$ work fine because their update equations do not involve any inverses of non-invertible matrices. However if $\HH_t\RR_t\HH_t^\top$ is non-diagonal and there are missing values, then the $\RR$ update equation  involves $\hatyt$. That will involve the inverse of $\HH_t\RR_{11}\HH_t^\top$  (section \ref{sec:exp.Y}), which might have zeros on the diagonal.  In that case, use the $\IR_t$ modification that deals with such zeros (equation \ref{eq:IRt.degen}).

\subsection{$\ZZ$ and $\aa$ update equations}
We need to deal with $\ZZ$ and $\aa$ elements that appear in rows where the diagonal of $\RR=0$.  These values will not appear in the likelihood function unless they also happen to also appear on the rows where the diagonal of $\RR$ is not 0 (because they are constrained to be equal for example).  However, in this case the $\ZZ^{(0)}$ and $\aa^{(0)}$ are logically constrained by the equation
$$\yy_t^{(0)}=\ZZ_t^{(0)}\E[\xx_t]+\aa_t^{(0)}.$$
Notice there is no $\ww_t$ since $\RR=0$ for these rows.  The $\E[\xx_t]$ is ML estimate of $\xx_t$ computed in the Kalman smoother from the parameter values at iteration $i$ of the EM algorithm, so there is no information in this equation for $\ZZ$ and $\aa$ at iteration $i+1$.  The nature of the smoother is that it will find the $\xx_t$ that is most consistent with the data.  For example if our $y=Zx+a$ equation looks like so
\begin{equation}
\begin{bmatrix}
0\\
2\\
\end{bmatrix}
=
\begin{bmatrix}
1\\
1\\
\end{bmatrix}
x,
\end{equation}
there is no $x$ that will solve this.  However $x=1$ is the closest (lowest squared error) and so this is the information in the data about $x$.  The Kalman filter will use this and the relative value of $\QQ$ and $\RR$ to come up with the estimated $x$.  In this case, $\RR=0$, so the information in the data will completely determine $x$ and the smoother would return $x=1$ regardless of the process equation.

The $\aa$ and $\ZZ$ update equations require that $\sum_{t=1}^T \DD_{t,a}^\top\Rm_t\DD_{t,a}$ and $\sum_{t=1}^T \DD_{t,z}^\top\Rm_t\DD_{t,z}$ are invertible.  If $\ZZ_t^{(0)}$ and $\aa_t^{(0)}$ are fixed, this will be satisfied, however the restriction is a little less restrictive than that since it is possible that $\Rm_t$ does not have zeros on the diagonal in the same places so that the sum over $t$ could be invertible while the individual values at $t$ are not.  The section on the summary of constraints has the test for this constraint.

The  update equations also
involve $\hatyt$, and the modified algorithm for $\hatyt$ when $\HH_t$ has all zero rows will be needed.  Other than that, the constrained update equations work (sections \ref{sec:constA} and \ref{sec:constZ}).  

\subsection{$\uu$ update equation}

Here I discuss the update for $\uu$, or more specifically $\uupsilon$ which appears in $\uu$, when $\GG_t$ or $\HH_t$ have zero rows.  I require that $\uu^{is}_t$ is not estimated.  All the $\uu^{is}_t$ are fixed values.  The $\uu_t^d$ may be estimated or more specifically there may be $\uupsilon$ in $\uu_t^d$ that are estimated;     $\uu_t^d = \ff_{u,t}^d+\DD_{u,t}^d\uupsilon$.

For the constrained $\uu$ update equation with deterministic $\xx$'s takes the following form.  It is similar to the unconstrained update equation except that that a part from the $\yy$ part of the likelihood now appears:
\begin{equation}
\begin{split}\label{eq:u.update.degen.2}
\pmb{\upsilon}_{j+1} = \bigg(\sum_{t=1}^T(\Delta_{t,2}^\top\Rm_t\Delta_{t,2} + \Delta_{t,4}^\top\Qm_t\Delta_{t,4})\bigg)^{-1} \times
 \bigg( \sum_{t=1}^T \big( \Delta_{t,2}^\top\Rm_t\Delta_{t,1} + \Delta_{t,4}^\top\Qm_t\Delta_{t,3} \big)  \bigg)\\
\end{split}
\end{equation}

Conceptually, I think the approach described here is the similar to the approach presented in section 4.2.5 of \citep{Harvey1989}, but it is more general because it deals with the case where some $\uu$ elements are shared (linear functions of some set of shared values), possibly across deterministic and stochastic elements.  Also, I present it here within the context of the EM algorithm, so solving for the maximum-likelihood $\uu$ appears in the context of maximizing $\Psi^+$ with respect to $\uu$ for the update equation at iteration $j+1$.

\subsubsection{$\uu^{(0)}$ is not estimated}
When $\uu^{(0)}$ is not estimated (since it is at some user defined value via $\DD_u$ and $\ff_u$), the part we are estimating, $\uu^+$, only appears in the $\xx$ part of the likelihood. The update equation for $\uu$ remains equation \eqref{eq:u.general.update1}.

\subsubsection{$\uu^d$ is estimated}
The derivation of the update equation proceeds as usual.  We need to take the partial derivative of $\Psi^+$ (equation \ref{eq:degen.logL.x0}) holding everything constant except $\uupsilon$, elements of which might appear in both $\uu_t^d$ and $\uu_t^s$ (but not $\uu_t^{is}$ since I require that $\uu_t^{is}$ has no estimated elements). 

The expected log-likelihood takes the following form, where $t_0$ is the time where the initial state is defined ($t=0$ or $t=1$):
\begin{equation}\label{eq:degen.logL.u}
\begin{split}
\Psi^+ =  
-\frac{1}{2}\sum_1^T (\Delta_{t,1} - \Delta_{t,2}\pmb{\upsilon})^\top\Rm_t(\Delta_{t,1} - \Delta_{t,2}\pmb{\upsilon}) - \frac{1}{2}\sum_1^T\log |\RR_t| \\
-\frac{1}{2}\sum_{t_0+1}^T (\Delta_{t,3} - \Delta_{t,4}\pmb{\upsilon})^\top\Qm_t(\Delta_{t,3} - \Delta_{t,4}\pmb{\upsilon}) - \frac{1}{2}\sum_{t_0+1}^T\log |\QQ_t| \\
- \frac{1}{2}(\XX_{t_0} - \xixi)^\top \LAMm(\XX_{t_0} - \xixi) - \frac{1}{2}\log |\LAM| - \frac{n}{2}\log 2\pi \\
\end{split}
\end{equation}
$\LAMm=\FF^\top\LAM^{-1}\FF$.  If $\xx_{t_0}$ is treated as fixed, $\FF$ is all zero and the line with $\LAMm$ drops out.  If some but not all $\xx_{t_0}$ are treated as fixed, then only the stochastic rows appear in the last line.  In any case, the last line does not contain $\uupsilon$, thus when we do the partial differentiation with respect to $\uupsilon$, this line drops out. 

The $\Delta$ terms are defined as:
\begin{equation}\label{eq:degen.u.update.Deltas}
\begin{split}
\Delta_{t,1}&=\hatyt - \ZZ_t(\II_m - \II_t^d)\hatxt - \ZZ_t\II_t^{d}(\BB^*_t\E[\XX_{t_0}] + \ff^*_t) - \aa_t \\
\Delta_{t,2}&= \ZZ_t\II_t^{d}\DD^*_t \\
\Delta_{t_0,3}&=0_{m\times 1} \\
\Delta_{t,3}&=\hatxt - \BB_t(\II_m - \II_{t-1}^d)\hatxtm - \BB_t\II_{t-1}^{d}(\BB^*_{t-1} \E[\XX_{t_0}] + \ff^*_{t-1}) - \ff_{t,u} \\
\Delta_{t_0,4}&=0_{m\times m} \DD_{1,u} \\
\Delta_{t,4}&=\DD_{t,u}+\BB_t\II_{t-1}^{d}\DD^*_{t-1} \\
\E[\XX_{t_0}]&=((\II_m-\II_\lambda^{(0)})\widetilde{\xx}_{t_0}+\II_\lambda^{(0)} \xixi)
\end{split}
\end{equation}
$\II^d_t$, $\BB_t^*$, $\ff_t*$, and $\DD_t^*$ are defined in equation \eqref{eq:xt.det.sum2}.  The values of these at $t_0$ is special so that the math works out. The expectation ($\E$) has been subsumed into the $\Delta$s since $\Delta_2$ and $\Delta_4$ do not involve $\XX$ or $\YY$, so terms like $\XX^\top\XX$ never appear.

Take the derivative of this with respect to $\uupsilon$ and arrive at:
\begin{equation}
\begin{split}
\label{eq:p.degen.update.u}
\uupsilon_{j+1} = \big(\sum_{t=1}^T\Delta_{t,4}^\top\Qm_t\Delta_{t,4} + \sum_{t=1}^T\Delta_{t,2}^\top\Rm_t\Delta_{t,2}\big)^{-1} \times
 \bigg(\sum_{t=1}^T\Delta_{1,4}^\top\Qm_t\Delta_{1,3} + \sum_{t=1}^T \Delta_{t,2}^\top\Rm_t\Delta_{t,1})\bigg)
\end{split}
\end{equation}

\subsection{$\xixi$ update equation}

\subsubsection{$\xixi$ is stochastic}
This means that none of the rows of $\FF$ (in $\FF\lambda$)  are zero, so $\II_\lambda^{(0)}$ is all zero and the update equation reduces to a constrained version of the classic $\xixi$ update equation:
\begin{equation}
\label{eq:p.degen.update.x0.0.a}
\pp_{j+1} = \big(\DD_\xi^\top\LAM^{-1}\DD_\xi\big)^{-1}\DD_\xi^\top\LAM^{-1}(\E[\XX_{t_0}] - \ff_\xi)
\end{equation}
\subsubsection{$\xixi^{(0)}$ is not estimated}
When $\xixi^{(0)}$ is not estimated (because you fixed it as some value), we do not need to take the partial derivative with respect to $\xixi^{(0)}$ since we will not be estimating it. Thus the update equation is unchanged from the constrained update equation.

\subsubsection{$\xixi^{(0)}$ is estimated}
Using the same approach as for  $\uu$ update equation, we take the derivative of \eqref{eq:degen.logL.x0} with respect to $\pp$ where $\xixi=\ff_\xi+\DD_\xi\pp$.  $\Psi^+$ will take the following form:
\begin{equation}\label{eq:degen.logL.x0.0}
\begin{split}
\Psi^+ &=   \\
&-\frac{1}{2}\sum_{t=1}^T (\Delta_{t,5} - \Delta_{t,6}\pp)^\top\Rm_t(\Delta_{t,5} - \Delta_{t,6}\pp) - \frac{1}{2}\sum_1^T\log |\RR_t| \\
& -\frac{1}{2}\sum_{t=1}^T (\Delta_{t,7} - \Delta_{t,8}\pp)^\top\Qm_t(\Delta_{t,7} - \Delta_{t,8}\pp) - \frac{1}{2}\sum_1^T\log |\QQ_t| \\
& -\frac{1}{2} (\E[\XX_{t_0}] - \ff_\xi - \DD_\xi\pp)^\top\LAMm(\E[\XX_{t_0}] - \ff_\xi - \DD_\xi\pp) - \frac{1}{2}\log |\LAM| \\
& - \frac{n}{2}\log 2\pi \\
\end{split}
\end{equation}

The $\Delta$'s are defined as follows using $ \E[\XX_{t_0}]=(\II_m-\II_l^{(0)})\widetilde{\xx}_{t_0}+\II_l^{(0)}\xixi$ where it appears in $\II_t^d\E[\XX_t]$.
\begin{equation}\label{eq:degen.x0.update.Deltas}
\begin{split}
\Delta_{t,5}&=\hatyt - \ZZ_t(\II_m - \II_t^d)\hatxt - \ZZ_t\II_t^{d}(\BB^*_t((\II_m-\II_\lambda^{(0)})\widetilde{\xx}_{t_0}+\II_\lambda^{(0)} \ff_\xi) + \uu^*_t) - \aa_t \\
\Delta_{t,6}&= \ZZ_t\II_t^{d}\BB^*_t\II_\lambda^{(0)}\DD_\xi \\
\Delta_{t_0,7}&=0_{m\times 1} \\
\Delta_{t,7}&=\hatxt - \BB_t(\II_m - \II_{t-1}^d)\hatxtm - \BB_t\II_{t-1}^{d}(\BB^*_{t-1} ((\II_m-\II_l^{(0)})\widetilde{\xx}_{t_0}+\II_\lambda^{(0)} \ff_\xi) + \uu^*_{t-1}) - \uu_t \\
\Delta_{t_0,8}&=0_{m\times m} \DD_\xi\\
\Delta_{t,8}&=\BB_t\II_{t-1}^{d}\BB^*_{t-1}\II_\lambda^{(0)}\DD_\xi\\
\uu^*_t=\ff^*_t+\DD^*_t\uupsilon
\end{split}
\end{equation}
The expectation can be pulled inside the $\Delta$s since the $\Delta$s in front of $\pp$ do not involve $\XX$ or $\YY$.

Take the derivative of this with respect to $\pp$ and arrive at:
\begin{equation}
\begin{split}
\label{eq:p.degen.update.x0.0.a}
\pp_{j+1} &= \big(\sum_{t=1}^T\Delta_{t,8}^\top\Qm_t\Delta_{t,8} + \sum_{t=1}^T\Delta_{t,6}^\top\Rm_t\Delta_{t,6} + \DD_\xi^\top\LAMm\DD_\xi\big)^{-1} \times\\
&\quad \bigg(\sum_{t=1}^T\Delta_{1,8}^\top\Qm_t\Delta_{1,7} + \sum_{t=1}^T \Delta_{t,6}^\top\Rm_t\Delta_{t,5} + \DD_\xi^\top\LAMm(\E[\XX_{t_0}] - \ff_\xi)\bigg)
\end{split}
\end{equation}

\subsubsection{When $\HH_t$ has 0 rows in addition to $\GG_t$}
When $\HH_t$ has all zero rows, some of the $\pp$ or $\uupsilon$ may constrained by the model, but these constraints do not appear in $\Psi^+$ since $\Rm_t$ zeros out those constraints.  For example, if $H_t$ is all zeros and $\xx_1 \equiv \xixi$, then $\xixi$ is constrained to equal $\ZZ^{-1}(\hatyone-\aa_1)$.    

The model needs to be internally consistent and we need to be able to estimate all the $\pp$ and the $\uupsilon$.  Rather than try to estimate the correct $\pp$ and $\uupsilon$ to ensure internal consistency of the model with the data when some of the $\HH_t$ have 0 rows, I test by running the Kalman filter with the degenerate variance modification (in particular the modification for $\FF$ with zero rows is critical) before starting the EM algorithm.  Then I test that $\hatyt-\ZZ_t\hatxt-\aa_t$ is all zeros.  If it is not, within machine accuracy, then there is a problem.  This is reported and the algorithm stopped\footnote{In some cases, it is easy to determine the correct $\xixi$.  For example, when $\HH_t$ is all zero rows, $t_0=1$ and there is no missing data at time $t=1$, $\xixi=\ZZ^*(\yy_1-\aa_1)$, where $\ZZ^*$ is the pseudoinverse. One would want to use the SVD pseudoinverse calculation in case $\ZZ$ leads to an under-constrained system (some of the singular values of $\ZZ$ are 0). }  

I also test that $\big(\sum_{t=1}^T\Delta_{t,8}^\top\Qm_t\Delta_{t,8} + \sum_{t=1}^T\Delta_{t,6}^\top\Rm_t\Delta_{t,6} + \DD_\xi^\top\LAMm\DD_\xi\big)$ is invertible to ensure that all the $\pp$ can be solved for, and I test that $\big(\sum_{t=1}^T\Delta_{t,4}^\top\Qm_t\Delta_{t,4} + \sum_{t=1}^T\Delta_{t,2}^\top\Rm_t\Delta_{t,2}\big)$ is invertible so that all the $\uupsilon$ can be solved for.  If errors are present, they should be apparent in iteration 1, are reported and the EM algorithm stopped.

\subsection{$\BB^{(0)}$ update equation for degenerate models}
I do not have an update equation for $\BB^{(0)}$ and for now, I side-step this problem by requiring that any $\BB^{(0)}$ terms are fixed.

\section{Kalman filter and smoother modifications for degenerate models}

\subsection{Modifications due to degenerate $\RR$ and $\QQ$}
[1/1/2012 note.  These modifications mainly have to do with inverses that appear in the Shumway and Stoffer's presentation of the Kalman filter.  Later I want to switch to Koopman's smoother algorithm which avoids these inverses altogether.]

In principle, when either $\GG_t\QQ_t$ or $\HH_t\RR_t$ has zero rows, the standard Kalman filter/smoother equations would still work and provide the correct state outputs and likelihood.  In practice however errors will be generated because under certain situations, one of the matrix inverses in the Kalman filter/smoother equations will involve a matrix with a zero on the diagonal and this will lead to the computer code throwing an error.  

When $\HH_t\RR_t$ has zero rows, problems arise in the Kalman update part of the Kalman filter.  
The Kalman gain is 
\begin{equation}\label{eq:KKt.2}
\KK_t = \VV_t^{t-1}(\ZZ_t^*)^\top(\ZZ_t^*\VV_t^{t-1}(\ZZ_t^*)^\top + \HH_t\RR_t^*\HH_t^\top)^{-1}
\end{equation}
Here, $\ZZ_t^*$ is the missing values modified $\ZZ_t$ matrix with the $i$-th rows zero-ed out if the $i$-th element of $\yy_t$ is missing (section \ref{sec:kalman.smoother}, equation \ref{eq:yaZ.miss}).  Thus if the $i$-th element of $\yy_t$ is missing and the $i$-th row of $\HH_t$ is zero, the $(i,i)$ element of $(\ZZ_t^*\VV_t^{t-1}(\ZZ_t^*)^\top + \HH_t\RR_t^*\HH_t^\top)$ will be zero also and one cannot take its inverse.  In addition, if the initial value $\xx_1$ is treated as fixed but unknown then $\VV_1^0$ will be a $m \times m$ matrix of zeros.  Again in this situation $(\ZZ_t^*\VV_t^{t-1}(\ZZ_t^*)^\top + \HH_t\RR_t^*\HH_t^\top)$ will have zeros at any $(i,i)$ elements where the $i$-th row of $\HH_t$ is also zero.

The first case, where zeros on the diagonal arise due to missing values in the data, can be solved using the  matrix which pulls out the rows and columns corresponding to the non-missing values ($\OMG_t^{(1)}$).  Replace $\big(\ZZ_t^*\VV_t^{t-1}(\ZZ_t^*)^\top + \HH_t\RR_t^*\HH_t^\top\big)^{-1}$ in equation \eqref{eq:KKt.2} with
\begin{equation}
(\OMG_t^{(1)})^\top\big(\OMG_t^{(1)}(\ZZ_t^*\VV_t^{t-1}(\ZZ_t^*)^\top + \HH_t\RR_t^*\HH_t^\top)(\OMG_t^{(1)})^\top\big)^{-1}\OMG_t^{(1)}
\end{equation}
Wrapping in $\OMG_t^{(1)}(\OMG_t^{(1)})^\top$ gets rid of all the zero rows/columns in $\ZZ_t^\prime\VV_t^{t-1}(\ZZ_t^\prime)^\top + \HH_t\RR_t^\prime\HH_t^\top$, and the matrix is reassembled with the zero rows/columns reinserted by wrapping in $(\OMG_t^{(1)})^\top\OMG_t^{(1)}$.  This works because $\RR_t^\prime$ is the missing values modified $\RR$ (section \ref{sec:missing}) and is block diagonal across the $i$ and non-$i$ rows/columns, and $\ZZ_t^\prime$ has the $i$-columns zero-ed out. Thus removing the $i$ columns and rows before taking the inverse has no effect on the product $\ZZ_t(...)^{-1}$. When $\VV_1^0=\mathbf{0}$, set $\KK_1=\mathbf{0}$ without computing the inverse (see equation \ref{eq:KKt.2} where $\VV_1^0$ appears on the left).

There is also a numerical issue to deal with.  When the $i$-th row of $\HH_t$ is zero, some of the elements of $\xx_t$ may be completely specified (fully known) given $\yy_t$.  Let's call these fully known elements of $\xx_t$, the $k$-th elements.  In this case, the $k$-th row and column of $\VV_t^t$ must be zero because given $y_t(i)$, $x_t(k)$ is known (is fixed) and its variance, $\VV_t^t(k,k)$, is zero.  Because  $\KK_t$ is computed using a numerical estimate of the inverse, the standard $\VV_t^t$ update equation (which uses $\KK_t$) will cause these elements to be close to zero but not precisely zero, and they may even be slightly negative on the diagonal. This will cause serious problems when the Kalman filter output is passed on to the EM algorithm. Thus after $\VV_t^t$ is computed using the normal Kalman update equation, we will want to explicitly zero out the $k$ rows and columns in the filter.  

When $\GG_t$ has zero rows, then we might also have similar numerical errors in $\JJ$ in the Kalman smoother.  The $\JJ$ equation is 
\begin{equation}\label{eq:Jt.2}
\begin{split}
\JJ_t &= \VV_{t-1}^{t-1}\BB_t^\top(\VV_t^{t-1})^{-1}\\
&\text{where }\VV_t^{t-1} = \BB_t \VV_{t-1}^{t-1} \BB_t^\top + \GG_t\QQ_t\GG_t^\top
\end{split}
\end{equation}
If there are zeros on the diagonals of ($\LAM$ and/or $\BB_t$) and zero rows in $\GG_t$ and these zeros line up, then if the $\BB_t^{(0)}$ and $\BB_T^{(1)}$ elements in $\BB_t$ are blocks\footnote{This means the following.  Let the rows where the diagonal elements in $\QQ$ equal zero be denoted $i$ and the the rows where there are non-zero diagonals be denoted $j$. The $\BB_t^{(0)}$ elements are the $\BB_t$ elements where both row and column are in $i$.  The $\BB_t^{(1)}$ elements are the $\BB$ elements where both row and column are in $j$.  If the $\BB_t^{(0)}$ and $\BB_t^{(1)}$ elements in $\BB$ are blocks, this means all the $\BB_t(i,j)$ are 0; no deterministic components interact with the stochastic components.}, there will be zeros on the diagonal of $\VV_t^t$.  Thus there will be zeros on the diagonal of $\VV_t^{t-1}$ and it cannot be inverted.  In this case, the corresponding elements of $\VV_t^T$ need to be zero since what's happening is that those elements are deterministic and thus have 0 variance.

We want to catch these zero variances in $\VV_t^{t-1}$ so that we can take the inverse.  Note that this can only happen when there are zeros on the diagonal of $\GG_t\QQ_t\GG_t^\top$ since $\BB_t\VV_{t-1}^{t-1} \BB_t^\top$ can never be negative on the diagonal since $\BB_t\BB_t^\top$ must be positive-definite and so is $\VV_{t-1}^{t-1}$.  The basic idea is the same as above.  We replace $(\VV_t^{t-1})^{-1}$ with:
\begin{equation}
(\OMG_{Vt}^+)^\top\big(\OMG_{Vt}^+(\VV_t^{t-1})(\OMG_{Vt}^+)^\top\big)^{-1}\OMG_{Vt}^+
\end{equation}
where $\OMG_{Vt}^+$ is a matrix that removes all the positive $\VV_t^{t-1}$ rows analogous to  $\OMG_t^{(1)}$.

\subsection{Modifications due to fixed initial states}
When the initial state of $\xx$ is fixed, then it is a bit like $\LAM=0$ although actually $\LAM$ does not appear in the model and $\xixi$ has a different interpretation.  

When the initial state of $\xx$ is treated as stochastic, then if $t_0=0$,  $\xixi$ is the expected value of $\xx_0$ conditioned on no data.  In the Kalman filter this means $\xx_0^0=\xixi$ and $\VV_0^0=\LAM$; in words, the expected value of $\xx_0$ conditioned on $\yy_0$ is $\xixi$ and the variance of $\xx_0^0$ conditioned on $\yy_0$ is $\LAM$.  When $t_0=1$, then $\xixi$ is the expected value of $\xx_1$ conditioned on no data.  In the Kalman filter this means $\xx_1^0=\xixi$ and $\VV_1^0=\LAM$.  Thus where $\xixi$ and $\LAM$ appear in the Kalman filter equations is different depending on $t_0$; the $\xx_t^t$ and $\VV_t^{t}$ initial condition versus the $\xx_t^{t-1}$ and $\VV_t^{t-1}$ initial condition.

When some or all of the $\xx_{t_0}$ are fixed, denoted the $\II_\lambda^{(0)}\xx_{t_0}$, the fixed values are not a random variables.  While technically speaking, the expected value of a fixed value does not exist, we can think of it as a random variable with a probability density function with all the weight on the fixed value.   Thus $\II_\lambda^{(0)}\E[\xx_{t_0}]=\xixi$ regardless of the data.  The data have no information for $\II_\lambda^{(0)}\xx_{t_0}$ since we fix $\II_\lambda^{(0)}\xx_{t_0}$ at $\II_\lambda^{(0)}\xixi$. If $t_0=0$, we initialize the Kalman filter as usual with $\xx_0^0=\xixi$ and $\VV_0^0=\FF\LAM\FF^\top$, where the fixed $\xx_{t_0}$ rows correspond to the zero row/columns in $\FF\LAM\FF^\top$.  The Kalman filter will return the correct expectations even when some of the diagonals of $\HH\RR\HH^\top$ or $\GG\QQ\GG^\top$ are 0---with the constraint that we have no purely deterministic elements in the model (meaning there are no errors terms from either $\RR$ or $\QQ$).  

When $t_0=1$, $\II_\lambda^{(0)}\xx_1^0$ and $\II_l^{(0)}\xx_1^1=\xixi$ regardless of the data and $\VV_1^0=\FF\LAM\FF^\top$ and $\VV_1^1=\FF\LAM\FF^\top$, where the fixed rows of $\xx_1$ correspond with the 0 row/columns in $\FF\LAM\FF^\top$.  We also set $\II_\lambda^{(0)}\KK_1$, meaning the rows of $\xx_1$ that are fixed, to all zero because $\KK_1$ is the information in $\yy_1$ regarding $\xx_1$ and there is no information in the data regarding the values of $\xx_1$ that are fixed to equal $\II_\lambda^{(0)}\xixi$.  

With $\VV_1^1$, $\xx_1^1$ and $\KK_1$ set to their correct initial values, the normal Kalman filter equations will work fine.  However it is possible for the data at $t=1$ to be inconsistent with the model if the rows of $\yy_1$ corresponding to any zero row/columns in $\ZZ_1\FF\LAM\FF^\top\ZZ_1^\top+\HH_1\RR_1\HH_1^\top$ are not equal to $\ZZ_1\xixi + \aa_1$.  Here is a trivial example, let the model be $x_t=x_{t-1}+w_t$, $y_t=x_t$, $x_1=1$.  Then if $y_1$ is anything except 1, the model is impossible.  Technically, the likelihood of $x_1$ conditioned on $Y_1=y_1$ does not exist since neither $x_1$ nor $y_1$ are realizations of a random variable (since they are fixed), so when the likelihood is computed using the innovations form of the likelihood, the $t=1$ does not appear, at least for those $\yy_1$ corresponding to any zero row/columns in $\ZZ_1\FF\LAM\FF^\top\ZZ_1^\top+\HH_1\RR_1\HH_1^\top$.  Thus these internal inconsistencies would neither provoke an error nor cause Inf to be returned for the likelihood.  In the MARSS package, the Kalman filter has been modified to return LL=Inf and an error.

\section{Summary of requirements for degenerate models}
Below are discussed the update equations for the different parameters.  Here I summarize the constraints that are scattered throughout these subsections.  These requirements are coded into the function MARSSkemcheck() in the MARSS package but some tests must be repeated in the function degen.test(), which tests if any of the $\RR$ or $\QQ$ diagonals can be set to zero if it appears they are going to zero.  A model that is allowed when $\RR$ and $\QQ$ are non-zero, might be disallowed if $\RR$ or $\QQ$ diagonals were to be set to zero.  degen.test() does this check.

\begin{itemize}
\item $(\II_m \otimes \II_r^{(0)}\ZZ_t\II_q^{(0)})\DD_{t,b}$,  is  all zeros; is all zeros. If there is a all zero row in $\HH_t$ and it is linked (through $\ZZ$) to a all zero row in $\GG_t$, then the corresponding $\BB_t$ elements are fixed instead of estimated.  Corresponding $\BB$ rows means those rows in $\BB$ where there is a non-zero column in $\ZZ$.  We need $\II_r^{(0)}\ZZ_t\II_q^{(0)}\BB_t$ to only specify fixed $\BB_t$ elements, which means $\vec(\II_r^{(0)}\ZZ_t\II_q^{(0)}\BB_t\II_m)$ only specifies fixed values.  This in turn leads to the condition above. MARSSkemcheck()
\item $(\II_1 \otimes \II_r^{(0)}\ZZ_t\II_q^{(0)})\DD_{t,u}$ is all zeros;  if there is a all zero row in $\HH_t$ and it is linked (through $\ZZ_t$) to a all zero row in $\GG_t$, then the corresponding $\uu_t$ elements are fixed instead of estimated.  MARSSkemcheck()
\item $(\II_m \otimes \II_r^{(0)})\DD_{t,z}$, where  is  all zeros; if $y$ has no observation error, then the corresponding $\ZZ_t$ rows are fixed values. $(\II_m \otimes \II_r^{(0)})$ is a diagonal matrix with 1s for the rows of $\DD_{t,z}$ that correspond to elements of $\ZZ_t$ on the $R=0$ rows. MARSSkemcheck()
\item $(\II_1 \otimes \II_r^{(0)})\DD_{t,a}$ is all zeros; if $y$ has no observation error, then the corresponding $\aa_t$ rows are fixed values. MARSSkemcheck()
\item $(\II_m \otimes \II_q^{(0)})\DD_{t,b}$ is all zeros.  This means $\BB^{(0)}$ (the whole row) is fixed.  While $\BB^d$ could potentially be estimated potentially, my derivation assumes it is not. MARSSkemcheck()
\item $(\II_1 \otimes \II_{q,t>m}^{is})\DD_{t,u}$ is all zeros.  This means $\uu^{is}$ is fixed.  Here $is$ is defined as those rows that are indirectly stochastic at time $m$, where $m$ is the dimension of $\BB$; it can take up to $m$ steps for the $is$ rows to be connected to the $s$ rows through $\BB$. MARSSkemcheck()
\item If $\uu^{(0)}$ or $\xixi^{(0)}$ are being estimated, then the adjacency matrices defined by $\BB_t \neq 0$ are not time-varying. This means that the locations of the 0s in $\BB_t$ are not changing over time. $\BB_t$ however may be time-varying. MARSSkemcheck()
\item $\II_q^{(0)}$ and $\II_r^{(0)}$ are time invariant (an imposed assumption).  This means that the location of the 0 rows in $\GG_t$ and $\HH_t$ (and thus in $\ww_t$ and $\vv_t$) are not changing through time.  It would be easy enough to allow $\II_r^{(0)}$ to be time varying, but to make my derivation easier, I assume it is time constant.
\item $\ZZ_t^{(0)}$ in $\E[\YY_t^{(0)}]=\ZZ_t^{(0)}\E[\XX_t]+\aa_t^{(0)}$ does not imply an over-determined system of equations.  Because the $\vv_t$ rows are zero for the ${(0)}$ rows of $\yy$, it must be possible for this equality to hold.  This means that $\ZZ_t^{(0)}$ cannot specify an over-determined system although an underdetermined system is ok.  MARSSkemcheck() checks by examining the singlular values of $\ZZ_t^{(0)}$ returned from the singlular value decomposition (svd).   The number of singlular values must not be less than $m$ (columns of $\ZZ$).  If it is less than $m$, it means the equation system is over-determined. Singular values equal to 0 are ok; it means the system is under-determined given only the observation equation, but that's ok because we also have the state equation will determine the under states and the Kalman smoother will presumably throw an error if the state process is under-determined (if that would even make sense...).
\item The state process cannot be over-determined via constraints imposed from the deterministic observation process ($\RR=0$) and the deterministic state process ($\QQ=0$).  If this is the case the Kalman gain equation (in the Kalman filter) will throw an error.  Checked in MARSS() via call to MARSSkf() before fitting call; degen.test(), in MARSSkem() will also test via MARSSkf call if some R or Q are attempted to be set to 0.  If B or Z changes during kem or optim iterations such that this constraint does not hold, then algorithm will exit with an error message.
\item The location of the 0s in $\BB$ are time-invariant.  The $\BB$ can be time-varying but not the location of 0s.  Also, I want $\BB$ to be such that once a row becomes indirectly stochastic is stays that way.  For example, if $\BB=\bigl[ \begin{smallmatrix}
0&1\\ 1&0 \end{smallmatrix} \bigr]$, then row 2 flips back and forth from being indirectly stochastic to deterministic.
\end{itemize}
The dimension of the identity matrices in the above constraints is given by the subscript on $\II$ except when it is implicit.

\section{Implementation comments}\label{sec:implementation}
The EM algorithm is a hill-climbing algorithm and like all hill-climbing algorithms it can get stuck on local maxima.  There are a number approaches to doing a pre-search of the initial conditions space, but a brute force  random Monte Carol search appears to work well \citep{Biernackietal2003}.  It is slow, but normally sufficient.  In my experience, Monte Carlo initial conditions searches become important as the fraction of missing data in the data set increases.  Certainly an initial conditions search should be done before reporting final estimates for an analysis.  However in our\footnote{``Our'' and ``we'' in this section means work and papers by E. E. Holmes and E.J. Ward.} studies on the distributional properties of parameter estimates, we rarely found it necessary to do an initial conditions search.

The EM algorithm will quickly home in on parameter estimates that are close to the maximum, but once the values are close, the EM algorithm can slow to a crawl.   Some researchers start with an EM algorithm to get close to the maximum-likelihood parameters and then switch to a quasi-Newton method for the final search.  In many ecological applications, parameter estimates that differ by less than 3 decimal places are for all practical purposes the same.  Thus we have not used the quasi-Newton final search.

Shumway and Stoffer (2006; chapter 6) imply in their discussion of the EM algorithm that both $\xixi$ and $\LAM$ can be  estimated, though not simultaneously.  Harvey (1989), in contrast, discusses that there are only two allowable cases for the initial conditions: 1) fixed but unknown and 2) a initial condition set as a prior. In case 1, $\xixi$ is $\xx_0$ (or $\xx_1$) and is then estimated as a parameter; $\LAM$ is held fixed at 0.  In case 2, $\xixi$ and $\LAM$ specify the mean and variance of $\XX_0$ (or $\XX_1$) respectively. Neither are estimated; instead, they are specified as part of the model.  

As mentioned in the introduction, misspecification of the prior on $\xx_0$ can have catastrophic and undetectable effects on your parameter estimates.  For many MARSS models, you will never see this problem.  However, if you are fitting models that imply a correlation structure between the hidden states (i.e. the variance-covariance matrix of the $\XX$'s is not diagonal), then your prior can definitely create problems if it does not have the same correlation structure as that implied by your MLE model.  A common default is to use a prior with a diagonal variance-covariance matrix.  This can lead to serious problems if the implied variance-covariance of the $\XX$'s is not diagonal.  A diffuse prior does not get around this since it has a correlation structure also even if it has infinite variance.  

One way you can detect that you have a problem is to start the EM algorithm at the outputs from a Newton-esque algorithm.  If the EM estimates diverge and the likelihood drops, you have a problem.  Here are a few suggestions for getting around the problem:
\begin{itemize}
	\item Treat $\xx_0$ as an estimated parameter and set $\VV_0$=0.  If the model is not stable going backwards in time, then treat $\xx_1$ as the estimated parameter; this will allow the data to constrain the $\xx_1$ estimate (since there is no data at $t=0$, $\xx_0$ has no data to constrain it).
	\item Try a diffuse prior, but first read the info in the KFAS R package about diffuse priors since MARSS uses the KFAS implementation.  In particular, note that you will still be imposing an information on the correlation structure using a diffuse prior; whatever $\VV_0$ you use is telling the algorithm what correlation structure to use.  If there is a mismatch between the correlation structure in the prior and the correlation structure implied by the MLE model, you will not be escaping the prior problem. But sometimes you will know your implied correlation structure.  For example, you may know that the $\xx$'s are independent or you may be able to solve for the stationary distribution a priori if your stationary distribution is not a function of the parameters you are trying to estimate.  Other times you are estimating a parameter that determines the correlation structure (like $\BB$) and you will not know a priori what the correlation structure is.
\end{itemize}
	
In some cases, the update equation for one parameter needs other parameters.  Technically, the Kalman filter/smoother should be run between each parameter update, however following \citet{GhahramaniHinton1996} the default MARSS algorithm skips this step (unless the user sets \verb@control$safe=TRUE@) and each updated parameter is used for subsequent update equations.  If you see warnings that the log-likelihood drops, then try setting \verb@control$safe=TRUE@.  This will increase computation time greatly.

\section{MARSS R package}
R code for the Kalman filter, Kalman smoother, and EM algorithm is provided as a separate R package, MARSS, available on CRAN (http://cran.r-project.org/web/packages/MARSS).  MARSS was developed by Elizabeth Holmes, Eric Ward and Kellie Wills and provides maximum-likelihood estimation and model-selection for both unconstrained and constrained MARSS models. The package contains a detailed user guide which shows various applications. In addition to model fitting via the EM algorithm, the package provides algorithms for bootstrapping, confidence intervals, auxiliary residuals, and model selection criteria.

\bibliography{./Manual}

\begin{thebibliography}{}

\bibitem[Biernacki et~al., 2003]{Biernackietal2003}
Biernacki, C., Celeux, G., and Govaert, G. (2003).
\newblock Choosing starting values for the {EM} algorithm for getting the
  highest likelihood in multivariate gaussian mixture models.
\newblock {\em Computational Statistics and Data Analysis}, 41(3-4):561--575.

\bibitem[Borman, 2009]{Borman2009}
Borman, S. (2009).
\newblock The expectation maximization algorithm - a short tutorial.

\bibitem[Ghahramani and Hinton, 1996]{GhahramaniHinton1996}
Ghahramani, Z. and Hinton, G.~E. (1996).
\newblock Parameter estimation for linear dynamical systems.
\newblock Technical Report CRG-TR-96-2, University of Totronto, Dept. of
  Computer Science.

\bibitem[Harvey, 1989]{Harvey1989}
Harvey, A.~C. (1989).
\newblock {\em Forecasting, structural time series models and the Kalman
  filter}.
\newblock Cambridge University Press, Cambridge, UK.

\bibitem[Henderson and Searle, 1979]{HendersonSearle1979}
Henderson, H.~V. and Searle, S.~R. (1979).
\newblock Vec and vech operators for matrices, with some uses in jacobians and
  multivariate statistics.
\newblock {\em The Canadian Journal of Statistics}, 7(1):65--81.

\bibitem[Johnson and Wichern, 2007]{JohnsonWichern2007}
Johnson, R.~A. and Wichern, D.~W. (2007).
\newblock {\em Applied multivariate statistical analysis}.
\newblock Prentice Hall, Upper Saddle River, NJ.

\bibitem[Koopman and Ooms, 2011]{KoopmanOoms2011}
Koopman, S. and Ooms, M. (2011).
\newblock {\em Forecasting economic time series using unobserved components
  time series models}, pages 129--162.
\newblock Oxford University Press, Oxford.

\bibitem[Koopman, 1993]{Koopman1993}
Koopman, S.~J. (1993).
\newblock Distrubance smoother for state space models.
\newblock {\em Biometrika}, 80(1):117--126.

\bibitem[McLachlan and Krishnan, 2008]{McLachlanKrishnan2008}
McLachlan, G.~J. and Krishnan, T. (2008).
\newblock {\em The {EM} algorithm and extensions}.
\newblock John Wiley and Sons, Inc., Hoboken, NJ, 2nd edition.

\bibitem[Roweis and Ghahramani, 1999]{RoweisGhahramani1999}
Roweis, S. and Ghahramani, Z. (1999).
\newblock A unifying review of linear gaussian models.
\newblock {\em Neural Computation}, 11:305--345.

\bibitem[Shumway and Stoffer, 2006]{ShumwayStoffer2006}
Shumway, R. and Stoffer, D. (2006).
\newblock {\em Time series analysis and its applications}.
\newblock Springer-Science+Business Media, LLC, New York, New York, 2nd
  edition.

\bibitem[Shumway and Stoffer, 1982]{ShumwayStoffer1982}
Shumway, R.~H. and Stoffer, D.~S. (1982).
\newblock An approach to time series smoothing and forecasting using the {EM}
  algorithm.
\newblock {\em Journal of Time Series Analysis}, 3(4):253--264.

\bibitem[Wu et~al., 1996]{Wuetal1996}
Wu, L. S.-Y., Pai, J.~S., and Hosking, J. R.~M. (1996).
\newblock An algorithm for estimating parameters of state-space models.
\newblock {\em Statistics and Probability Letters}, 28:99--106.

\bibitem[Zuur et~al., 2003]{Zuuretal2003a}
Zuur, A.~F., Fryer, R.~J., Jolliffe, I.~T., Dekker, R., and Beukema, J.~J.
  (2003).
\newblock Estimating common trends in multivariate time series using dynamic
  factor analysis.
\newblock {\em Environmetrics}, 14(7):665--685.

\end{thebibliography}
\bibliographystyle{apalike}

\end{document}